%% file: ANA-HIGG-2020-24-PAPER.tex
\documentclass[PAPER, atlasdraft=false,cernpreprint,UKenglish, texmf, orcidlogo]{atlasdoc}
\usepackage[subfigure]{atlaspackage}
\usepackage{atlasbiblatex}

\usepackage{atlasphysics}
\usepackage{multirow}
\usepackage{comment}
\usepackage{amsmath}

\graphicspath{{logos/}{figures/}}

\usepackage{atlascontribute}
\usepackage{ANA-HIGG-2020-24-PAPER-defs}
\usepackage{PMGRefs-defs}
\AtlasTitle{Measurement of the associated production of a top-antitop-quark pair and a Higgs boson  decaying into a $b\bar{b}$ pair in $pp$ collisions at $\sqrt{s}=13$~TeV using the ATLAS detector at the LHC}

\AtlasAbstract{%
This paper reports the measurement of Higgs boson production in association with a $t\bar{t}$ pair in the $H\rightarrow b\Bar{b}$ decay channel. The analysis uses 140~fb$^{-1}$ of 13 \TeV proton--proton collision data collected with the ATLAS detector at the Large Hadron Collider. The final states with one or two electrons or muons are employed.
An excess of events over the expected background is found with an observed (expected) significance of 4.6 (5.4) standard deviations. The $t\bar{t}H$ cross-section is $\sigma_{\ttH} = \xstotal~\text{fb} = \xsstatsyst~\text{fb}$ for a Higgs boson mass of 125.09 \GeV, consistent with the prediction of the Standard Model of $507^{+35}_{-50}$ \text{fb}.
The cross-section is also measured differentially in bins of the Higgs boson transverse momentum within the simplified template cross-section framework.
}

\AtlasRefCode{HIGG-2020-24}

\PreprintIdNumber{CERN-EP-2024-194}

\AtlasJournal{Eur. Phys. J. C}
\AtlasJournalRef{Eur. Phys. J. C 85 (2025) 210}
\AtlasDOI{10.1140/epjc/s10052-025-13740-x}

\hypersetup{pdftitle={ATLAS document},pdfauthor={The ATLAS Collaboration}}

\begin{document}

\maketitle

\tableofcontents

%
%

%

%

\section{Introduction}
\label{sec:intro}
After the discovery of the Higgs boson~\cite{Englert:1964aa,Higgs:1964aa,Guralnik:1964aa}
in 2012 by the ATLAS~\cite{HIGG-2012-27} and CMS~\cite{CMS-HIG-12-028} collaborations, attention has turned to detailed measurements of its properties and couplings as a means of testing the predictions of the Standard Model (SM)~\cite{Glashow:1961aa,Weinberg:1967aa,Salam:1969aa}.
The Higgs boson coupling to the top quark, the heaviest particle in the
SM, is of special interest as it could be very sensitive to effects of physics beyond the SM (BSM)~\cite{Englert:2014uua}. It is indirectly constrained assuming no BSM contributions to loop-induced processes from measurements of gluon--gluon fusion Higgs boson production and decay into $\gamma\gamma$~\cite{HIGG-2021-23,CMS-HIG-22-001}.
The
Higgs boson production in association with a pair of top quarks (\ttH),
where the top quark couples to the Higgs boson at tree level,
provides a possibility for a direct measurement of the top-quark's Yukawa coupling
without assumptions about the potential presence of BSM
physics~\cite{Ng:1983jm,Kunszt:1984ri,Dawson:2002tg,Beenakker:2001rj}. It was observed
by the ATLAS and CMS collaborations using several Higgs boson decay modes ~\cite{HIGG-2018-13,CMS-HIG-17-035}.

The decay into two $b$-quarks is predicted to have the largest branching fraction of about 58\%~\cite{deFlorian:2016spz} and has the advantage of allowing for the reconstruction of the Higgs boson four-momentum from the kinematics of its decay products. Furthermore, the \ttHbb channel involves only fermionic Higgs boson couplings in the production and decay, leading to an enhanced sensitivity for probing them. However, this final state is affected by a large irreducible background arising from the \ttbar production in associations with jets (\ttjets), in particular when the jets originate from $b$- or $c$-quarks, which is challenging to predict theoretically.

The ATLAS Collaboration has measured the \ttHbb production at
$\sqs = 13$~\TeV\ in the final state with at least one lepton using the full Run 2 dataset collected in proton--proton ($pp$) collisions at the Large Hadron Collider (LHC) \cite{Evans:2008zzb} between 2015 and 2018, corresponding to an integrated luminosity of
139~\ifb~\cite{HIGG-2020-23}. The measured signal strength, defined as the ratio of the measured cross-section to that predicted by the SM, was found to be  $0.35^{+0.36}_{-0.34}$, and corresponds to an observed (expected) significance of 1.0 (2.7) standard deviations. The signal strength was also measured differentially in five intervals of the Higgs boson transverse momentum (\ptH)
in the simplified template cross-section framework (STXS) ~\cite{deFlorian:2016spz}, probing potential
\ptH-dependent deviations from the SM expectation.

The CMS Collaboration also recently released a measurement of the \ttHbb production
using 138 \ifb of data collected at $\sqs = 13$~\TeV\ in 2016-2018~\cite{CMS-HIG-19-011}.
The analysis achieved an observed (expected) significance of 1.3 (4.1) standard deviation
and measured a signal strength of $0.33 \pm 0.26$. Additionally, the \ttH production
rate was determined in intervals of \ptH.

This paper presents a re-analysis of the full Run 2 dataset collected at $\sqrt{s}=13$~\TeV with the ATLAS detector
in final states with one or two leptons, referred to in the following as single-lepton and dilepton channels,
and supersedes the result of Ref.~\cite{HIGG-2020-23}.
Compared with the previous result, this analysis has an increased acceptance by selecting events with  less restrictive requirements on the number of jets identified as originating from $b$-hadrons ($b$-jets), resulting in an increased efficiency to select \ttH signal.
It utilises a revised  treatment of the different flavour components of the \ttjets background, and in particular, of the \ttbar production in association with $b$-jets, the main background and the dominant source of the systematic uncertainty in the previous \ttHbb measurement. For the modelling of this background, a new sample of Monte Carlo (MC) simulated events with improved settings was produced, and a corresponding set of systematic uncertainties was developed~\cite{ATL-PHYS-PUB-2022-006,ATL-PHYS-PUB-2022-026}. This measurement also uses improved analysis techniques: an advanced $b$-jet identification and an improved event categorisation. In particular, unlike the previous result, where the event categorisation into the background-dominated and signal-rich categories was made based on the number of jets and the number of $b$-jets, this analysis uses a multiclass neural network to define regions enriched in different components of the main background, the \ttjets production processes, and the signal. A more powerful multivariate discriminant is also employed to separate the signal from background and to reconstruct \ptH.
These improvements lead to better overall sensitivity and allow for a  measurement of the \ttHbb production cross-section in six bins of \ptH, 0\,--\,60~\GeV, 60\,--\,120~\GeV, 120\,--\,200~\GeV, 200\,--\,300~\GeV, 300\,--\,450~\GeV, and $\geq$\,450~\GeV, as obtained from the MC event record before the Higgs boson decays, in the STXS formalism.


%
%

\section{ATLAS detector}
\label{sec:atlas2a}
\newcommand{\AtlasCoordFootnote}{%
ATLAS uses a right-handed coordinate system with its origin at the nominal interaction point (IP)
in the centre of the detector and the \(z\)-axis along the beam pipe.
The \(x\)-axis points from the IP to the centre of the LHC ring,
and the \(y\)-axis points upwards.
Cylindrical coordinates \((r,\phi)\) are used in the transverse plane,
\(\phi\) being the azimuthal angle around the \(z\)-axis.
The pseudorapidity is defined in terms of the polar angle \(\theta\) as \(\eta = -\ln \tan(\theta/2)\).
Angular distance is measured in units of \(\Delta R \equiv \sqrt{(\Delta\eta)^{2} + (\Delta\phi)^{2}}\).}

The ATLAS experiment~\cite{PERF-2007-01} at the LHC is a multipurpose particle detector
with a forward--backward symmetric cylindrical geometry and a near \(4\pi\) coverage in solid angle.\footnote{\AtlasCoordFootnote}
It consists of an inner tracking detector (ID) surrounded by a thin superconducting solenoid
providing a \qty{2}{\tesla} axial magnetic field, electromagnetic and hadronic calorimeters, and a muon spectrometer.
The inner tracking detector covers the pseudorapidity range \(|\eta| < 2.5\).
It consists of silicon pixel, silicon microstrip, and transition radiation tracking detectors.
Lead/liquid-argon (LAr) sampling calorimeters provide electromagnetic (EM) energy measurements
with high granularity within the region \(|\eta|< 3.2\).
A steel/scintillator-tile hadronic calorimeter covers the central pseudorapidity range (\(|\eta| < 1.7\)).
The endcap and forward regions are instrumented with LAr calorimeters
for EM and hadronic energy measurements up to \(|\eta| = 4.9\).
The muon spectrometer (MS) surrounds the calorimeters and is based on
three large superconducting air-core toroidal magnets with eight coils each.
The field integral of the toroids ranges between \num{2.0} and \qty{6.0}{\tesla\metre}
across most of the detector.
The muon spectrometer includes a system of precision tracking chambers up to \(|\eta| = 2.7\) and fast detectors for triggering up to \(|\eta| = 2.4\).
The luminosity is measured mainly by the LUCID--2~\cite{LUCID2} detector, which is located close to the beampipe.
A two-level trigger system is used to select events~\cite{TRIG-2016-01}.
The first-level trigger is implemented in hardware and uses a subset of the detector information
to accept events at a rate below \qty{100}{\kHz}.
This is followed by a software-based trigger that
reduces the accepted event rate to \qty{1}{\kHz} on average
depending on the data-taking conditions.
A software suite~\cite{SOFT-2022-02} is used in data simulation, in the reconstruction
and analysis of real and simulated data, in detector operations, and in the trigger and data acquisition
systems of the experiment.


%

\section{Data and Monte Carlo simulation samples}
\label{sec:mcsamples}

This analysis was performed using the $pp$ collision data collected by the ATLAS detector between 2015 and 2018 at $\sqrt{s} = \SI{13}{\TeV}$.
After the application of data-quality requirements~\cite{DAPR-2018-01},
the dataset corresponds to an integrated luminosity of 140~\ifb.
Samples of simulated events were produced to model the different signal and background processes. Additional samples were produced to estimate the modelling uncertainties for each process.
The effects of the additional \pp\ collisions in the same or a nearby bunch crossing (pile-up) were modelled by overlaying minimum bias events generated with \pythia~\cite{Sjostrand:2007gs}
using the A3 set of tunable parameters~\cite{ATL-PHYS-PUB-2016-017} onto the simulated hard-scatter event.
The MC events were weighted to reproduce the distribution of the average number of interactions per bunch crossing observed in the data.
The MC samples were processed using the full ATLAS detector simulation~\cite{SOFT-2010-01} based on \GEANT~\cite{Agostinelli:2002hh}. Some alternative samples
used to evaluate the modelling uncertainties were produced using fast simulation, where the full \GEANT simulation of the calorimeter response is replaced by a detailed parameterisation of the shower shapes. For the observables used in this analysis, both simulations were found to give similar modelling. All MC samples were  reconstructed using the same software as for collider data. Corrections were applied to the simulated events so that the selection efficiencies, energy scales and energy resolutions of the physics objects closely match those determined from data control samples.

All samples generated with \powhegbox~\cite{Nason:2004rx,Frixione:2007nw,Frixione:2007vw,Alioli:2010xd} and \MGNLO~\cite{Alwall:2014hca} were interfaced to \pythia~\cite{Sjostrand:2014zea}
to simulate the parton shower (PS), fragmentation, and underlying event with the
A14 tune~\cite{ATL-PHYS-PUB-2014-021} and the \NNPDF[2.3lo]~\cite{Ball:2014uwa}
parton distribution function (PDF) set. Some alternative samples use
the \herwigseven~\cite{Bahr:2008pv,Bellm:2015jjp} PS model with the H7UE set
of tuned parameters~\cite{Bellm:2015jjp} and the \MMHT[lo]
PDF set~\cite{Harland-Lang:2014zoa}. Samples using \pythia and \herwigseven have heavy-flavour hadron decays modelled by EvtGen~\cite{Lange:2001uf}.
The masses of the top quark, \mtop, and of the Higgs boson, $m_\mathrm{H}$, are set to $172.5~\GeV$ and $125~\GeV$, respectively.

The nominal \ttH\ signal sample was generated at next-to-leading order (NLO)
in the five flavour scheme (5FS) using the \powhegbox~\cite{Hartanto:2015uka}
generator with the \NNPDF[3.0nlo]  ~\cite{Ball:2014uwa} PDF set.
The \hdamp\ parameter\footnote{The \hdamp\ parameter
controls the transverse momentum \pt\ of the first additional emission
beyond the leading-order Feynman diagram
in the PS and therefore regulates the  high-\pt\ emission against which the \ttbar\ system recoils.}
was set to $3/4\cdot (m_t + m_{\bar{t}} + m_H)=352.5$~\GeV,
and the functional form of the renormalisation and factorisation scales was set to $\sqrt[3]{m_\text{T}(t)\cdot m_\text{T}(\bar{t}) \cdot m_\text{T}(H)}$, where
$m_\text{T}=\sqrt{m^2+p_\text{T}^2}$ is the transverse mass of a
generated particle, $m$ is its mass, and $p_\text{T}$ is its transverse
momentum. It is normalised to the theoretical prediction
of Ref.~\cite{deFlorian:2016spz} computed at NLO in QCD.\footnote{A new theoretical
computation of the \ttH production at next-to-next-to-leading order in QCD was
published recently in Ref.~\cite{Catani_2023}.}

An alternative \ttH sample generated with the same \powhegbox set-up,
but interfaced to \herwigseven is used to evaluate uncertainties due to the choice of parton shower, hadronisation and underlying event model.
The uncertainty related to the matching between the matrix element (ME)
generator and the PS is accessed by changing the definition of the
hardness of the \powheg emission calculated by \pythia via the
parameter \pthard from the  value provided by \powheg to the \pt of the \powheg
emission following Refs.~\cite{ATL-PHYS-PUB-2023-029,Hoche:2021mkv}.

Several MC samples with different accuracy in ME generator and with different PS models are used in this analysis to model the main \ttjets\ background.
This background is categorised according to the flavour of the additional jets in the event, excluding  jets from top-quark or $W$ boson decays, using the same procedure as described in Ref.~\cite{HIGG-2013-27}.
Generator-level particle jets are reconstructed from stable particles (mean lifetime $\tau>3\times10^{-11}$ seconds, excluding muons and neutrinos)
using the anti-$k_t$ algorithm ~\cite{Cacciari:2008gp} with a radius parameter $R=0.4$, and are required to have transverse momentum $\pt>15$~\GeV\ and $|\eta|<2.5$.
The flavour of a jet is determined by counting $b$- or $c$-hadrons within $\Delta R<0.4$ of the jet axis.
Jets matched to exactly one $b$-hadron, with \pt above 5~\GeV, are labelled single-$b$-jets,
while those matched to two or more $b$-hadrons are labelled $B$-jets
(with no \pt\ requirement on the second $b$-hadron); single-$c$- and $C$-jets are
defined analogously, only considering jets not already defined as single-$b$- or $B$-jets.
Events that have a single-$b$- or $B$-jet, are labelled as \ttb and \ttB respectively, events
with two or more $b$-jets are labelled as \ttbb. These three categories together are collectively referred to as \ttbin. Events with no single-$b$- or $B$-jet but at least one single-$c$- or $C$-jet are labelled as \ttcin. Finally, events not containing any heavy-flavour jets aside from those from top-quark or $W$ boson decays are labelled as \ttlight.

The \ttlight and \ttcin contributions are modelled by a MC sample produced with the HVQ programme~\cite{Frixione:2007nw} in the \powhegbox~
generator at NLO in QCD in the five flavour scheme (5FS) with the \NNPDF[3.0nlo] PDF set.
The \hdamp\ parameter was set to
1.5~\mtop~\cite{ATL-PHYS-PUB-2014-021,ATL-PHYS-PUB-2016-020}.
An additional sample was generated with the \hdamp parameter increased by a factor of two to evaluate the uncertainty in the modelling of \ttlight and \ttcin stemming from the choice of the \hdamp value.

To predict the \ttbin\ background with the highest available precision, the \ttbbMC\ MC sample is generated at NLO, where the additional $b$-quarks are included in the ME. The four flavour scheme (4FS) was used for this sample.
It was produced with the \powhegboxRes~\cite{Jezo:2018yaf} generator and \openloops~\cite{Buccioni:2019sur,Cascioli:2011va,Denner:2016kdg}, using the implementation of this process in \powhegboxRes\ ~\cite{ttbbPowheg},
with the \NNPDF[3.1nnlonf4]~\cite{Ball_2017} PDF set interfaced to \pythia. Based on the studies of Ref.~\cite{ATL-PHYS-PUB-2022-006}, the factorisation scale is set to $\frac{1}{2}\sum_{i=t,\bar{t},b,\bar{b},j} m_{\mathrm{T},i}$,
the renormalisation scale is set to $\frac{1}{2} \cdot \sqrt[4]{m_{\text{T}}(t)\cdot m_{\text{T}}(\bar{t})\cdot m_{\text{T}}(b)\cdot m_{\text{T}}(\bar{b})}$, and the \hdamp\ parameter is set to $H_{\mathrm{T}}/2$. The \powheg\ internal parameter $h_\text{bzd}$ that regulates the damping function together with the parameter \hdamp, was set to 5.
The choice of model used for the recoil in the initial state  parton shower was found to impact significantly the \ttbin background predictions~\cite{ATL-PHYS-PUB-2022-026}. The corresponding uncertainty is evaluated by changing the \pythia\ parameter from a global recoil to a dipole recoil while keeping the rest of the settings identical to the nominal \ttbbMC sample.

Similar to the \ttH signal sample, alternative samples were generated to assess the PS and matching uncertainties in the modelling of \ttlight, \ttcin and \ttbin by replacing in the corresponding nominal set-ups the \pythia shower model by \herwig and producing samples with a varied \pthard parameter.

For an independent pseudo-data test, additional samples of \ttbar events were produced to model \ttlight and \ttcin events with the \SHERPA~\cite{Bothmann:2019yzt} generator. The NLO-accurate matrix elements for up to one additional parton, and LO-accurate matrix elements for up to four additional partons were calculated with the {\textsc{Comix}}~\cite{Gleisberg:2008fv} and \OPENLOOPS~libraries. They were matched with the \SHERPA PS~\cite{Schumann:2007mg} with the default set of tuned parameters using the \MEPSatNLO
prescription~\cite{Hoeche:2011fd,Hoeche:2012yf,Catani:2001cc,Hoeche:2009rj} with a matching scale of
\qty{30}{\GeV}. Furthermore, \ttbin  events were simulated using \SHERPA and \openloops with \ttbbMC  ME calculated at NLO accuracy using {\textsc{Comix}} in the 4FS using the same functional form of the factorisation and renormalisation scales as in the nominal \ttbbMC set-up.

Single-top-quark production processes, i.e. $tW$ associated production,
t-channel and s-channel production, were modelled using the \powhegbox~\cite{Re:2010bp,Alioli:2009je,Frixione:2008yi} generator at NLO in QCD.
The t-channel process was generated in the 4FS with the
\NNPDF[3.0nlonf4] PDF set, while for $tW$ and s-channel processes the 5FS \NNPDF[3.0nlo] PDF was used.
The $tW$ MC sample was generated using the factorisation and renormalisation scales set to $H_\text{T}/2$ with $H_\text{T}$ defined as a sum of the transverse mass of the $W$ boson, the top quark and the transverse momentum of an additional parton. The overlap between $tW$ and \ttbar\ production~\cite{Frixione:2008yi} was removed using the diagram-removal scheme~\cite{ATL-PHYS-PUB-2021-042}. An alternative $tW$ MC sample implementing the diagram subtraction scheme~\cite{Frixione:2008yi} was produced to evaluate the systematic uncertainty due to the $tW$ and \ttbar\ interference treatment.

The events in the nominal \ttW sample were simulated using the \SHERPA~\cite{Gleisberg:2008ta} generator with the \NNPDF[3.0nnlo] PDF set. The ME was calculated for up to one additional parton at NLO and up to two partons at LO using \textsc{Comix} and \OPENLOOPS, and merged with the \SHERPA PS using the \MEPSatNLO prescription with a merging scale of 30~\GeV. The alternative sample was generated with \mgamc with up to one additional parton in the final state at NLO accuracy in the strong coupling, using the \NNPDF[3.1nnlo]~\cite{Ball_2017} PDF set. The different jet multiplicities were merged using the FxFx NLO ME and PS merging prescription~\cite{Frederix:2012ps} with a merging scale of 30~\GeV.
Background events from $t\bar{t}Z/\gamma^\ast$ and the rare processes \tZq, \tWZ, \tHjb, \tWH and \tttt were simulated at
NLO in QCD using the \MGNLO generator. The alternative sample used to evaluate the PS uncertainty in the \ttZ background was generated with the same set-up as the nominal but interfaced to \herwigseven.

The production of \Vjets events (where $V=W$ or $Z$) was simulated with the \SHERPA generator using NLO-accurate matrix elements for up to two partons and LO-accurate matrix elements for up to four partons. Samples of diboson final states ($VV$) were also simulated with the \SHERPA  generator.

All MC samples corresponding to small backgrounds were normalised to the most precise available theoretical predictions closely following Ref.~\cite{HIGG-2020-23}. The normalisation of the \ttW background was updated to the most recent prediction of Ref.~\cite{Buonocore_2023}.
The \ttlight and \ttcin components were  normalised to the \ttbar cross-section computed at next-to-next-to-leading order (NNLO) in QCD including the resummation of next-to-next-to-leading-logarithmic (NNLL) soft-gluon terms~\cite{Czakon:2011xx} while the \ttbin normalisation is taken from the \ttbbMC MC simulation.


\section{Objects and event selection}
\label{sec:selection}
Events are selected using single-lepton triggers with variable electron and
muon transverse momentum thresholds, and various identification and isolation criteria
depending on the lepton flavour and the data-taking period~\cite{TRIG-2018-01,TRIG-2018-05}. The lowest \pt threshold at trigger level used for muons is $20$~\GeV ($26$~\GeV), while for electrons the threshold is $24$~\GeV ($26$~\GeV) in 2015 (2016\,--\,2018).
Events are required to have at least one vertex with at least two associated ID tracks with $\pt > 0.5$~\GeV.
In each event, the primary vertex is defined as the reconstructed vertex having the highest scalar
sum of squared \pt of associated tracks~\cite{ATL-PHYS-PUB-2015-026}
among the vertices consistent with the average beam-spot position.

Electron candidates are reconstructed from energy deposits in the electromagnetic calorimeter matched to tracks reconstructed in the ID system and are required to satisfy the $\textit{MediumLH}$ identification criterion~\cite{EGAM-2018-01}. They are required to have $\pt>10$~\GeV and $\left|\eta\right|<2.47$, excluding the calorimeter barrel-endcap transition region ($1.37 < \left|\eta\right| < 1.52$).
Muon candidates are reconstructed from tracks in the MS that are associated with tracks from the ID and are required to satisfy the $\textit{Loose}$ identification criterion~\cite{MUON-2018-03} and to have $\pt>10$~\GeV, $\left|\eta\right|<2.5$.
Electron (muon) candidates must be associated with the primary vertex
of the event: the transverse impact parameter divided by its estimated
uncertainty, ${|\dzero|/\sigma(\dzero)}$, is required to be less than five
(three) for electron (muon) candidates. The longitudinal impact parameter
must satisfy ${|\zzsth|<0.5}$~mm for both lepton flavours.

Jets are reconstructed from particle flow objects~\cite{PERF-2015-09} using the anti-$k_t$ jet clustering algorithm in the \textsc{FastJet} implementation~\cite{Fastjet} with a radius parameter $R=0.4$. Jets are required to satisfy the $\textit{Tight}$ criterion of the \textit{jet vertex tagger} (JVT) algorithm~\cite{PERF-2014-03} to mitigate the contribution from pile-up jets, and to have $\pt>25$~\GeV and $\left|\eta\right|<2.5$. The jet energy scale (JES) and resolution are calibrated using simulations with in situ corrections obtained from data~\cite{JETM-2018-05}. Events containing jets originating from non-collision sources or detector noise are removed.

Jets containing $b$-hadrons, referred to as \bjets, are identified using the DL1r \btag algorithm~\cite{ATLAS:2022qxm} that uses a neural network based on the distinctive features of $b$-hadron decays, primarily the impact parameters of tracks and the displaced vertices reconstructed in the ID. Additional input to this network is provided by discriminating variables constructed
by a recurrent neural network that exploits the spatial and kinematic correlations between tracks originating from the same $b$-hadron. A multivariate $b$-tagging discriminant value is calculated for each jet.
The $b$-tagged jets are required to satisfy the working point (WP) corresponding to an efficiency of 70\% or 85\% for identifying $b$-quark initiated jets in \ttbar simulated events for the single-lepton and dilepton channels, respectively.
For the 70\% (85\%) WP the rejection factors against light-quark/gluon jets and $c$-quark jets are 625 and 12 (40 and 3), respectively.
To fully exploit the $b$-tagging information of an event, each jet is
assigned a $b$-tagging score that defines if a jet satisfies a given WP but
fails to satisfy the adjacent tighter one.
In addition to the standard JES calibration, $b$-tagged jets satisfying the 85\% WP receive additional flavour-specific corrections to their four-vectors to improve their energy measurement (scale and resolution) following the \emph{muon-in-jet} procedure described in Ref.~\cite{HIGG-2016-29}.

Hadronically decaying $\tau$-leptons ($\tau_{\kern0.1em\relax\text{had}}$) are distinguished from jets using their track multiplicity and a multivariate discriminant based on their shower shapes in calorimeter and on tracking information~\cite{ATL-PHYS-PUB-2019-033}. They are required to have $\pt > 25$~\GeV and $\left|\eta\right| < 2.5$, and to pass the \textit{Medium} $\tau$-lepton identification working point.

The missing transverse momentum vector, with magnitude \met, is defined as the negative sum of the transverse momenta of the reconstructed and calibrated physical objects, plus a `soft term’ built from all other tracks associated with the primary vertex~\cite{JETM-2020-03} and not matched to a reconstructed object.

Targeting event topologies with a Higgs boson decaying into collimated hadronic final states, reclustered (RC) jets~\cite{Nachman:2014kla} are reconstructed from the selected jets, using the anti-$k_t$ jet clustering algorithm with a radius parameter $R=1.0$. RC jets are required to have an invariant mass $M>50$~\GeV, $\pt>200$~\GeV, $\left|\eta\right|<2.0$, have at least two constituent small-$R$ jets (subjets) and have an angular distance $\Delta R>1.0$ from all electrons.

An overlap removal procedure is applied to avoid the double counting of detector signatures. Electron candidates sharing a track with a muon candidate are first removed. Jets found within a $\Delta R$ = 0.2 cone of an electron are removed and electrons within a $\Delta R = 0.4$ cone of a remaining jet are rejected. Jets with less than three associated tracks and within
$\Delta R = 0.2$ of a muon and muons within $\Delta R = 0.4$ of a jet with more than two associated tracks are rejected. A $\tau_{\kern0.1em\relax\text{had}}$ candidate is rejected if it is separated by $\Delta R < 0.2$ from any selected electron or muon. No overlap removal is performed between jets and $\tau_{\kern0.1em\relax\text{had}}$ candidates.

Events are selected if they contain exactly one lepton candidate in the single-lepton channel
and exactly two lepton candidates with opposite electric charges in the dilepton channel.
At least one lepton should have $\pt>27$~\GeV and be matched to the corresponding object at the trigger level. In the events with two electrons, the second lepton is required to have $\pt>15$~\GeV while in the other two dilepton channels, $e\mu$ and $\mu\mu$, it must have $\pt>10$~\GeV. In events with two electrons or two muons, the dilepton invariant mass is required to be above $15$~\GeV, to suppress contribution from the decays of heavy-flavour resonances and low-mass Drell–Yan processes,
and be outside a window of $\pm$~8~\GeV centred at the $Z$ boson mass.
To maintain orthogonality with other \ttH decay channels, events are vetoed if they contain one or more (two or more) $\tau_{\kern0.1em\relax\text{had}}$ candidates in the dilepton (single-lepton) channel.

To reduce the non-prompt and mis-identified lepton background contribution,
additional lepton identification and isolation requirements are applied.
Electrons (muons) are required to satisfy the $\textit{TightLH}$ ($\textit{Medium}$)
identification criteria and the tight isolation criteria based on the calorimeter and
tracking information for electrons~\cite{EGAM-2018-01} and on tracking information
only for muons~\cite{MUON-2018-03}.
Events for which the leptons fail to meet these requirements are removed.

In the single-lepton channel two event categories are defined, referred in the following as resolved and boosted categories. The latter is designed to select events in which the Higgs boson is produced with high transverse momentum as a collimated large-$R$ jet. Events that do not satisfy the boosted category selection are assigned to the resolved one. In the single-lepton resolved channel, events are selected if they contain at least five jets, at least three of which satisfy the 70\% \btag WP. In the single-lepton boosted channel, events are required to have at least one large-$R$ jet and at least four small-$R$ jets, including those contained within the large-$R$ jet, at least three of which must satisfy the 85\% \btag WP.
In the dilepton channel, events are selected if they contain at least three jets
satisfying the 85\% \btag WP, and among these at least two jets that satisfy the 70\% \btag WP.
These requirements are henceforth referred to as ``preselection'' and have an acceptance of 6.3\% for selecting \ttH events with \hbb decay, an acceptance that is more than a factor of three larger than in the previous analysis~\cite{HIGG-2020-23}.
The corresponding acceptance in the signal regions is 2.1\%.


\section{Background modelling}
\label{sec:background}
The largest background from \ttjets production is modelled by MC simulation with the data-driven corrections described in Section~\ref{sec:ttjets}.
Small backgrounds include single top production, \ttW, \ttZ, \tttt, rare top-quark processes and non-top-quark processes such as \Vjets and diboson production. All of them are estimated from simulations.
The contribution from the background arising from non-prompt or mis-identified leptons is determined from data in the single-lepton channel (see Section \ref{sec:misid}),
and from MC simulation in the dilepton channel, where this background arises primarily from \ttbar\ events with one prompt lepton and is very small.

\subsection{Modelling of \ttjets background}
\label{sec:ttjets}
The \ttbar\ sample used to model the \ttlight\ and \ttcin\ contributions is generated at NLO as described in Section~\ref{sec:mcsamples}, with up to one additional parton from the ME calculation. All jets not originating from the decay chain of one of the top quarks, i.e. additional jets, are produced by the parton shower. The dedicated measurement~\cite{TOPQ-2018-15} demonstrated deficiencies in modelling the number of additional jets in the \ttbar\ events  and the scalar sum of the transverse momenta of the top quarks.
Moreover, the rate of \ttbar\ production in association with $c$- and $b$-jets was observed to be underestimated in the previous analysis~\cite{HIGG-2020-23}. The new \ttbbMC MC sample produced for this analysis with lower value of the renormalisation scale than the one used in Ref.~\cite{HIGG-2020-23} predicts larger inclusive \ttbin cross-section, which is expected to match data well. Nevertheless, the rate of the components of the \ttbin background defined in Section~\ref{sec:mcsamples}, \ttB, \ttb and \ttbb, could still be mismodelled. Thus, to obtain a reliable estimate of the \ttjets\ background, the \ttbar\ MC events are reweighted  using data. The corrections applied include a rescaling of the
\ttjets flavour components followed by a kinematic reweighting and are described
in the next paragraph. Any possible residual mismodelling is accounted for by
the systematic uncertainties in the profile-likelihood fit used for the extraction of the signal strength which is described in Section \ref{sec:strategy}.

Initial scaling factors for the five \ttjets flavour components,
\ttlight, \ttcin, \ttB, \ttb and \ttbb, are obtained separately for the
single-lepton and dilepton channels from the profile-likelihood fit to data.
The fit is performed in background-dominated control regions including all
systematic uncertainties (see Section~\ref{sec:systematics}) prior to the signal
extraction fit described in Section~\ref{sec:strategy}.
These scaling factors have similar values to those shown in Table~\ref{tab:norm_factors}.
Following the flavour components rescaling, a reweighting is used to mitigate the kinematic mismodelling of the scalar sum of the \pt's of the reconstructed leptons and jets, \HT, observed for the \ttlight and \ttcin components of the inclusive \ttjets \powheg+ \pythia sample. The \ttbin component is excluded from the reweighting since it is simulated by a dedicated \ttbbMC MC sample. Thus the reweighting derived in the regions dominated by \ttlight and \ttcin might not be applicable to \ttbin. A dedicated reweighting was investigated for the \ttbin component but found to have a negligible effect and is not used in the analysis.

The reweighting corrects the distributions of \HT in exclusive jet multiplicity bins from $\Nj = 5$ to $\Nj \ge 8$ in the single-lepton channel and from $\Nj = 3$ to $\Nj \ge 6$ in the dilepton channel.
It is derived in \ttbar enriched regions selected by using looser $b$-tagging requirements after subtracting from data all contributions except for \ttlight and \ttcin and taking the ratio of data to the sum of \ttlight and \ttcin yields predicted by MC in each \HT bin. These regions are orthogonal to the signal and control regions of the analysis and contain less that 0.2\% of signal.

The reweighting factors are also derived in the same way for each systematic variation affecting the \ttlight and \ttcin predictions, such that all systematic variations match the \HT distribution in data.
\subsection{Non-prompt or mis-identified lepton background}
\label{sec:misid}
A data-driven method, referred to as ``fake factor''
method~\cite{EGAM-2019-01}, based on the measurement of lepton selection efficiencies using different identification and isolation criteria, is used to estimate the non-prompt or mis-identified (fake) lepton background in the single-lepton channel.

The fake rate is measured in a fake-dominated region selected by requiring exactly one lepton with loose identification and isolation criteria, at least two jets, at least two $b$-tagged jets satisfying the 70\% WP, and a scalar sum of missing transverse energy and the leptonically-decaying $W$ boson mass below 60~\GeV.
It is parameterised as a function of the leading jet \pt and the lepton $\left|\eta\right|$.
The expected number of events arising from the fake lepton  background is determined by applying the measured lepton fake rate
to data events satisfying the selection requirements of each analysis region except that the lepton is required to pass loose identification and isolation criteria and to fail the tighter requirements.
%
%
%


\section{Signal extraction}
\label{sec:strategy}

After the preselection, events are classified into non-overlapping background-dominated or control regions (CR) and signal-enriched regions (SR) with higher signal-to-background ratios. The CRs provide stringent constraints on the normalisation of the backgrounds and the systematic uncertainties in a combined fit with the signal regions.

To define the regions in the single-lepton and dilepton channels, a multiclass classification neural network based on the permutation-invariant transformer architecture
with attention mechanism~\cite{vaswani2023attention} is trained to predict the probability of an event to be from the \ttH signal or from one of the five \ttjets background categories introduced in Section~\ref{sec:mcsamples}: \ttb, \ttB, \ttbb, \ttcin, \ttlight jets.
The probability $p_i$ of a network class $i$ is converted into a discriminant $d_i$ in order to maximise the separation between the class and all the other classes ($i\neq j$), to yield a similar number of events in the \ttb, \ttB, and inclusive \ttbb control regions, and to maximise sensitivity. They are defined as
\begin{equation}
d_i = \frac{p_i}{\sum_{j \neq i} p_j \cdot \hat{N}_{ij}}.
\end{equation}
Here, the denominator of each discriminant is a weighted average of all the remaining class probabilities $p_{j \neq i}$.
The weights $\hat{N}_{ij}={N_j}/{\sum_{k \neq i} N_k}$ are the respective fractions of event yields $N_j$ relative to the total yield of all the remaining classes $k \neq i$. These yields are determined from MC simulation after the preselection and incorporate the \ttjets\ scaling factors described in Section~\ref{sec:ttjets}.

In the single-lepton (dilepton) channel, events fulfilling $d_{\ttH}>4.072$ ($d_{\ttH}>9.031$) are assigned to the signal region. The thresholds are defined to maximize the ratio of signal to the square root of the background in the inclusive signal regions.
All other events are assigned to the \ttjets background category with the highest value of the discriminant.
By far the dominant Higgs boson decay mode in the signal region is $H\rightarrow b\bar{b}$ with a fraction of more than 97\% (94\%) of the \ttH events in the single-lepton (dilepton) channel followed by the $H\rightarrow WW$ decay.
This motivates the training of a second neural network, based on the same
architecture, to identify the two jets most likely originating from the Higgs boson decay, to reconstruct \ptH that is used to further classify the signal region events into the six reconstructed STXS regions.
The performance of this reconstruction network to correctly identify both Higgs boson decay products decreases with increasing jet multiplicity and generally increases with increasing true \ptH.
Both Higgs boson decay products are correctly identified for 51\% (58\%) of \ttHbb events in the single-lepton resolved (dilepton) channel.
The boundaries of the regions are optimised to maximise fraction of the events with the corresponding truth \ptH, listed in Figure~\ref{fig:transformer_flowchart}, separately per \ttbar decay channel to capture potential differences due to the final states and separate neural networks.
This optimisation leads, in particular, to improved purity in the highest truth \ptH bins, the region of the phase space where this analysis provides a particularly valuable contribution to the global Higgs boson combination.
The optimal reconstructed \ptH boundaries are found to be
[0, 60, 114, 186, 270, 402, $\infty$]~\GeV in the dilepton channel and
[0, 60, 114, 192, 282, 408, $\infty$]~\GeV in the single-lepton resolved channel. They are referred to in the following as STXS 1\,--\,6 regions.

The inclusive control region dominated by the \ttbb background in the single-lepton channel is further split into three categories encompassing STXS regions 1\,--\,2, 3\,--\,4, and 5\,--\,6, to ensure a good control of the \ttbb background modelling over the bins of reconstructed Higgs boson \pt. These regions are henceforth referred to as \ttbb CR 1, CR 2, and CR 3. The split is not applied in the dilepton channel due to limited statistics.

Events fulfilling the boosted selection criteria are assigned to dedicated boosted signal and control regions.
RC jets are used as input to a multiclass deep neural network (DNN), trained to distinguish high-\pt Higgs boson candidates decaying into collimated final states from top quarks and multijet backgrounds following the same strategy as in the previous analysis~\cite{HIGG-2020-23}.
An event is flagged as containing a boosted Higgs boson candidate if one of the RC jets has a high probability of originating from a Higgs boson, as estimated by the DNN. Boosted Higgs boson candidates are required to have $\pt \geq 300$~\GeV, a mass consistent with the Higgs boson mass window of $100$~\GeV to $140$~\GeV, contain at least two subjets, of which exactly two are required to satisfy the \btag 85\% WP, and have a DNN score above $0.4$.
At least two small-$R$ jets that do not form the Higgs boson candidate are required to satisfy the \btag 77\% WP.
The flowchart in Figure \ref{fig:transformer_flowchart} summarises the event classification strategy.

\begin{figure}[t!]
\centering
\includegraphics[width=\linewidth]{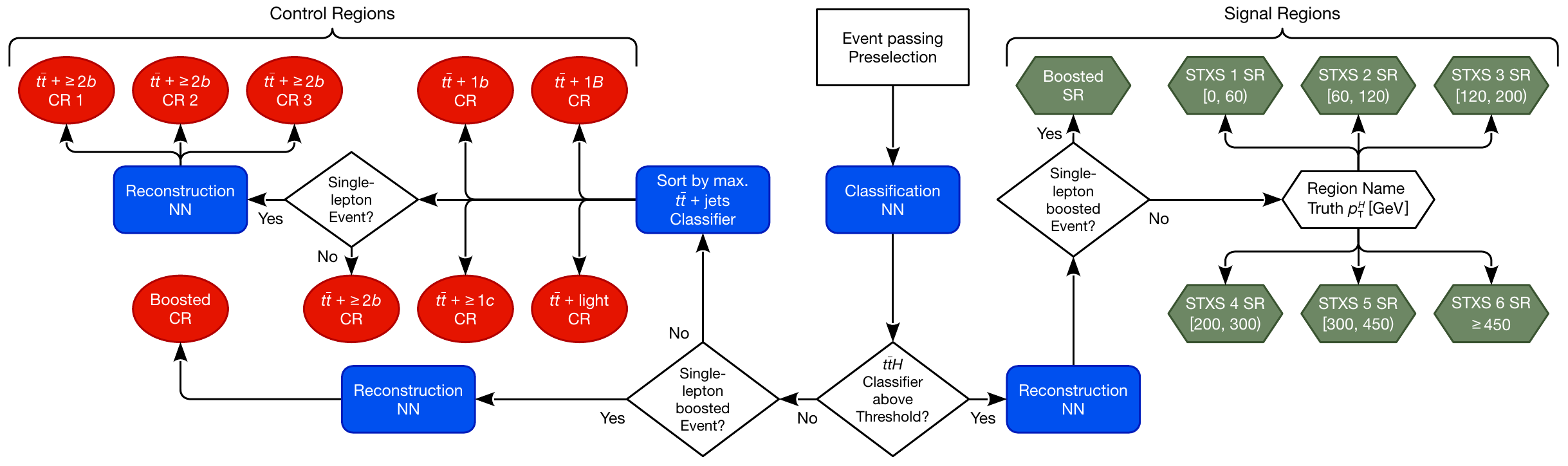}
\caption{Flowchart depicting the event classification and Higgs boson \pt reconstruction strategy used in the analysis.}
\label{fig:transformer_flowchart}
\end{figure}

The transformer networks are trained using low-level features of the reconstructed jets, electrons, muons, and the missing transverse momentum of the event. For each reconstructed object, the kinematic features used are the $x$-, $y$- and $z$-component of the momentum ($p_x$, $p_y$, $p_z$), the energy, the $\pt$, the mass, the pseudorapidity, the azimuthal angle and its sine and cosine. Supplying redundant four-vector information is observed to improve network performance.
Additionally, the DL1r pseudo-continuous \btag score is included for jets, and the electric charge
and a variable indicating if it is an electron or a muon is included for leptons.

The multiclass event-classification network is trained using events sampled from the nominal and alternative \ttH and \ttjets background samples to profit from an increased number of events in training.
The classification network yields an area under the receiver-operator characteristic curve (AUC) of $0.753$ ($0.774$) for discriminating between \ttH and the \ttbin background for events satisfying the preselection in the single-lepton resolved (dilepton) channel. The Higgs boson \pt reconstruction network is trained using only the \ttH samples.
Figure~\ref{fig:stxs_migration_matrix} shows for each bin of truth \ptH the fraction of \ttH events assigned to each of the STXS signal regions.

\begin{figure}[t!]
\centering
\subfigure[]{\includegraphics[width=0.32\textwidth]{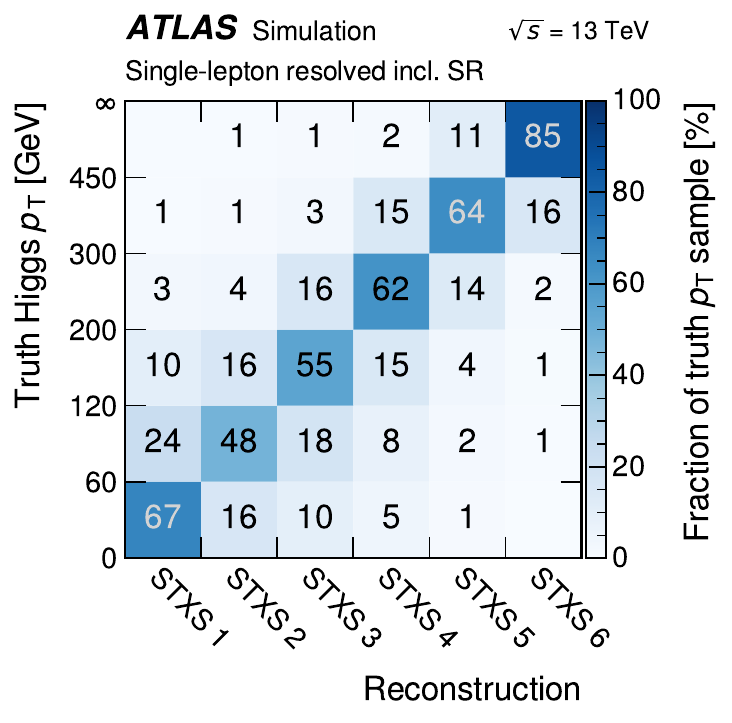}}
\hfill
\subfigure[]{\includegraphics[width=0.32\textwidth]{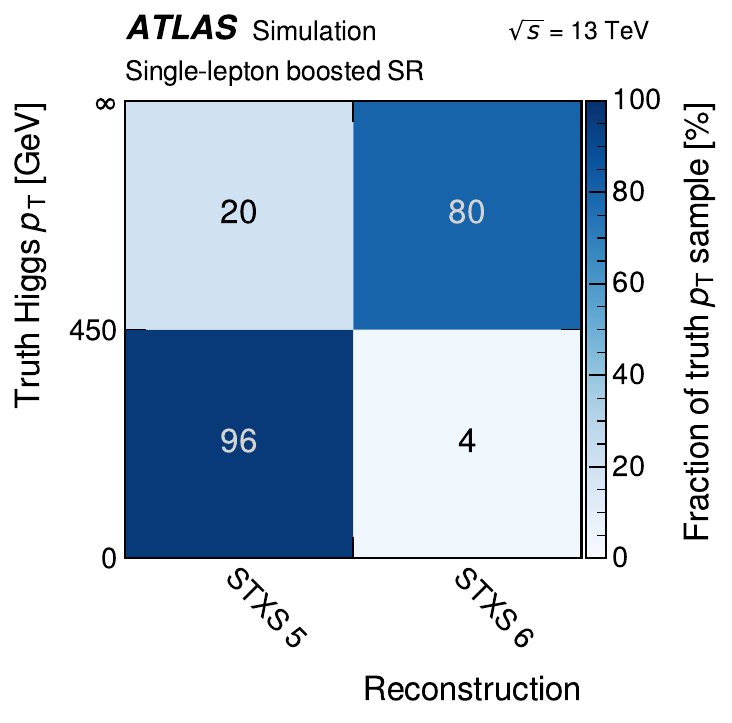}}
\hfill
\subfigure[]{\includegraphics[width=0.32\textwidth]{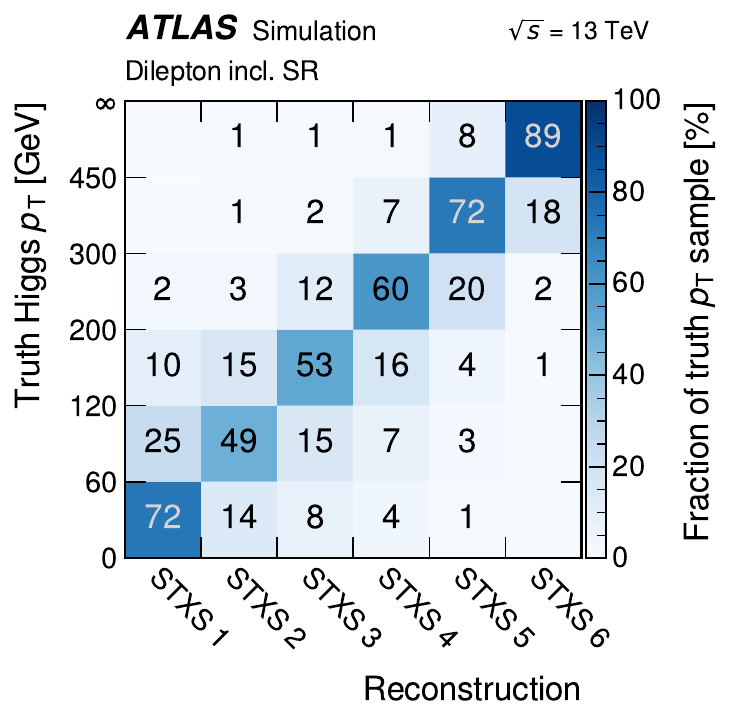}}
\caption{STXS migration matrices for the \ttH signal regions in (a) the single-lepton resolved, (b) the single-lepton boosted, and (c) the dilepton channel.
The yield fractions of the predicted STXS bins are normalised per truth STXS sample.
\label{fig:stxs_migration_matrix}}
\end{figure}

In the single-lepton channel, a total of 15 analysis regions are defined, including six STXS signal regions, two boosted regions (signal and control) and seven control regions targeting the different components of the \ttjets background. In the dilepton channel, a total of 11 analysis regions are defined, including the six STXS signal regions and five \ttjets control regions. The expected yields in the single-lepton (dilepton) signal and control regions are summarised in Tables \ref{tab:yields_lj_sr_prefit} and \ref{tab:yields_lj_cr_prefit} (\ref{tab:yields_dil_sr_prefit} and \ref{tab:yields_dil_cr_prefit}) after applying the data-driven corrections to the \ttjets background discussed in Section~\ref{sec:ttjets}.

The \ttH inclusive signal strength, $\mu_{\ttH}$, defined as the ratio of the measured \ttH cross-section including all Higgs boson decay modes to the corresponding SM prediction, or the signal strengths in each of the six STXS bins, $\mu_i$, and the normalisation factors of the components of the \ttjets background are determined via a binned likelihood fit to the distributions in all signal and control regions defined above. The \ttH signal sample is split up according to the true \ptH in each event, and the $\mu_i$ act on the respective component. In the resolved signal and control regions, the corresponding discriminant distribution is used in the likelihood fit,
while in the boosted regions, event yields in two bins of the reconstructed \ptH distribution,  300\,--\,450~\GeV and $\geq$\,450~\GeV,  are used in the fit.

The \ttb, \ttcin and \ttlight normalisation factors are chosen to float independently in the single-lepton and dilepton channels, while the normalisations of the \ttbb and \ttB backgrounds are scaled in a correlated way between the two channels.
This choice is made based on studies of the different correlation assumptions between the normalisation factors in the combined fit. The chosen configuration  provides enough freedom for the fit model and minimises pulls and constraints of the nuisance parameters in the fit to data.

The inclusive measurement of the signal strength and cross-section is performed in the full phase space. The measurement of these parameters in STXS bins is carried out for \ttH events with a Higgs boson produced centrally within $\left|y\right| \leq 2.5$ at truth level, while the remaining forward events are considered as background.

The statistical model is based on a likelihood function built with HistFactory~\cite{Cranmer:1456844} as the product over every bin of the Poisson probability for the observed data, given the SM prediction. The value of each nuisance parameter, describing the systematic uncertainties for both signal and background processes (see Section~\ref{sec:systematics}), is constrained by a Gaussian penalty term present in the likelihood function, while all normalisation factors are unconstrained.
The statistical uncertainty arising from the limited number of simulated events is included in the likelihood in the form of additional nuisance parameters with Poisson constraint terms.
The maximum-likelihood fit is performed with the RooFit package~\cite{RooFit}. A
test statistic based on the profile-likelihood ratio is used to assess the compatibility of the observed data with the background-only hypothesis ($\mu=0$) \cite{Cowan:2010js}.

Tests of the statistical model fitted to pseudo-data constructed with the \SHERPA \ttjets samples were performed to help inform the choice of fit variables and uncertainty model, with the goal of minimising bias, maximising robustness and optimising sensitivity.

%

%

\begin{table}[t!]
\setlength{\extrarowheight}{3pt}
\begin{center}
\caption{Expected yields in the single-lepton signal regions before the fit to data. The data-driven corrections to the \ttjets background are applied and all uncertainties are included, except the ones associated with the \ttjets background normalisation factors, which are not defined pre-fit. The ``Other'' category includes the $s$- and $t$-channel single-top, \tZq and \tWZ processes. The \ttH signal is normalised to the SM \ttH cross-section. For presentation purposes, uncertainties are symmetrised.
}
\scalebox{0.6}{
\begin{tabular}{lccccccc}
\toprule \toprule
& \text{STXS 1 SR} & \text{STXS 2 SR} & \text{STXS 3 SR} & \text{STXS 4 SR} & \text{STXS 5 SR} & \text{STXS 6 SR} & \text{Boosted SR}\\
\midrule
\ttH truth \ptH 0\,--\,60 \GeV   &  77 $\pm$ 12& 18.7  $\pm$ 3.2 & 11.1  $\pm$ 2.3 & 5.7  $\pm$ 1.0 & 1.66  $\pm$ 0.2 & 0.42  $\pm$ 0.09 & 0.58  $\pm$ 0.13 \\
\ttH truth \ptH 60\,--\,120 \GeV   & 49 $\pm$ 5& 99 $\pm$ 13& 37 $\pm$ 6& 15.5  $\pm$ 2.2 & 4.4  $\pm$ 0.5 & 1.25  $\pm$ 0.19 & 1.54  $\pm$ 0.25\\
\ttH truth \ptH 120\,--\,200 \GeV   &  22.7  $\pm$ 2.2 & 37 $\pm$ 4& 125 $\pm$ 16& 34 $\pm$ 5& 8.1  $\pm$ 1.3 & 2.0  $\pm$ 0.4 & 2.31  $\pm$ 0.28 \\
\ttH truth \ptH 200\,--\,300 \GeV   &  4.0  $\pm$ 0.5 & 5.5  $\pm$ 0.7 & 22.5  $\pm$ 2.9 & 88 $\pm$ 12& 19.4  $\pm$ 3.0 & 2.7  $\pm$ 0.4 & 3.3  $\pm$ 0.6 \\
\ttH truth \ptH 300\,--\,450 \GeV   & 0.39  $\pm$ 0.11 & 0.64  $\pm$ 0.16 & 1.4  $\pm$ 0.4 & 7.6  $\pm$ 1.3 & 32 $\pm$ 5& 8.0  $\pm$ 1.3 & 20 $\pm$ 3\\
\ttH truth \ptH ${\geq}$\,450 \GeV   &  <0.1 & <0.1 & 0.14  $\pm$ 0.06 & 0.21  $\pm$ 0.09 & 1.30  $\pm$ 0.31 & 9.8  $\pm$ 1.8 & 6.9  $\pm$ 1.3 \\
\ttH $\left|y\right| > 2.5$   & 0.16  $\pm$ 0.06 & 0.10  $\pm$ 0.04 & <0.1 & <0.1 & <0.1 & <0.1 & <0.1 \\
\ttbb   & 870  $\pm$ 80& 870  $\pm$ 90& 980 $\pm$ 100& 640 $\pm$ 60& 270 $\pm$ 40& 92 $\pm$ 21& 103 $\pm$ 8 \\
\ttb   & 210 $\pm$ 60& 210 $\pm$ 50& 250 $\pm$ 80& 180 $\pm$ 60& 78 $\pm$ 28 & 25 $\pm$ 9& 17 $\pm$ 5 \\
\ttB   & 64 $\pm$ 19& 73 $\pm$ 24& 100 $\pm$ 40& 87 $\pm$ 30& 46 $\pm$ 19& 22 $\pm$ 11& 11 $\pm$ 4 \\
\ttcin   & 210 $\pm$ 50& 210 $\pm$ 50& 250 $\pm$ 50& 190 $\pm$ 50& 88 $\pm$ 24& 27 $\pm$ 13& 22 $\pm$ 8 \\
\ttlight   & 42 $\pm$ 16& 45 $\pm$ 12& 68 $\pm$ 22& 64 $\pm$ 19& 31 $\pm$ 11& 12 $\pm$ 11& 3.2  $\pm$ 3 \\
\ttZ   & 36 $\pm$ 6& 38 $\pm$ 6& 53 $\pm$ 8 & 43 $\pm$ 8& 24 $\pm$ 5& 7.2  $\pm$ 1.1 & 5.8  $\pm$ 1.2 \\
\tW   & 20 $\pm$ 8& 35 $\pm$ 12& 60 $\pm$ 33& 55  $\pm$ 32& 33 $\pm$ 25& 17 $\pm$ 15& 5 $\pm$ 4\\
\Wjets   &  10 $\pm$ 6& 12 $\pm$ 8& 30 $\pm$ 15& 29  $\pm$ 14 & 23 $\pm$ 12& 12 $\pm$ 6& 3.7  $\pm$ 1.9 \\
\ttW   &  4.2  $\pm$ 0.5 & 5.9  $\pm$ 0.8 & 9.8  $\pm$ 1.6 & 8.9  $\pm$ 1.9 & 5.6  $\pm$ 0.7 & 3.2  $\pm$ 0.5 & 1.06  $\pm$ 0.26 \\
\Zjets   &  2.5  $\pm$ 1.1 & 5.1  $\pm$ 2& 6.5  $\pm$ 2.6 & 7.4  $\pm$ 2.7 & 3.1  $\pm$ 1.1 & 1.7  $\pm$ 0.6 & 0.5  $\pm$ 0.2 \\
Diboson   & 1.3  $\pm$ 0.8 & 2.5  $\pm$ 1.40 & 4.4  $\pm$ 2.3 & 4.7  $\pm$ 2.7 & 3.1  $\pm$ 1.6 & 1.4  $\pm$ 0.7 & 0.7  $\pm$ 0.4 \\
\tttt   &  5.1  $\pm$ 2.2 & 6.7  $\pm$ 3& 11 $\pm$ 5& 9 $\pm$ 4& 4.8  $\pm$ 2.1 & 2.5  $\pm$ 1.1 & 1.1  $\pm$ 0.5 \\
\tHjb   &  0.77  $\pm$ 0.22 & 1.06  $\pm$ 0.21 & 1.61  $\pm$ 0.32 & 1.58  $\pm$ 0.28 & 1.11  $\pm$ 0.30 & 0.26  $\pm$ 0.12 & 0.72  $\pm$ 0.14 \\
\tWH   & 1.68  $\pm$ 0.23 & 2.16  $\pm$ 0.27 & 3.8  $\pm$ 0.4 & 4.1  $\pm$ 0.5 & 2.22  $\pm$ 0.25 & 1.03  $\pm$ 0.12 & 1.01  $\pm$ 0.11 \\
Other  &  11.3  $\pm$ 1.2 & 11 $\pm$ 1.5 & 14.4  $\pm$ 2& 11.9  $\pm$ 1.5 & 7.3  $\pm$ 1.1 & 2.43  $\pm$ 0.35 & 2.2  $\pm$ 0.6 \\
Fakes   & 29  $\pm$ 12& 29  $\pm$ 15& 27 $\pm$ 12& 31 $\pm$ 19& 22 $\pm$ 13& 3.8  $\pm$ 2.5 & 8 $\pm$ 5 \\
\hline
Total  & 1670 $\pm$ 160& 1710 $\pm$ 170& 2060 $\pm$ 210& 1520 $\pm$ 150& 710 $\pm$ 80& 250 $\pm$ 40& 221 $\pm$ 19 \\
\hline
Data   & 1672 & 1657 & 2016 & 1441 & 676 & 241 & 216 \\
\bottomrule \bottomrule
\end{tabular}
}
\label{tab:yields_lj_sr_prefit}
\end{center}
\end{table}


\begin{table}[t!]
\setlength{\extrarowheight}{3pt}
\begin{center}
\caption{Expected yields in the single-lepton control regions before the fit to data. The data-driven corrections to the \ttjets background are applied and all uncertainties are included, except the ones associated with the \ttjets background normalisation factors, which are not defined pre-fit. The ``Other'' category includes the $s$- and $t$-channel single-top, \tZq and \tWZ processes. The \ttH signal is normalised to the SM \ttH cross-section. For presentation purposes, uncertainties are symmetrised.
}
\scalebox{0.55}{
\begin{tabular}{lcccccccc}
\toprule \toprule
& \ttlight ~\text{CR} & \ttcin ~\text{CR} & \ttb ~\text{CR} & \ttB ~\text{CR} & \ttbb ~\text{CR 1}& \ttbb~\text{CR 2} & \ttbb ~\text{CR 3} & \text{Boosted CR} \\
\midrule
\ttH truth \ptH 0\,--\,60 \GeV   & 56  $\pm$ 6& 59  $\pm$ 7& 89  $\pm$ 11& 46  $\pm$ 6& 80  $\pm$ 10& 12.5  $\pm$ 1.7 & 2.09  $\pm$ 0.26 & 0.54  $\pm$ 0.14 \\
\ttH truth \ptH 60\,--\,120 \GeV   & 86  $\pm$ 5& 98  $\pm$ 8& 133 $\pm$ 12& 92 $\pm$ 13& 112  $\pm$ 11 & 33 $\pm$ 5& 4.5  $\pm$ 0.9 & 1.15  $\pm$ 0.23 \\
\ttH truth \ptH 120\,--\,200 \GeV   & 59 $\pm$ 4& 76 $\pm$ 7& 76 $\pm$ 8& 88 $\pm$ 11& 41 $\pm$ 4& 68 $\pm$ 8& 5.4  $\pm$ 1.1 & 1.36  $\pm$ 0.28  \\
\ttH truth \ptH 200\,--\,300 \GeV  & 23 $\pm$ 2.5 & 31.9  $\pm$ 3.5 & 21.2  $\pm$ 3.1 & 41 $\pm$ 5& 7.3  $\pm$ 1.1 & 34 $\pm$ 4& 6.7  $\pm$ 1.1 & 1.42  $\pm$ 0.32 \\
\ttH truth \ptH 300\,--\,450 \GeV   & 7.7  $\pm$ 1.0 & 10 $\pm$ 1.5 & 4.7  $\pm$ 0.8 & 14.3  $\pm$ 2& 1.12  $\pm$ 0.27 & 4.4  $\pm$ 0.8 & 9.8  $\pm$ 1.5 & 4.8  $\pm$ 0.7 \\
\ttH truth \ptH ${\ge}$\,450 \GeV  & 1.67  $\pm$ 0.27 & 2.4  $\pm$ 0.5 & 0.78  $\pm$ 0.19 & 4.1  $\pm$ 0.7 & 0.18  $\pm$ 0.05 & 0.35  $\pm$ 0.12 & 3 $\pm$ 0.7 & 1.75  $\pm$ 0.32  \\
\ttH $\left|y\right| > 2.5$   & 0.98  $\pm$ 0.08 & 0.72  $\pm$ 0.17 & 1.38  $\pm$ 0.22 & 0.57  $\pm$ 0.15 & 0.37  $\pm$ 0.05 & 0.188  $\pm$ 0.029 & <0.1 & <0.1 \\
\ttbb   & 2320 $\pm$ 280& 3800 $\pm$ 400& 5400 $\pm$ 600& 3500 $\pm$ 400& 7200 $\pm$ 600& 4900 $\pm$ 500& 1140 $\pm$ 170& 141 $\pm$ 10 \\
\ttb   & 3900 $\pm$ 400& 4600 $\pm$ 700& 10600 $\pm$ 1700 & 5100 $\pm$ 600& 2100 $\pm$ 500& 1120 $\pm$ 320& 230 $\pm$ 100& 54 $\pm$ 19 \\
\ttB   & 890 $\pm$ 80& 1390 $\pm$ 270& 1780 $\pm$ 270& 5100 $\pm$ 400& 530 $\pm$ 190& 380 $\pm$ 140& 100 $\pm$ 50& 32 $\pm$ 7 \\
\ttcin   & 17000 $\pm$ 3300& 17900 $\pm$ 2600& 7800 $\pm$ 1000& 5400 $\pm$ 1300& 2500 $\pm$ 500& 1220 $\pm$ 270& 240 $\pm$ 70& 133 $\pm$ 35 \\
\ttlight   & 31800 $\pm$ 3500& 5900 $\pm$ 1100& 4000 $\pm$ 600& 2000 $\pm$ 500& 700 $\pm$ 180& 320 $\pm$ 110& 49 $\pm$ 26& 34 $\pm$ 8 \\
\ttZ   & 152 $\pm$ 20& 171 $\pm$ 25& 151 $\pm$ 21& 146 $\pm$ 22& 122 $\pm$ 18& 117 $\pm$ 15& 36 $\pm$ 5& 5.7  $\pm$ 0.9 \\
\tW   & 1040 $\pm$ 200& 750 $\pm$ 220& 820 $\pm$ 240& 560 $\pm$ 190& 350 $\pm$ 140& 360 $\pm$ 150& 130 $\pm$ 90& 10 $\pm$ 6\\
\Wjets   & 370 $\pm$ 180& 370 $\pm$ 180& 230 $\pm$ 120& 170 $\pm$ 80& 370 $\pm$ 190& 250 $\pm$ 130& 90 $\pm$ 40& 5.6  $\pm$ 2.9  \\
\ttW   & 100 $\pm$ 12& 114  $\pm$ 17& 43 $\pm$ 7& 44 $\pm$ 5& 25.6  $\pm$ 3.5 & 26 $\pm$ 5& 10.5  $\pm$ 1.6 & 1.97  $\pm$ 0.33\\
\Zjets   & 110  $\pm$ 40& 110  $\pm$ 40& 110  $\pm$ 40& 64 $\pm$ 23& 100 $\pm$ 40& 55  $\pm$ 20& 16 $\pm$ 6& 0.9  $\pm$ 0.4  \\
Diboson   & 37 $\pm$ 19& 39 $\pm$ 20& 23 $\pm$ 12& 21 $\pm$ 11& 24 $\pm$ 13& 23 $\pm$ 12& 10 $\pm$ 5& 1.1  $\pm$ 0.7  \\
\tttt   & 2.9  $\pm$ 1.2 & 24 $\pm$ 10& 7.1  $\pm$ 3& 19 $\pm$ 8& 21 $\pm$ 9& 33 $\pm$ 14 & 17 $\pm$ 7& 1.4  $\pm$ 0.6  \\
\tHjb   & 8.2  $\pm$ 1.4 & 5.4  $\pm$ 0.9 & 8.1  $\pm$ 1.6 & 3.5  $\pm$ 0.6 & 5.4  $\pm$ 1.1 & 3.6  $\pm$ 0.7 & 0.53  $\pm$ 0.15 & 0.65  $\pm$ 0.20 \\
\tWH   & 7.1  $\pm$ 0.7 & 6.7  $\pm$ 0.7 & 9.6  $\pm$ 1& 5.7  $\pm$ 0.6 & 5.4  $\pm$ 0.6 & 5.8  $\pm$ 0.7 & 1.78  $\pm$ 0.2 & 0.44  $\pm$ 0.07 \\
Other  & 370 $\pm$ 40& 370 $\pm$ 40& 303 $\pm$ 35& 228  $\pm$ 22& 179 $\pm$ 22& 111  $\pm$ 12& 27.4  $\pm$ 2.8 & 3.9  $\pm$ 0.9  \\
Fakes   & 810 $\pm$ 300& 800 $\pm$ 400& 880 $\pm$ 350& 340 $\pm$ 140& 660 $\pm$ 300& 280 $\pm$ 130& 57  $\pm$ 32& 6.3  $\pm$ 3.3  \\
\hline
Total  & 59000 $\pm$ 6000& 37000 $\pm$ 4000& 32500 $\pm$ 2900& 22900 $\pm$ 2200& 15200 $\pm$ 1700 & 9400 $\pm$ 1000& 2190 $\pm$ 270& 440  $\pm$ 50\\
\hline
Data   & 61954 & 36528 & 32887 & 23245 & 15595 & 9397 & 2097 & 426  \\
\bottomrule \bottomrule
\end{tabular}
}
\label{tab:yields_lj_cr_prefit}
\end{center}
\end{table}


\begin{table}[t!]
\setlength{\extrarowheight}{3pt}
\begin{center}
\caption{Expected yields in the dilepton signal regions before the fit to data. The data-driven corrections to the \ttjets background are applied and all uncertainties are included, except the ones associated with the \ttjets background normalisation factors, which are not defined pre-fit. The ``Other'' category includes the \tZq, \tWZ and diboson processes. The \ttH signal is normalised to the SM \ttH cross-section. For presentation purposes, uncertainties are symmetrised.
}
\scalebox{0.6}{
\begin{tabular}{lcccccc}
\toprule \toprule
& \text{STXS 1 SR} & \text{STXS 2 SR} & \text{STXS 3 SR} & \text{STXS 4 SR} & \text{STXS 5 SR} & \text{STXS 6 SR} \\
\midrule
\ttH truth \ptH 0\,--\,60 \GeV   & 10.6 $\pm$ 1.5 & 2.03 $\pm$ 0.31 & 1.14 $\pm$ 0.17 & 0.63 $\pm$ 0.11 & 0.19 $\pm$ 0.07 & 0.046 $\pm$ 0.023 \\
\ttH truth \ptH 60\,--\,120 \GeV   & 6.5 $\pm$ 0.7 & 13.0 $\pm$ 1.8 & 4.1 $\pm$ 0.7 & 1.9 $\pm$ 0.4 & 0.7 $\pm$ 0.11 & 0.13 $\pm$ 0.07 \\
\ttH truth \ptH 120\,--\,200 \GeV   & 3.06 $\pm$ 0.34 & 4.7 $\pm$ 0.5 & 16.3 $\pm$ 2.0 & 5.0 $\pm$ 0.7 & 1.36 $\pm$ 0.19 & 0.24 $\pm$ 0.07\\
\ttH truth \ptH 200\,--\,300 \GeV   & 0.5 $\pm$ 0.07 & 0.72 $\pm$ 0.08 & 2.5 $\pm$ 0.29 & 13.0 $\pm$ 1.7 & 4.4 $\pm$ 0.6 & 0.41 $\pm$ 0.06\\
\ttH truth \ptH 300\,--\,450 \GeV   & <0.1 & <0.1 & 0.197 $\pm$ 0.033 & 0.82 $\pm$ 0.18 & 8.0 $\pm$ 1.1 & 1.98 $\pm$ 0.32 \\
\ttH truth \ptH ${\geq}$\,450 \GeV   & <0.1 & <0.1 & <0.1 & <0.1 & 0.26 $\pm$ 0.08 & 2.8 $\pm$ 0.5\\
\ttH $\left|y\right| > 2.5$   & <0.1 & <0.1 & <0.1 & <0.1 & <0.1 & <0.1\\
\ttbb   &  81 $\pm$ 8 & 76 $\pm$ 7 & 81 $\pm$ 13& 59 $\pm$ 4& 36.2 $\pm$ 3.0 & 11.7 $\pm$ 2.1\\
\ttb   & 14 $\pm$ 5& 15 $\pm$ 6& 20 $\pm$ 9& 17 $\pm$ 5& 11 $\pm$ 6& 2.9 $\pm$ 1.9\\
\ttB   &  4.4 $\pm$ 2.6 & 4.8 $\pm$ 3.2 & 5.7 $\pm$ 2.3 & 6.2 $\pm$ 1.9 & 6.6 $\pm$ 2.9 & 3.5 $\pm$ 2.3\\
\ttcin   & 11.3 $\pm$ 2.4 & 11.7 $\pm$ 2.6 & 15.4 $\pm$ 3.4 & 12.5 $\pm$ 2.6 & 7.1 $\pm$ 2.6 & 3.0 $\pm$ 1.7\\
\ttlight   & 1.1 $\pm$ 0.7 & 0.9 $\pm$ 0.5 & 1.4 $\pm$ 0.8 & 1.0 $\pm$ 0.5 & 1.0 $\pm$ 0.4 & 0.37 $\pm$ 0.30\\
\ttZ   &  4.6 $\pm$ 0.7 & 4.6 $\pm$ 1.1 & 7.2 $\pm$ 1.1 & 6.2 $\pm$ 1.3 & 5.3 $\pm$ 1.1 & 2 $\pm$ 0.6\\
\tW   & 1.7 $\pm$ 0.8 & 3$\pm$ 1.8 & 6 $\pm$ 4& 8 $\pm$ 5& 7 $\pm$ 5& 3.1 $\pm$ 2.9 \\
\ttW   & 0.65 $\pm$ 0.11 & 1.18 $\pm$ 0.15 & 2.1 $\pm$ 0.23 & 2.64 $\pm$ 0.3 & 2.7 $\pm$ 0.4 & 1.2 $\pm$ 0.5\\
\Zjets   &  1.8 $\pm$ 0.9 & 1.9 $\pm$ 0.9 & 3.4 $\pm$ 1.3 & 3.8 $\pm$ 1.4 & 3.3 $\pm$ 1.2 & 1.8 $\pm$ 0.7\\
\tttt   & 0.9 $\pm$ 0.4 & 1.1 $\pm$ 0.5 & 1.8 $\pm$ 0.7 & 1.7 $\pm$ 0.7 & 1.3 $\pm$ 0.6 & 0.65 $\pm$ 0.28\\
\tWH   & 0.23 $\pm$ 0.04 & 0.31 $\pm$ 0.04 & 0.57 $\pm$ 0.07 & 0.7 $\pm$ 0.08 & 0.66 $\pm$ 0.07 & 0.24 $\pm$ 0.03\\
Other & <0.1 & <0.1 & 0.28 $\pm$ 0.14 & 0.42 $\pm$ 0.19 & 0.23 $\pm$ 0.08 & 0.29 $\pm$ 0.13\\
Fakes   &  1.7 $\pm$ 0.9 & 2.2 $\pm$ 1.1 & 3.2 $\pm$ 1.6 & 2.9 $\pm$ 1.5 & 2.6 $\pm$ 1.3 & 1.1 $\pm$ 0.6 \\
\hline
Total  & 144 $\pm$ 13& 144 $\pm$ 13& 173 $\pm$ 20& 143 $\pm$ 12& 100 $\pm$ 11& 38 $\pm$ 6\\
\hline
Data   & 150 & 149 & 161 & 149 & 76 & 35 \\
\bottomrule \bottomrule
\end{tabular}
}
\label{tab:yields_dil_sr_prefit}
\end{center}
\end{table}


\begin{table}[t!]
\setlength{\extrarowheight}{3pt}
\begin{center}
\caption{Expected yields in the dilepton control regions before the fit to data. The data-driven corrections to the \ttjets background are applied and all uncertainties are included, except the ones associated with the \ttjets background normalisation factors, which are not defined pre-fit. The ``Other'' category includes the \tZq, \tWZ and diboson processes. The \ttH signal is normalised to the SM \ttH cross-section. For presentation purposes, uncertainties are symmetrised.
}
\scalebox{0.6}{
\begin{tabular}{lccccc}
\toprule \toprule
& \ttlight ~\text{CR} & \ttcin ~\text{CR} & \ttb ~\text{CR} & \ttB ~\text{CR} & \ttbb ~\text{CR} \\
\midrule
\ttH truth \ptH 0\,--\,60 \GeV   & 11.9 $\pm$ 1.4 & 19.7 $\pm$ 2.3 & 19.8 $\pm$ 2.3 & 9.9 $\pm$ 1.4 & 26.0 $\pm$ 3.0  \\
\ttH truth \ptH 60\,--\,120 \GeV   & 17.0 $\pm$ 1.3 & 34.5 $\pm$ 2.9 & 29.5 $\pm$ 3.0 & 20.4 $\pm$ 2.5 & 42.0 $\pm$ 4.0  \\
\ttH truth \ptH 120\,--\,200 \GeV   & 10.0 $\pm$ 0.9 & 27.1 $\pm$ 2.7 & 17.0 $\pm$ 2.1 & 19.2 $\pm$ 2.4 & 34.0 $\pm$ 3.5 \\
\ttH truth \ptH 200\,--\,300 \GeV   & 3.1 $\pm$ 0.35 & 11.1 $\pm$ 1.3 & 4.6 $\pm$ 0.7 & 9.2 $\pm$ 1.2 & 14.9 $\pm$ 1.9 \\
\ttH truth \ptH 300\,--\,450 \GeV   & 0.94 $\pm$ 0.15 & 3.5 $\pm$ 0.5 & 0.94 $\pm$ 0.19 & 3.3 $\pm$ 0.5 & 5.4 $\pm$ 0.8  \\
\ttH truth \ptH ${\geq}$\,450 \GeV  & 0.33 $\pm$ 0.07 & 0.65 $\pm$ 0.12 & 0.116 $\pm$ 0.033 & 0.97 $\pm$ 0.15 & 1.33 $\pm$ 0.26 \\
\ttH $\left|y\right| > 2.5$   & 0.31 $\pm$ 0.05 & 0.25 $\pm$ 0.04 & 0.4 $\pm$ 0.05 & 0.127 $\pm$ 0.028 & 0.166 $\pm$ 0.018 \\
\ttbb   & 495 $\pm$ 35 & 860 $\pm$ 60 & 1080 $\pm$ 200 & 650 $\pm$ 100 & 2700 $\pm$ 200 \\
\ttb   & 1410 $\pm$ 130& 1790 $\pm$ 230& 3600 $\pm$ 700& 1540$ \pm$ 180& 1200 $\pm$ 400\\
\ttB   & 310 $\pm$ 40& 460 $\pm$ 40& 540$\pm$ 210& 1560 $\pm$ 230& 310 $\pm$ 110\\
\ttcin   & 5600 $\pm$ 500& 7700 $\pm$ 600& 2210 $\pm$ 280& 1590 $\pm$ 330& 1070 $\pm$ 170\\
\ttlight   & 9200 $\pm$ 1500& 2800 $\pm$ 400& 590 $\pm$ 130& 410 $\pm$ 90& 190 $\pm$ 70\\
\ttZ   & 43 $\pm$ 6 & 81 $\pm$ 12 & 35 $\pm$ 6 & 36 $\pm$ 5 & 92 $\pm$ 12\\
\tW   & 300 $\pm$ 60& 210 $\pm$ 50& 220 $\pm$ 70& 150 $\pm$ 50& 300 $\pm$ 140\\
\ttW   & 37 $\pm$ 5& 72 $\pm$ 7& 12.1 $\pm$ 1.3 & 14.3 $\pm$ 3.3 & 29.8 $\pm$ 3.5 \\
\Zjets   & 340 $\pm$ 130& 260 $\pm$ 100& 250 $\pm$ 90& 93 $\pm$ 34& 320 $\pm$ 120\\
\tttt   & 1.6 $\pm$ 0.7 & 10$\pm$ 4& 1$\pm$ 0.4 & 4.8 $\pm$ 2& 28 $\pm$ 12\\
\tWH   & 0.98 $\pm$ 0.11 & 2.24 $\pm$ 0.22 & 2.69 $\pm$ 0.29 & 1.42 $\pm$ 0.14 & 4.0 $\pm$ 0.4 \\
Other & 5.6 $\pm$ 2.4 & 8.1 $\pm$ 3.1 & 4.4 $\pm$ 2.0 & 2.1 $\pm$ 0.7 & 12 $\pm$ 5\\
Fakes   & 110 $\pm$ 50& 120 $\pm$ 60& 40 $\pm$ 20& 40 $\pm$ 20& 80 $\pm$ 40 \\
\hline
Total  & 18000 $\pm$ 2000& 14400 $\pm$ 1100& 8700 $\pm$ 900& 6200 $\pm$ 500& 6400 $\pm$ 700\\
\hline
Data   & 18557 & 14361 & 8624 & 5830 & 6448 \\
\bottomrule \bottomrule
\end{tabular}
}
\label{tab:yields_dil_cr_prefit}
\end{center}
\end{table}



%
\section{Systematic uncertainties}
\label{sec:systematics}

Multiple sources of systematic uncertainty, arising from detector effects, theoretical assumptions and the limited number of events in the MC simulations, are considered in the analysis. They affect the categorisation of events as well as the normalisation and shape of the distributions used in the signal extraction fit.

All the sources of experimental uncertainty, except the uncertainty in the integrated luminosity, affect both the normalisations and the shapes of the distributions in all simulated samples. The uncertainty in the integrated luminosity is 0.83\%~\cite{DAPR-2021-01}, obtained using the LUCID-2 detector~\cite{LUCID2} for the primary luminosity measurements, complemented by measurements using the inner detector and calorimeters. The uncertainty in the pile-up modelling is obtained by varying the pile-up reweighting in the simulation within its uncertainties.

The correction factors applied to the simulated samples to improve the description of the lepton reconstruction, identification and isolation efficiencies, momentum scale and resolution, and
lepton trigger efficiencies are varied within their uncertainties~\cite{EGAM-2018-01, MUON-2018-03}  to estimate the corresponding systematic uncertainty.

The JES uncertainty~\cite{JETM-2018-05} accounts for contributions from jet-flavour composition, $\eta$-intercalibration, punch-through, single-particle response, calorimeter response to different jet flavours, and pile-up, resulting in 31 uncorrelated JES uncertainty components. The jet energy resolution was measured separately for data and MC using two in situ techniques~\cite{JETM-2018-05}. The  systematic uncertainty is defined as the quadratic difference between the jet energy resolutions for data and simulation and split into 13 uncorrelated uncertainty components. The uncertainty associated with the JVT discriminant is obtained by varying the efficiency correction factors~\cite{PERF-2014-03}.

The uncertainties associated with the $b$-tagging algorithm calibration are derived separately for \bjets, $c$-jets and light-flavour jets, as a function of the jet \pt, and decomposed into several uncorrelated components for each category, corresponding to the number of \pt bins multiplied by the number of DL1r pseudo-continuous scores~\cite{FTAG-2018-01,FTAG-2020-08,FTAG-2019-02}. This yields a total of 45 components for \bjets and 20 each for $c$- and light jets. For jets with a \pt above the \pt threshold where the \btag algorithm is calibrated, high-\pt extrapolation uncertainties derived using MC simulation are included.

The uncertainty in \MET results from the propagation of the uncertainties in the energy scales and resolutions of photons, leptons and jets, and from the modelling of its soft term~\cite{JETM-2020-03}.

For the \ttH signal, two cross-section uncertainties are applied accounting for the effect of varying the PDF and $\alpha_S$ and for missing higher order terms in the fixed order perturbative QCD calculations. They amount to $\pm$3.6\% and $\pm 9.2$\%, respectively~\cite{deFlorian:2016spz}. The systematic uncertainty in the \ttH cross-section
includes theory uncertainties due to migrations of events between the truth Higgs boson \pt bins~\cite{ATL-PHYS-PUB-2023-031}. An uncertainty of 2.2\% is assigned to the $H\rightarrow b\Bar{b}$ branching fraction~\cite{deFlorian:2016spz}.
Two uncertainties due to missing higher order terms in the MC simulation are estimated by varying independently the renormalisation and factorisation scales in the ME of the nominal MC sample by a factor of two up and down.
The uncertainties in the amount of initial- and final-state QCD radiation (ISR and FSR) predicted by the PS are estimated by varying the scale in $\alpha_S^{ISR}$ according to the values given by \textit{var3c} in the \pythia A14 tune and by varying the scale in $\alpha_S^{FSR}$ by a factor two up and down. To assess the uncertainties associated with the PS, hadronisation and underlying event, the nominal \ttH sample is compared with the alternative \powheg{}+\herwigseven sample, while the uncertainty due to the NLO matching procedure is estimated with \powheg{}+\pythia with a varied \pthard parameter value.

All the modelling uncertainties in the \ttjets background have independent nuisance parameters for the \ttb, \ttB, \ttbb, \ttcin and \ttlight processes. These systematic variations are normalised to conserve the total nominal event count after the preselection for each process.
Uncertainties due to missing higher order terms in the perturbative QCD calculations, in the amount of ISR and FSR as well as the uncertainties associated with the PS and hadronisation, and with the NLO matching procedure are estimated in the same way as for the \ttH signal.

For \ttbin, the choice of recoil scheme in the ISR PS has a sizeable effect in the normalisations and shapes of the distributions used in the analysis within the detector acceptance~\cite{ATL-PHYS-PUB-2022-006}. The uncertainty associated with the choice of the global recoil scheme is assessed by comparing the nominal sample with an alternative sample produced with the dipole recoil scheme.
For \ttlight and \ttcin, the uncertainty in the choice of the \hdamp parameter is estimated by using the alternative \powhegbox{}+\pythia sample with $\hdamp = 3~\mtop$.

Uncertainties related to the \HT reweighting procedure described in Section \ref{sec:ttjets} are assigned independently in the single-lepton and dilepton channels. The \ttlight and \ttcin normalisation factors are varied within their uncertainties to estimate the corresponding systematic uncertainty. An additional uncertainty is estimated by comparing the distributions with and without the weights derived in the \HT reweighting procedure, in bins of jet multiplicity.

A $\pm5\%$ normalisation uncertainty is considered for the cross-sections of the t- and s- single-top production modes~\cite{Kidonakis:2011wy,Kidonakis:2010tc}.
The normalisation uncertainty of the $tW$ background is 3.7\%~\cite{Kidonakis:2021vob}.
Modelling uncertainties in the $tW$ production due to the choice of the PS and hadronisation model, the NLO matching and the \hdamp parameter choice are evaluated in the same way as for the \ttlight and \ttcin backgrounds. The uncertainty associated with the interference between $tW$ and \ttbar\ production at NLO is assessed by comparing the nominal \powhegbox{}+\pythia\ sample produced using the diagram removal scheme to an alternative sample produced with the same generator but using the diagram subtraction scheme.

The total uncertainty in the theoretical $\ttbar W$ cross-section
computed at NNLO in the QCD and at NLO in the electro-weak interactions is 7.4\% ~\cite{Buonocore_2023}.
The modelling uncertainty in the $\ttbar W$ background is evaluated by comparing the nominal \sherpa simulation with the sample produced using \mgamc.
For the $\ttbar Z$ production, the uncertainty in the predicted cross-section at NLO in QCD is 12\% \cite{deFlorian:2016spz}.
The modelling uncertainty in the $\ttbar Z$ background is evaluated by comparing
the nominal \mgpythia sample with the sample where \pythia is replaced by
\herwigseven to simulate the PS and hadronisation.

A normalisation uncertainty of $_{-15\%}^{+70\%}$ is assigned for the \tttt background. The down variation corresponds to the theory uncertainty
covering effects from varying the factorisation and renormalisation scales, the PDFs and $\alpha_{\textrm{S}}$~\cite{vanbeekveld2022threshold} while the up variation covers the measured \tttt cross-section~\cite{TOPQ-2021-08}.
For \tZq, the total normalisation uncertainty is 7.9\% ~\cite{Alwall:2014hca}.
A normalisation uncertainty of 15.4\% (9.2\%) calculated using \MGNLO at NLO
is applied to the \tHjb\ (\tWH) background normalisation~\cite{HIGG-2020-03}. For $tWZ$, an uncertainty of $\pm50\%$ is used~\cite{Alwall:2014hca}.

The treatment of \Vjets uncertainties follows the previous
analysis~\cite{HIGG-2020-03-fixed}. An uncertainty of 40\% is assumed for the \Wjets cross-section, with an additional 30\% normalisation uncertainty used for $W$ boson production in association with heavy-flavour jets, taken as uncorrelated between events with two and events with more than two truth-level heavy-flavour jets.
These uncertainties are based on variations of the factorisation and renormalisation scales and of the matching parameters in the \sherpa\ samples.
An uncertainty of 35\% is applied to the \Zjets\ normalisation to account for both the variations of the scales and matching parameters in the \sherpa\ samples and the uncertainty in the correction factor for
the $Z$ boson production accompanied by two $b$-jets extracted from data.
A 50\% normalisation uncertainty is used for
the diboson background, which includes uncertainties in the inclusive cross-section and additional jet production~\cite{Grazzini_2016,ATL-PHYS-PUB-2016-002,STDM-2018-03}.

A 50\% normalisation uncertainty is assigned to the fake-lepton background estimate in the single-lepton and dilepton channels separately. An additional uncertainty obtained by using an alternative parameterisation of the fake rate (Section \ref{sec:misid}) is considered in the single-lepton channel.


\section{Results}
\label{sec:result}

An excess of events over the expected background is found with an observed~(expected) significance of 4.6~(5.4) standard deviations, in a combined profile-likelihood fit to data in all signal and control regions.
The measured \ttH signal strength for $m_H=125.09$~\GeV \cite{ATLAS:2015yey} is
\begin{equation}
\mu_{\ttH} = \mutotal = \mustatsyst.
\end{equation}
The total statistical uncertainty is defined as the uncertainty in $\mu_{\ttH}$ when all nuisance parameters associated with the systematic uncertainties are fixed to their best-fit values. The total systematic uncertainty is then defined as the difference in quadrature between the total and statistical uncertainties.
The systematic uncertainty in the signal strength $\mu_{\ttH}$ includes the theoretical uncertainties in the SM \ttH cross-section described in Section~\ref{sec:systematics}.
The measured \ttH signal strength in the single-lepton channel is
\begin{equation}
\mu_{\ttH}  = \mustatsystlj
\end{equation}
and in the dilepton channel is
\begin{equation}
\mu_{\ttH} = \mustatsystdil.
\end{equation}
The combined $\mu_{\ttH}$ is converted into an inclusive cross-section using the SM theoretical cross-section of \xstheory\,fb for $m_\texttt{H}=125.09$~\GeV and excluding its uncertainty~\cite{deFlorian:2016spz}.
The resulting cross-section is
\begin{equation}
\sigma_{\ttH} = \xstotal ~\text{fb} = \xsstatsyst~\text{fb},
\end{equation}
with a relative uncertainty of $^{+25\%}_{-22\%}$, consistent with the SM prediction. The sensitivity is driven by the single-lepton channel, where the systematic component of the uncertainty dominates. The sensitivity of the dilepton channel is limited by the size of the dataset.

The measured values of the \ttjets background normalisation factors, which are free
parameters of the fit, are listed in Table~\ref{tab:norm_factors}.
They are compatible between the single-lepton and the dilepton channels within two standard deviations. For \ttb, the measured value of the normalisation factor is slightly larger in the dilepton channel than in the single-lepton channel.

\begin{table}[b!]
\centering
\caption{Best-fit values of the \ttjets normalisation factors obtained from the fit to data with a single inclusive \ttH signal strength parameter. Independent normalisation factors for \ttb, \ttcin and \ttlight components are used in the dilepton and single-lepton channels. Before the fit, the \ttlight and \ttcin components are normalised to the \ttbar cross-section computed at NNLO in QCD including the resummation of NNLL soft-gluon terms~\cite{Czakon:2011xx} while the \ttbin normalisation is taken from the \ttbbMC MC simulation.
}
\setlength{\extrarowheight}{10pt}
\begin{tabular}{lccccc}
\cmidrule{1-6}\morecmidrules\cmidrule{1-6}
Normalisation factor & \ttlight & \ttcin & \ttb & \ttB & \ttbb \\  \cline{1-6}
Single-lepton  & $0.78^{+0.08}_{-0.08}$  &  $1.51^{+0.19}_{-0.18}$ & $1.06^{+0.10}_{-0.10}$  & \multirow{2}{*}{$1.15^{+0.15}_{-0.14}$} & \multirow{2}{*}{$0.94^{+0.08}_{-0.08}$} \\
Dilepton  & $0.88^{+0.11}_{-0.10}$ & $1.36^{+0.10}_{-0.10}$ &
$1.24^{+0.09}_{-0.09}$ &  & \\ \cmidrule{1-6}
\morecmidrules\cmidrule{1-6}
\end{tabular}
\label{tab:norm_factors}
\end{table}


Compared with the previous analysis using the same dataset, the current analysis selects 64\% (29\%) new events in the single-lepton (dilepton) SR that did not enter the selection of the previous analysis. This is consistent with the increase of the overall acceptance by a factor of three.
The statistical correlation between the two analyses is estimated using a booststrap method to be 19\%, assuming that the systematic uncertainties are independent. This assumption is justified by the fact that the systematic model of the most important \ttbin background is different between the two analyses and the experimental uncertainties are updated.
Based on this, the probability that the current result is compatible with the result of Ref.~\cite{HIGG-2020-23} is estimated as 21\%. %

Figure \ref{fig:POI_STXS_combined} shows the measured values of
\muSM\ in bins of the Higgs boson \pt, obtained from a combined profile-likelihood fit to data with a free \ttH signal strength parameter for each bin. The result of this measurement is compatible with the SM prediction with a p-value of 89\%, taking into account theoretical uncertainties in the SM \ttH cross-section. The measured value of the inclusive signal strength is also shown.
The measurement of \muSM\ in the [300,~450)~\GeV and [450,~$\infty$)~\GeV \ptH bins is limited by statistical uncertainties. In the remaining bins, there is a similar contribution from statistical and systematic uncertainties. Defining boosted regions improves sensitivity in the [450,~$\infty$)~\GeV \ptH bin by about 15\%, compared with the scenario where only the resolved event selection is applied to these events.
The correlations between the parameters of interest are shown in Figure \ref{fig:POI_STXS_correlations}. They do not exceed 30\% and are larger in the lower Higgs boson transverse momentum region.

The fitted values of \ttjets background normalisation factors are consistent with those obtained in the fit with a single $\mu_{\ttH}$ parameter shown in Table~\ref{tab:norm_factors}.

\begin{figure}[htbp]
\centering
\includegraphics[width=\textwidth]{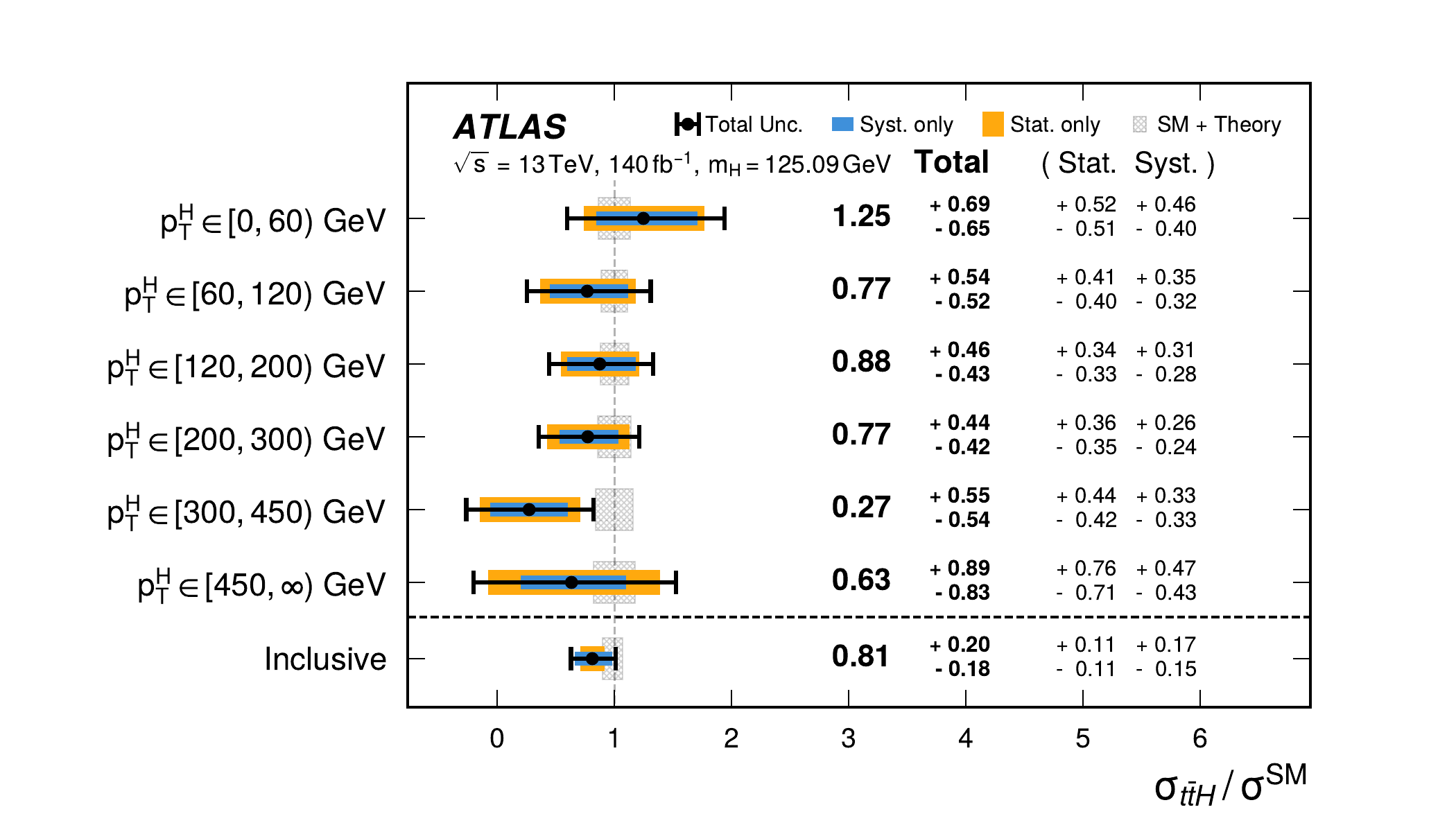}
\caption{The \ttH cross-sections measured in bins of truth  Higgs boson \pt for a Higgs boson rapidity $|y| \leq 2.5$, and measured inclusively in the full phase space, normalised to their SM predictions, as obtained from a combined profile-likelihood fit to data in all signal and control regions. The uncertainties are shown separately for the measurement and the prediction.
}
\label{fig:POI_STXS_combined}
\end{figure}

\begin{figure}[htbp]
\centering
\includegraphics[width=.8\textwidth]{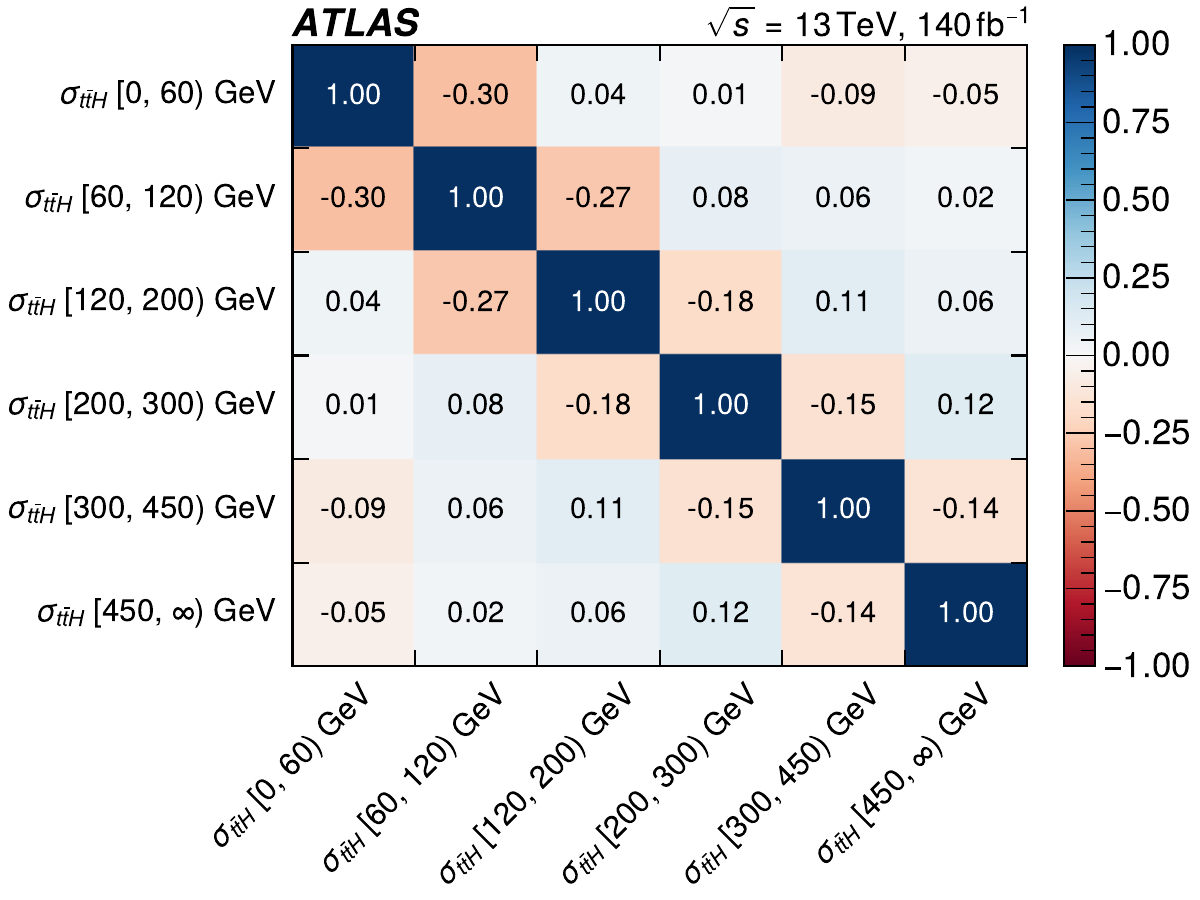}
\caption{Post-fit correlations between the measured values of the \ttH cross-section, $\sigma_{\ttH}$, in bins of truth \ptH.}
\label{fig:POI_STXS_correlations}
\end{figure}

The absolute and relative systematic uncertainties in the measurement of $\sigma_{\ttH}$,
grouped in categories, are shown in Table \ref{tab:grouped_impact},
and the effect of individual nuisance parameters ranked according to their impact on the inclusive $\sigma_{\ttH}/\sigma^{\text{SM}}$ is shown in Figure \ref{fig:ranking_mu_inclusive}. The largest impact originates from the signal modelling, followed by the \ttbb background modelling. In both cases, the dominant effects arise from the modelling of the FSR and the choice of the PS model.
Among the experimental uncertainties, the largest effects come from
the $b$-jet tagging and the jet energy scale.

Among the eight highest ranked nuisance parameters in Figure~\ref{fig:ranking_mu_inclusive}, the largest constraints are observed for the \ttbb uncertainties arising from the dipole shower model and the choice of matching algorithm. These constraints predominantly originate from the single-lepton channel, and more specifically the SRs and the \ttbb CR where the \ttbb background contribution is large. The constraints are more pronounced when the \ttbb CR is split into three in the single-lepton channel to better control the \ttbb background modelling in bins of the reconstructed Higgs boson \pt.

\begin{table}[htbp]
\centering
\caption{A list of the absolute and relative uncertainties in the measured $\sigma_{\ttH}$ grouped in categories. The contributions from different sources of uncertainty are evaluated after the fit. The quoted values are obtained by repeating the fit, while fixing the set of nuisance parameters of the sources corresponding to each category to their best-fit values, and subtracting in quadrature the resulting uncertainty from the total uncertainty of the nominal fit presented in the last row.
The total uncertainty is different from the sum in quadrature of the different components due to correlations between nuisance parameters in the fit. The \ttH and \ttbin radiation uncertainty categories include the renormalisation and factorisation scales, ISR and FSR uncertainties. The ``\ttH theory'' category includes STXS-related theoretical uncertainties and uncertainty in the \Hbb branching fraction.
The ``Minor background modelling'' category includes uncertainties in the fake-lepton background and in minor backgrounds as defined in the text. The total statistical uncertainty includes uncertainties in the normalisation factors.
}
\renewcommand{\arraystretch}{1.1}
\setlength{\tabcolsep}{12pt}
\begin{tabular}{lcc cc}
\toprule \toprule
Uncertainty source & \multicolumn{2}{c}{$\Delta\sigma_{\ttH}$ (fb)} & \multicolumn{2}{c}{$\Delta\sigma_{\ttH}/\sigma_{\ttH}$ (\%)}\\
\midrule
Process modelling \\
\hspace{2em} \ttH modelling \\
\hspace{3em} \ttH radiation & +35 & $-$21 & +9 &  $-$5 \\
\hspace{3em} \ttH parton shower & +32 & $-$19 & +8 &  $-$5    \\
\hspace{3em} \ttH matching & <0.1 & $-$0.3 & <0.1 &  $-$0.1 \\
\hspace{3em} \ttH theory & +25 & $-$17 & +6 &  $-$4 \\
\hspace{2em} \ttbin modelling \\
\hspace{3em} \ttbin radiation & \multicolumn{2}{c}{$\pm 31$} &  \multicolumn{2}{c}{$\pm 8$}  \\
\hspace{3em} \ttbin parton shower
& \multicolumn{2}{c}{$\pm 29$} & \multicolumn{2}{c}{$\pm 7$} \\
\hspace{3em} \ttbin matching & \multicolumn{2}{c}{$\pm 19$} & \multicolumn{2}{c}{$\pm 5$} \\
\hspace{2em} \ttcin modelling & \multicolumn{2}{c}{$\pm 18$} & \multicolumn{2}{c}{$\pm 4$} \\
\hspace{2em} \ttlight modelling & \multicolumn{2}{c}{$\pm 5$} & \multicolumn{2}{c}{$\pm 1$}\\
\hspace{2em} $tW$ modelling & \multicolumn{2}{c}{$\pm 16$} &  \multicolumn{2}{c}{$\pm 4$} \\
\hspace{2em} Minor background modelling & \multicolumn{2}{c}{$\pm 19$} & \multicolumn{2}{c}{$\pm 5$}\\
Flavour tagging & \multicolumn{2}{c}{$\pm 36$} & \multicolumn{2}{c}{$\pm 9$} \\
Jet modelling & \multicolumn{2}{c}{$\pm 22$} & \multicolumn{2}{c}{$\pm 5$} \\
Monte-Carlo statistics & \multicolumn{2}{c}{$\pm 17$} & \multicolumn{2}{c}{$\pm 4$}\\
Other instrumental & \multicolumn{2}{c}{$\pm 10$} & \multicolumn{2}{c}{$\pm 2$}\\
\hline
Total systematic uncertainty & +85 & $-$75 & +21  & $-$18 \\
\midrule
Normalisation factors & \multicolumn{2}{c}{$\pm 21$} & \multicolumn{2}{c}{$\pm 5$}\\
\midrule
Total statistical uncertainty & \multicolumn{2}{c}{$\pm 54$} & \multicolumn{2}{c}{$\pm 13$} \\
\midrule \midrule
Total uncertainty & +101 & $-$92 & +25 & $-$22\\
\hline
\bottomrule
\end{tabular}
\label{tab:grouped_impact}
\end{table}


\begin{figure}[t!]
\centering
\includegraphics[width=0.7\textwidth]{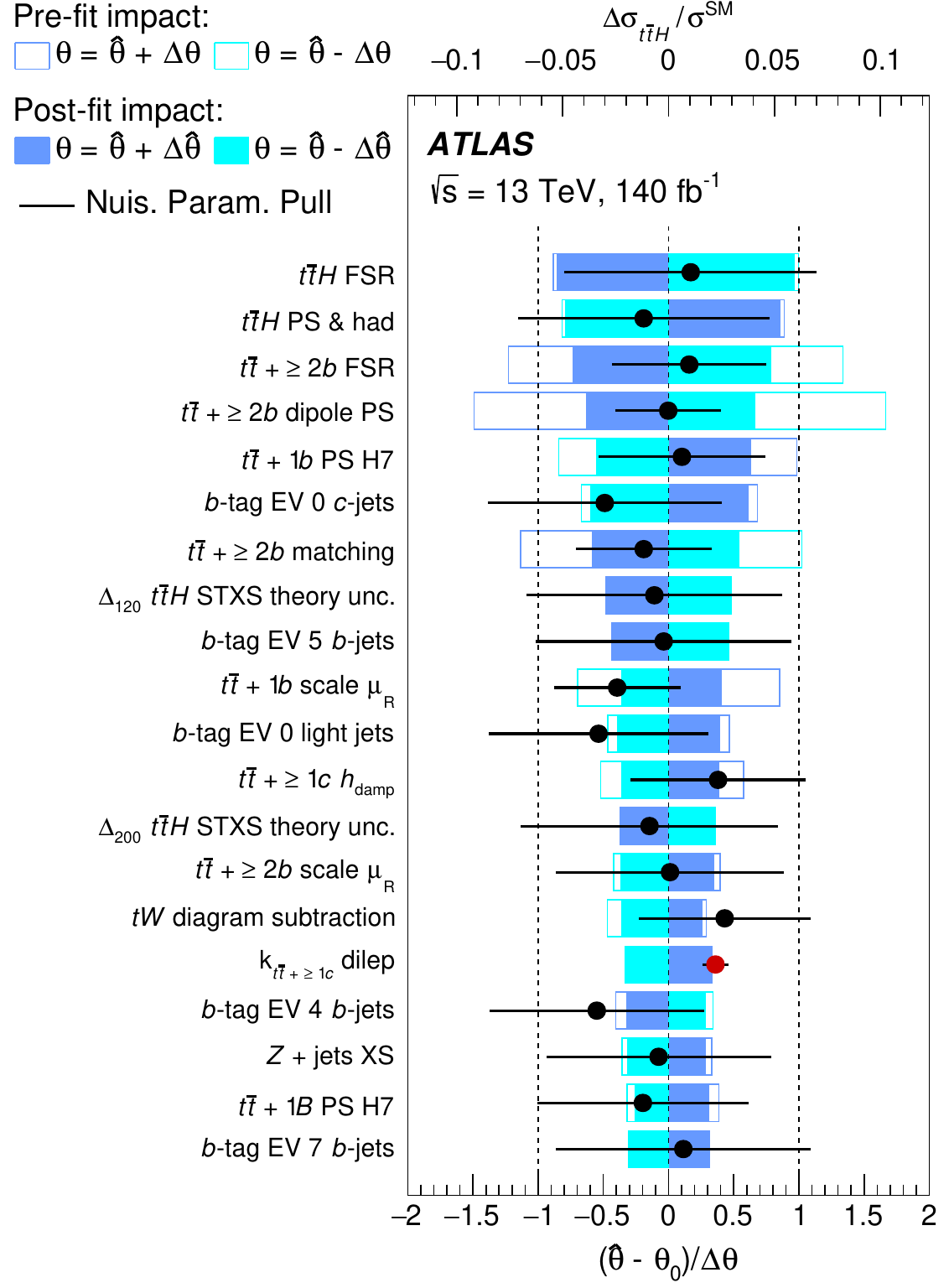}
\caption{
Ranking of the 20 modelling and experimental systematic uncertainties with the largest post-fit impact on the inclusive cross-section.
The empty (filled) blue rectangles correspond to the pre-(post-)fit impact on $\sigma_{\ttH}/\sigma^{\text{SM}}$ and refer to the upper scale of the plot.
The impact of each nuisance parameter, $\Delta\sigma_{\ttH}/\sigma^{\text{SM}}$, is computed by comparing the nominal best-fit value of $\sigma_{\ttH}/\sigma^{\text{SM}}$ with the result of the fit when fixing the considered nuisance parameter to its best-fit value, $\hat{\theta}$, shifted by its pre-fit (post-fit) uncertainties $\pm\Delta\theta$ ($\pm\Delta\hat{\theta}$).
The black markers show the pulls of the nuisance parameters relative to their nominal values, $\theta_{0}=0$.
The red marker shows the pull of the \ttcin normalisation factor in the dilepton channel relative to its nominal value, $\theta_{0}=1$.
No pre-fit uncertainty is defined for the normalisation factor as it is a free parameter of the fit.
The pulls and their relative post-fit errors, $\Delta\hat{\theta}/\Delta\theta$, refer to the lower scale of the plot.
}
\label{fig:ranking_mu_inclusive}
\end{figure}

The observed yields in all signal and control regions are compared to the post-fit predictions in Figures \ref{fig:post_summary_SR} and  \ref{fig:post_summary_CR}. In each region, all bins are aggregated into a single bin.
Figures \ref{fig:lj_postfit_SRs} (\ref{fig:dil_postfit_SRs}) and \ref{fig:lj_postfit_CRs} (\ref{fig:dil_postfit_CRs}) show a more detailed comparison in the single-lepton (dilepton) signal and control regions. The signal and background predictions in all post-fit distributions are obtained by setting the free parameters and the nuisance parameters to their best-fit values. In all post-fit plots, the uncertainty band includes all uncertainties and their correlations. The discriminant output in the plots is rescaled to be between zero and one using a logistic function.

The global goodness of fit~\cite{Baker:1983tu,cousins2023lectures} is 87\% for the \muSM\ measurement in \ptH bins, highlighting that good post-fit modelling is achieved.

Figure~\ref{fig:SoverB} shows the event yield in data compared with the post-fit signal~($S$) and total background~($B$)
predictions. It is ordered by the signal-to-background ratio of the bins from all the fitted regions. The predictions are shown for the best-fit signal strength and for the SM prediction. The observed data shows an excess over the background compatible with the best-fit signal strength $\mu_{\ttH}=0.81$
in the high \mbox{$\log_{10}\left(S/B\right)$} region.

\begin{figure}[t!]
\centering
\subfigure[]{\includegraphics[width=0.45\textwidth]{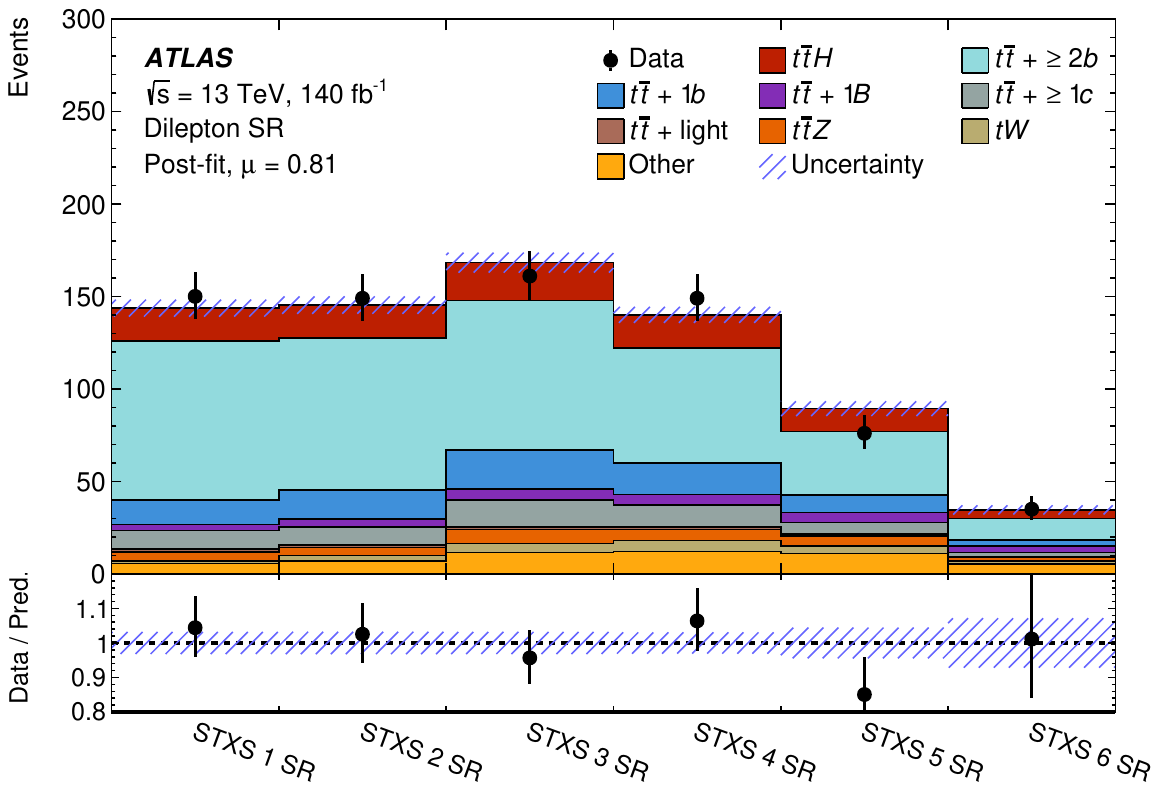}}\label{fig:post_summary_SR_1}
\subfigure[]{\includegraphics[width=0.45\textwidth]{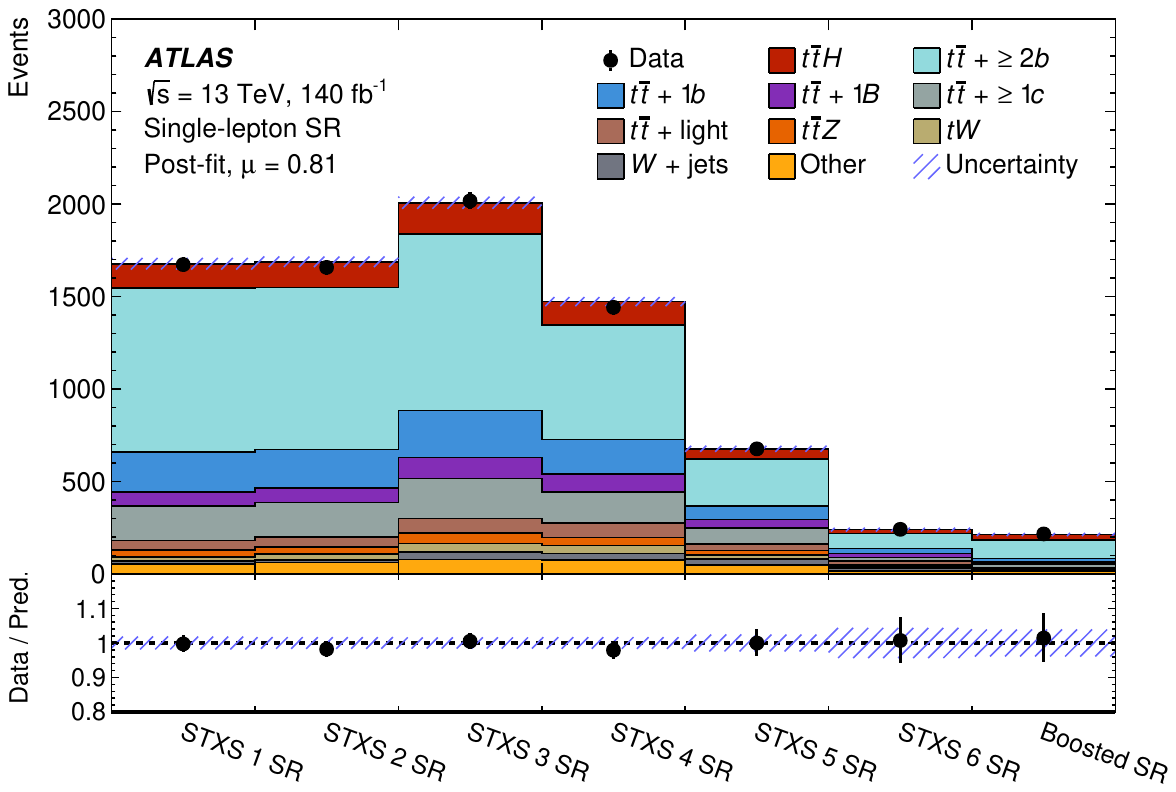}}\label{fig:post_summary_SR_2}
\caption{
Post-fit summary of the yields in the (a) dilepton and (b) single-lepton signal regions, with all regions aggregated into a single bin each. The uncertainty band includes all uncertainties and their correlations.
\label{fig:post_summary_SR}
}
\end{figure}
\begin{figure}[t!]
\centering
\subfigure[]{\includegraphics[width=0.45\textwidth]{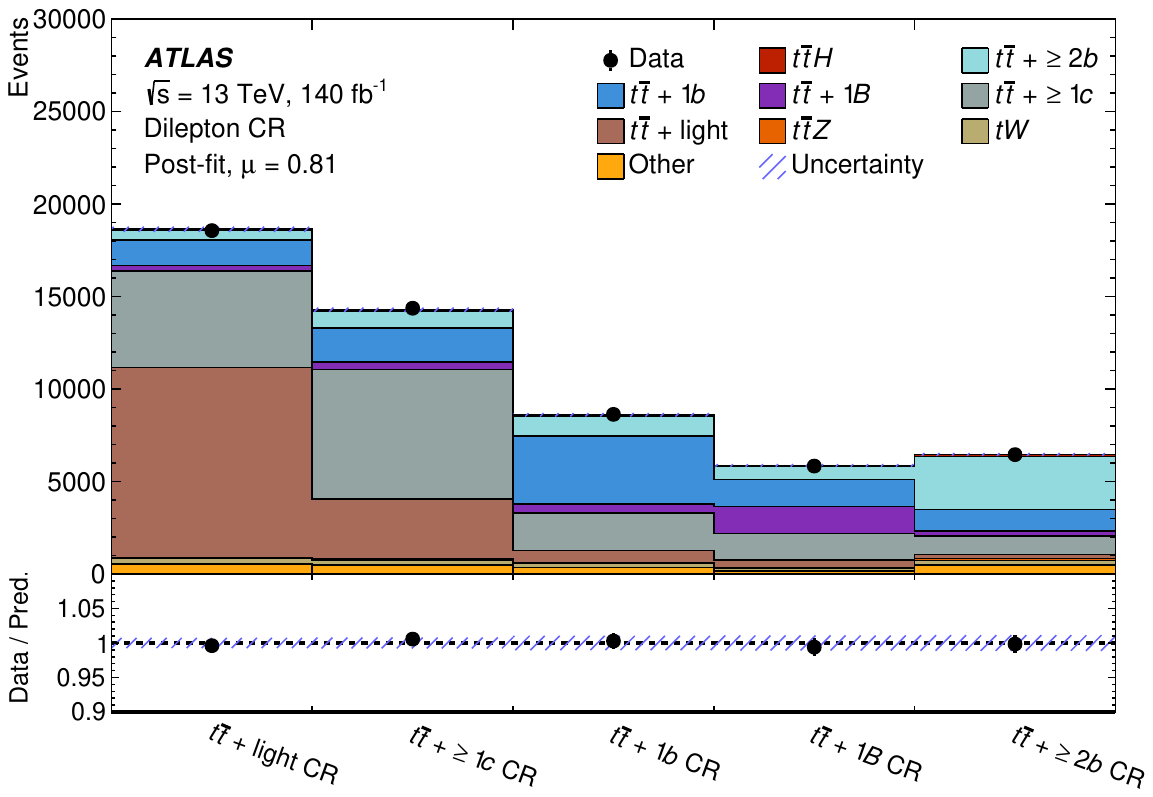}}\label{fig:post_summary_CR_1}
\subfigure[]{\includegraphics[width=0.45\textwidth]{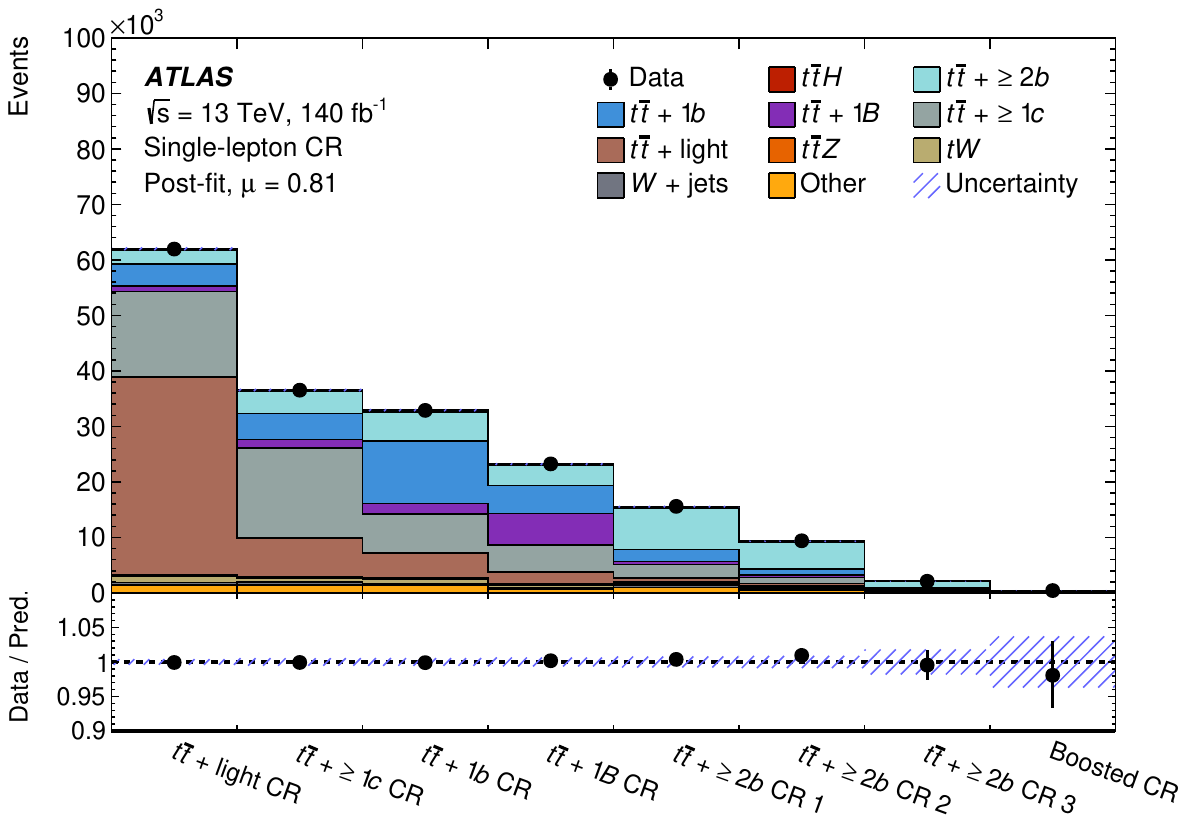}}\label{fig:post_summary_CR_2}
\caption{
Post-fit summary of the yields in the (a) dilepton and (b) single-lepton control regions, with all regions aggregated into a single bin each. The uncertainty band includes all uncertainties and their correlations.
\label{fig:post_summary_CR}
}
\end{figure}
%

\begin{figure}[t!]
\centering
\subfigure[]{\includegraphics[width=0.3\textwidth]{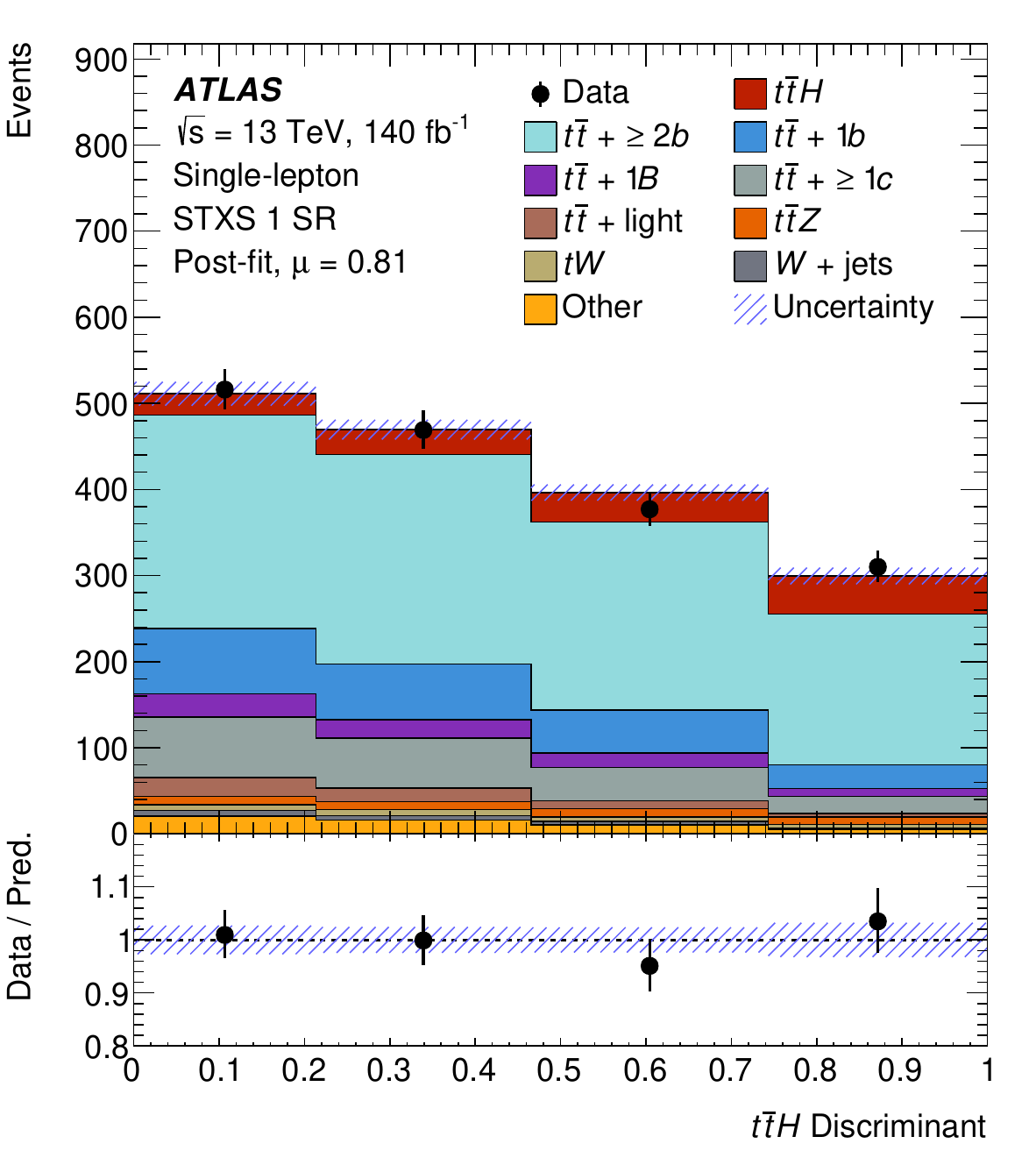}}\label{fig:lj_postfit_SRs_1}
\subfigure[]{\includegraphics[width=0.3\textwidth]{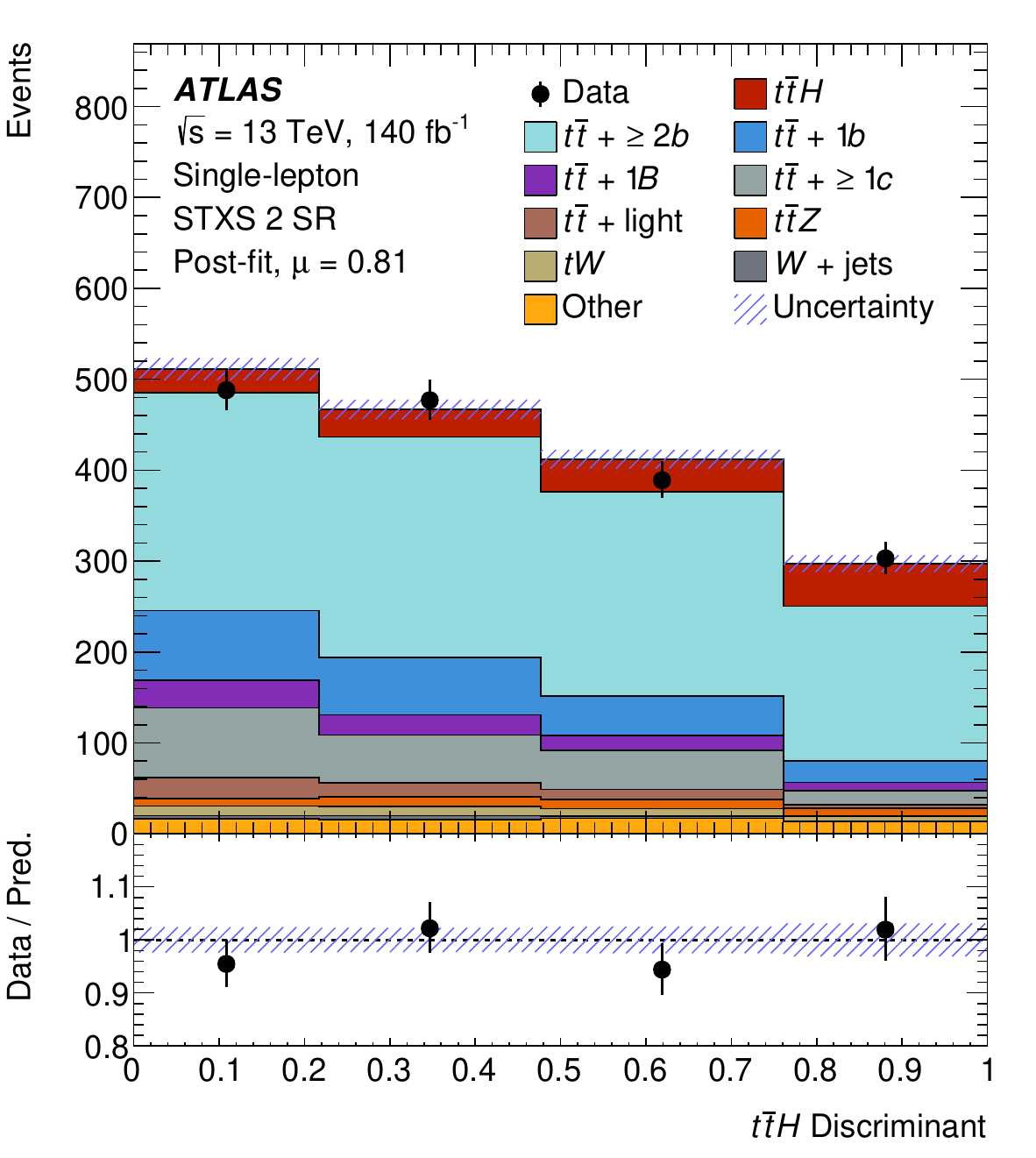}}\label{fig:lj_postfit_SRs_2}
\subfigure[]{\includegraphics[width=0.3\textwidth]{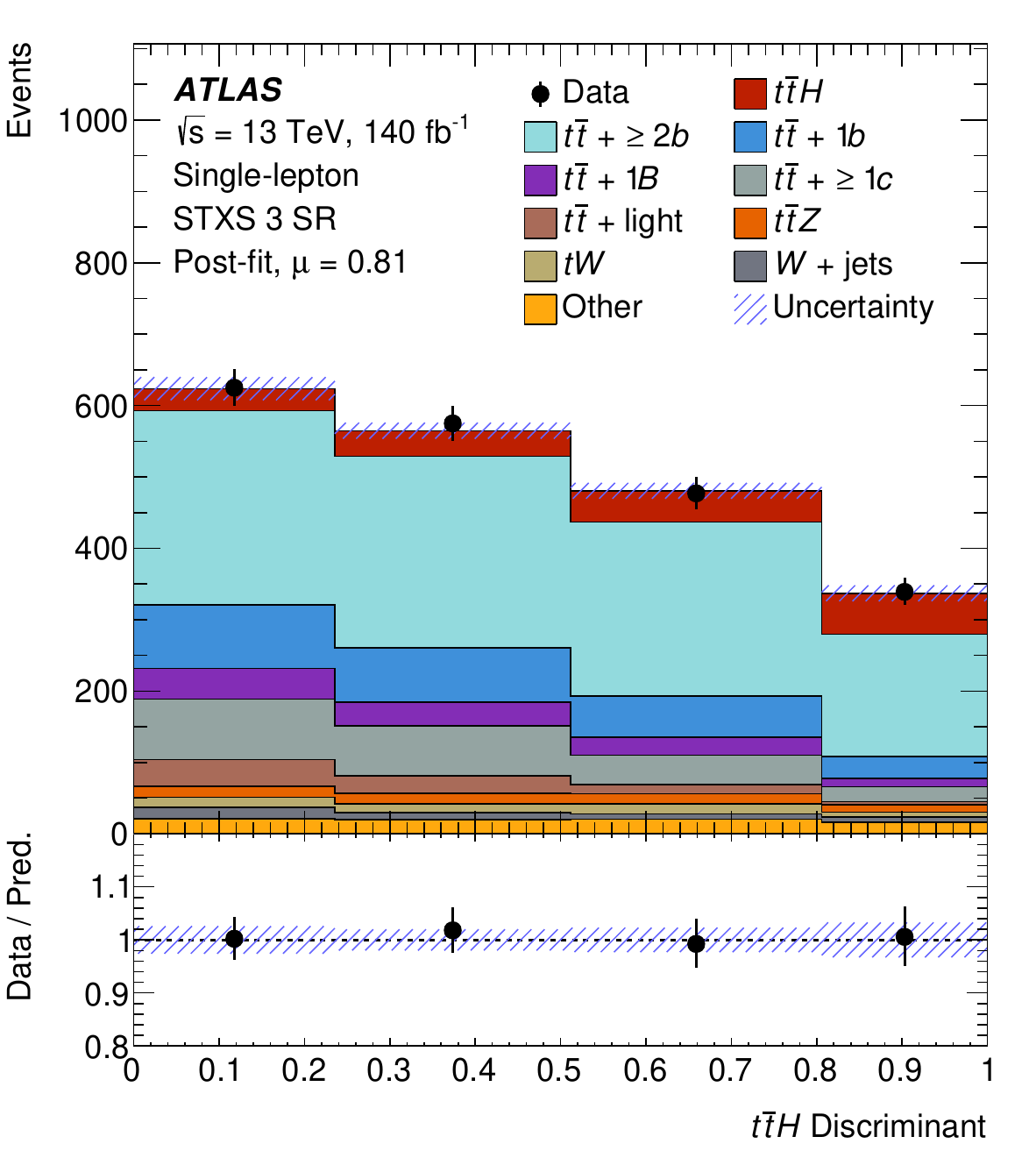}}\label{fig:lj_postfit_SRs_3}\\
\subfigure[]{\includegraphics[width=0.3\textwidth]{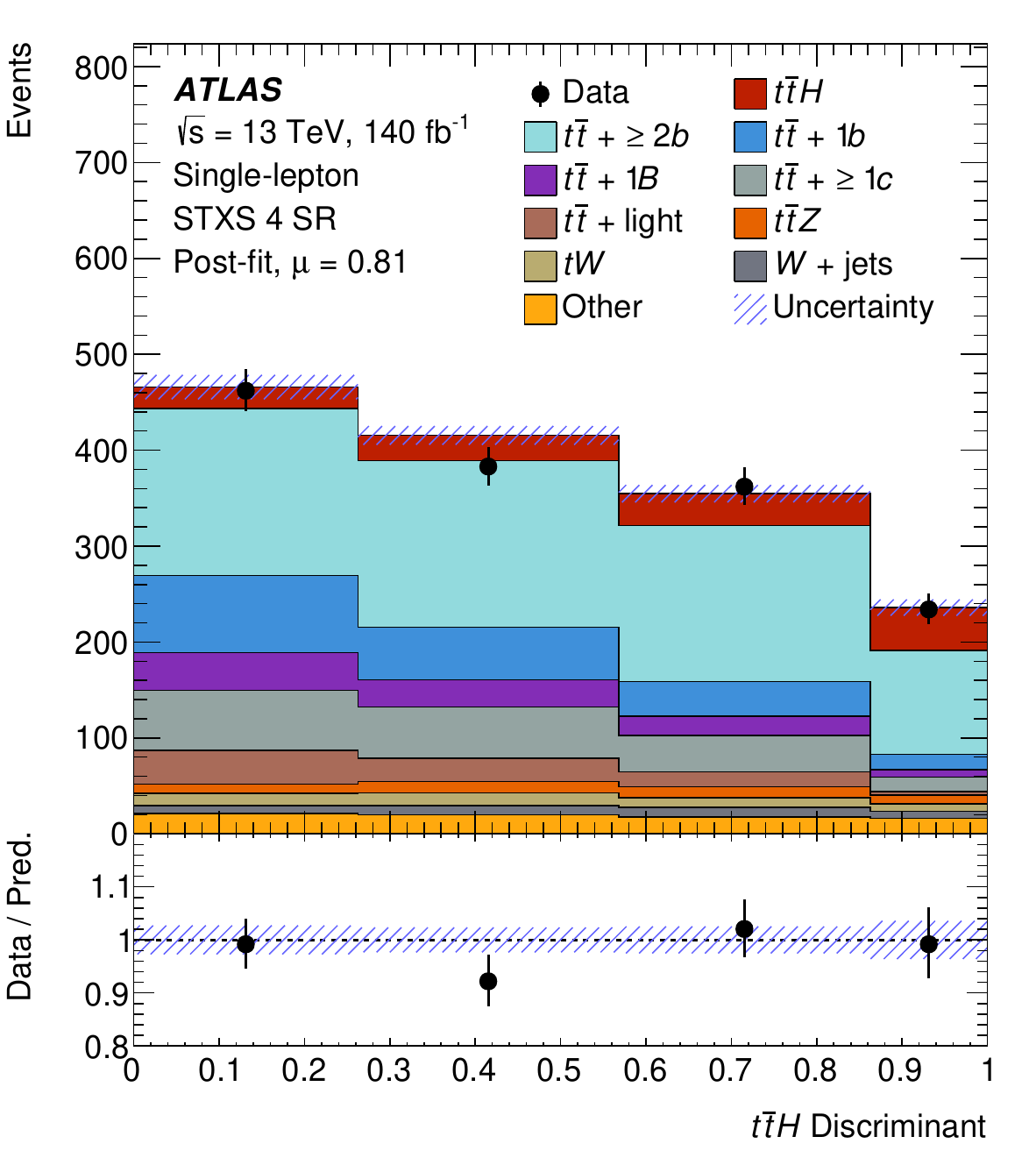}}\label{fig:lj_postfit_SRs_4}
\subfigure[]{\includegraphics[width=0.3\textwidth]{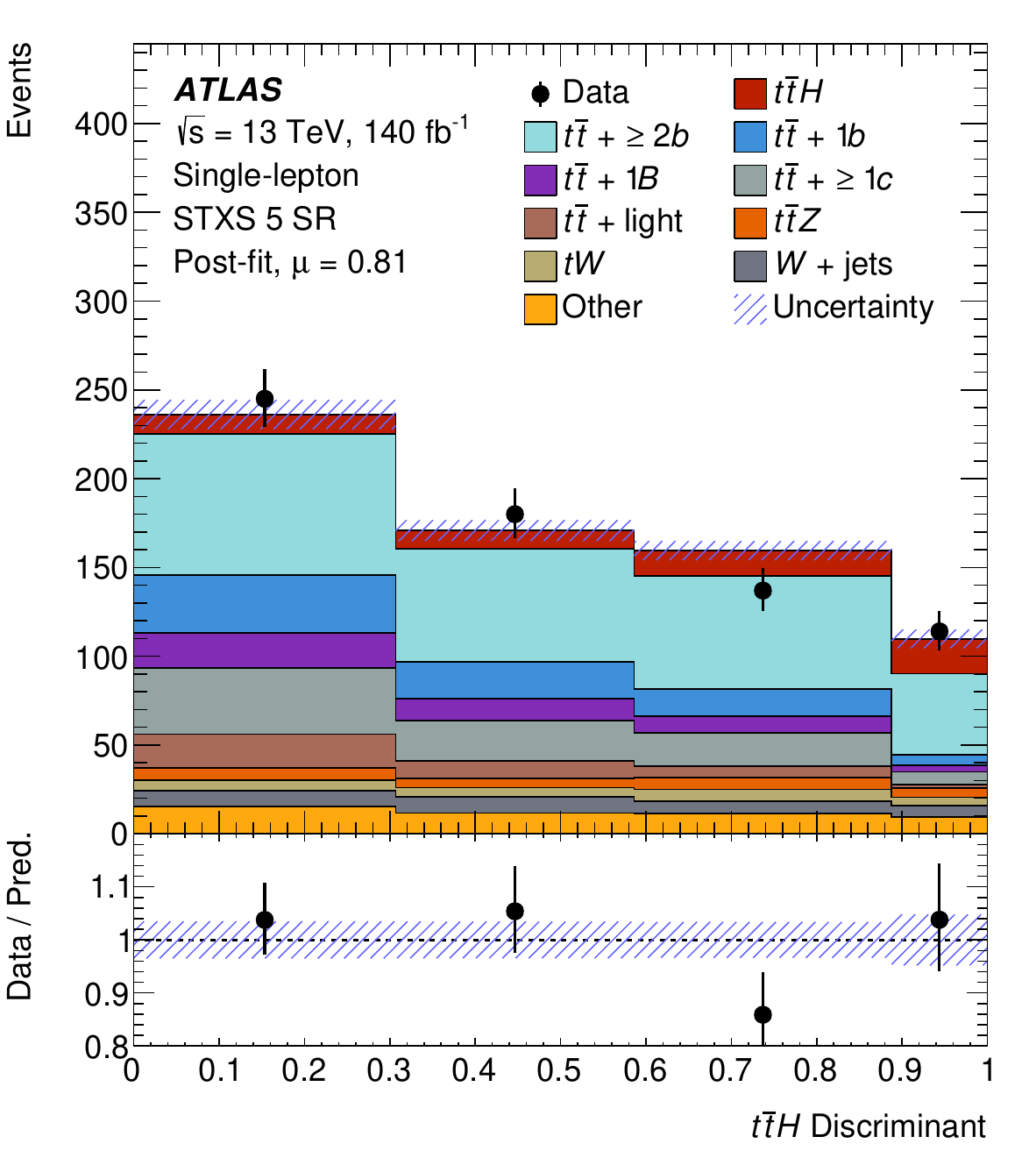}}\label{fig:lj_postfit_SRs_5}
\subfigure[]{\includegraphics[width=0.3\textwidth]{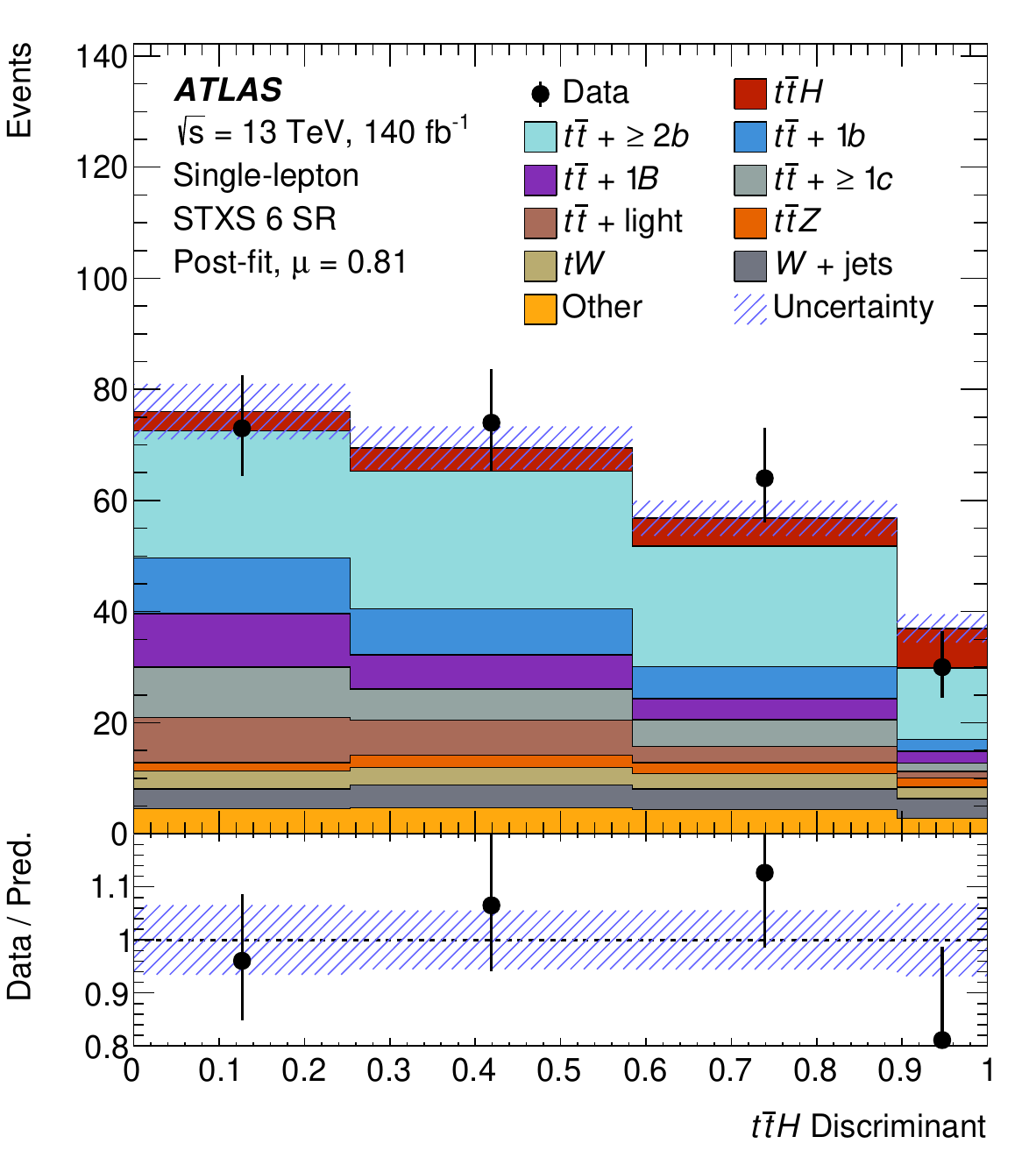}}\label{fig:lj_postfit_SRs_6}\\
\subfigure[]{\includegraphics[width=0.3\textwidth]{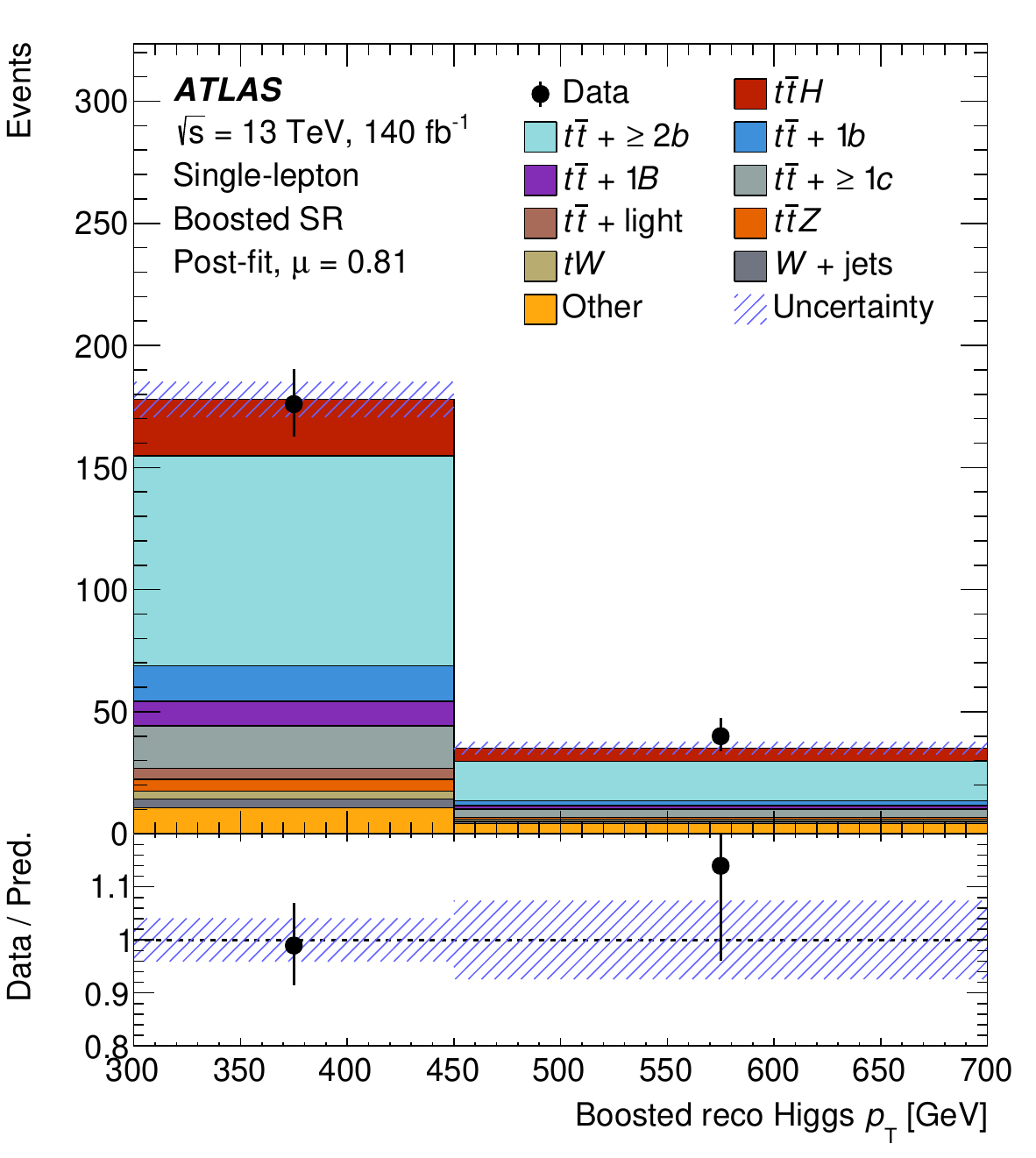}}\label{fig:lj_postfit_SRs_boost}
\caption{
Post-fit data/prediction comparisons of the signal region distributions entering the fit in the single-lepton channel.
The discriminants are rescaled to lie between zero and one using a logistic function.
The signal and background normalisation parameters and the nuisance parameters are set to their best-fit values. %
The uncertainty band includes all uncertainties and their correlations.
For the reconstructed Higgs boson \pt in the boosted signal region, the first (last) bin includes the underflow (overflow) contributions.
\label{fig:lj_postfit_SRs}
}
\end{figure}


\begin{figure}[t!]
\centering
\subfigure[]{\includegraphics[width=0.3\textwidth]{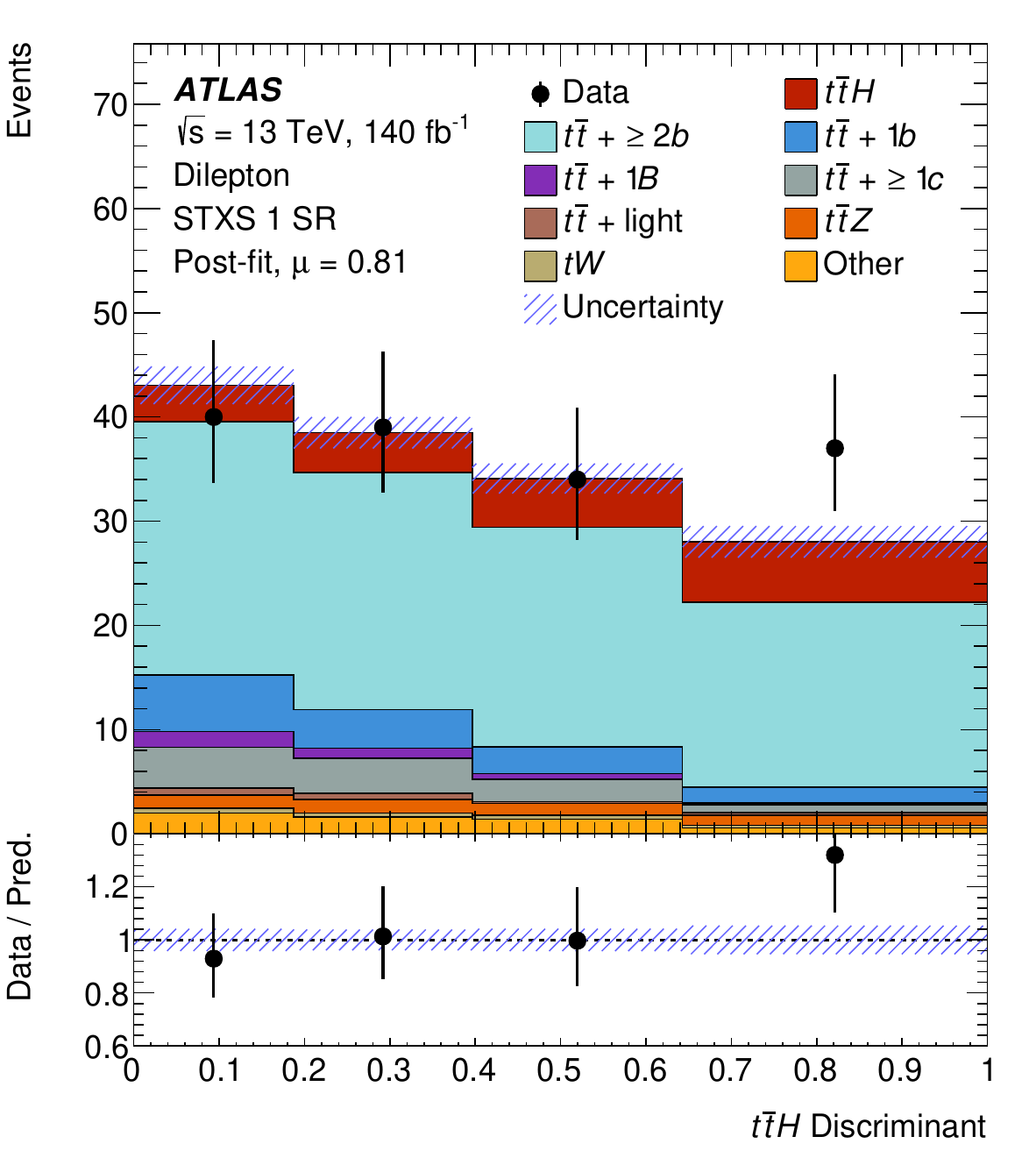}}\label{fig:dil_postfit_SRs_1}
\subfigure[]{\includegraphics[width=0.3\textwidth]{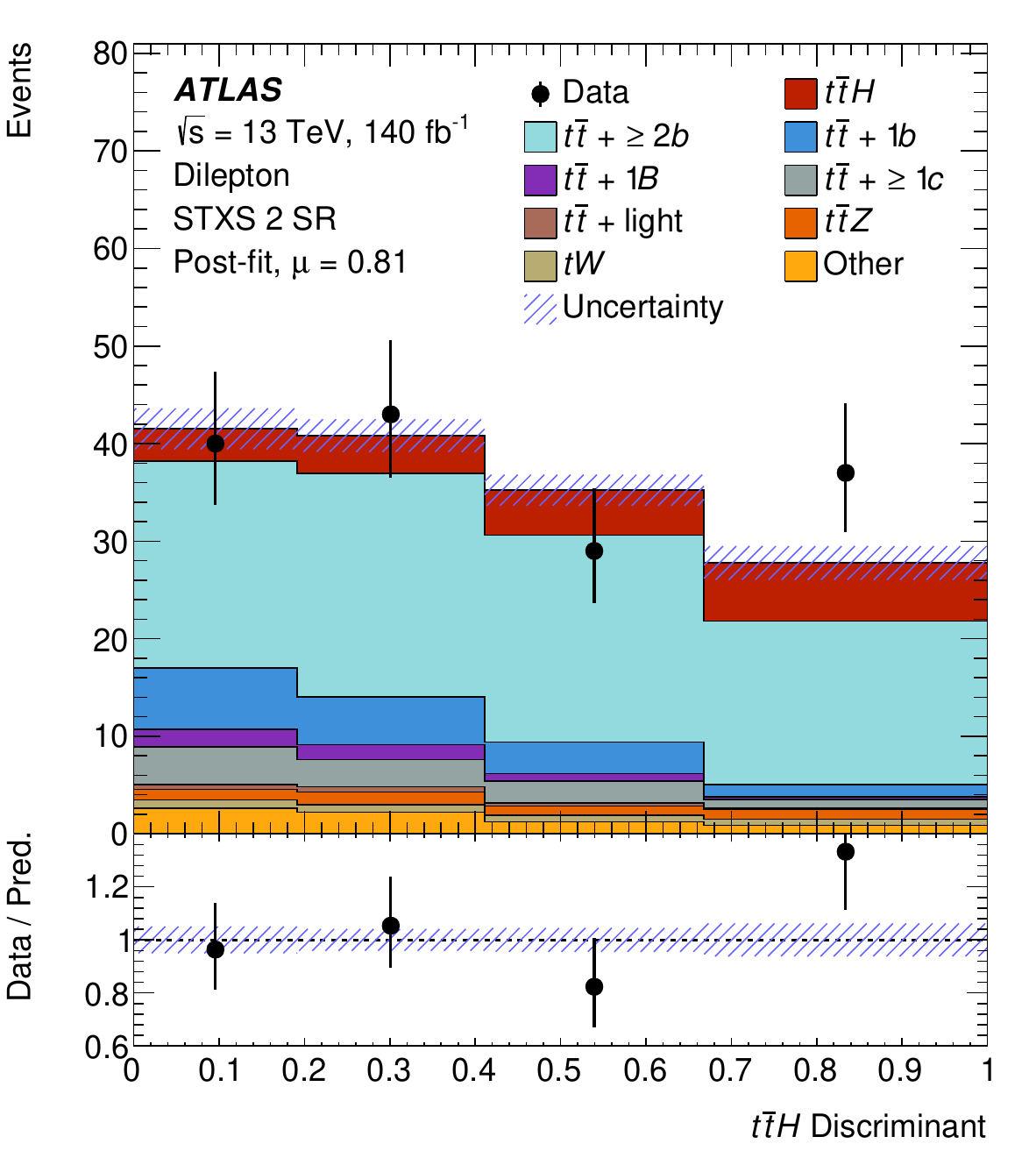}}\label{fig:dil_postfit_SRs_2}
\subfigure[]{\includegraphics[width=0.3\textwidth]{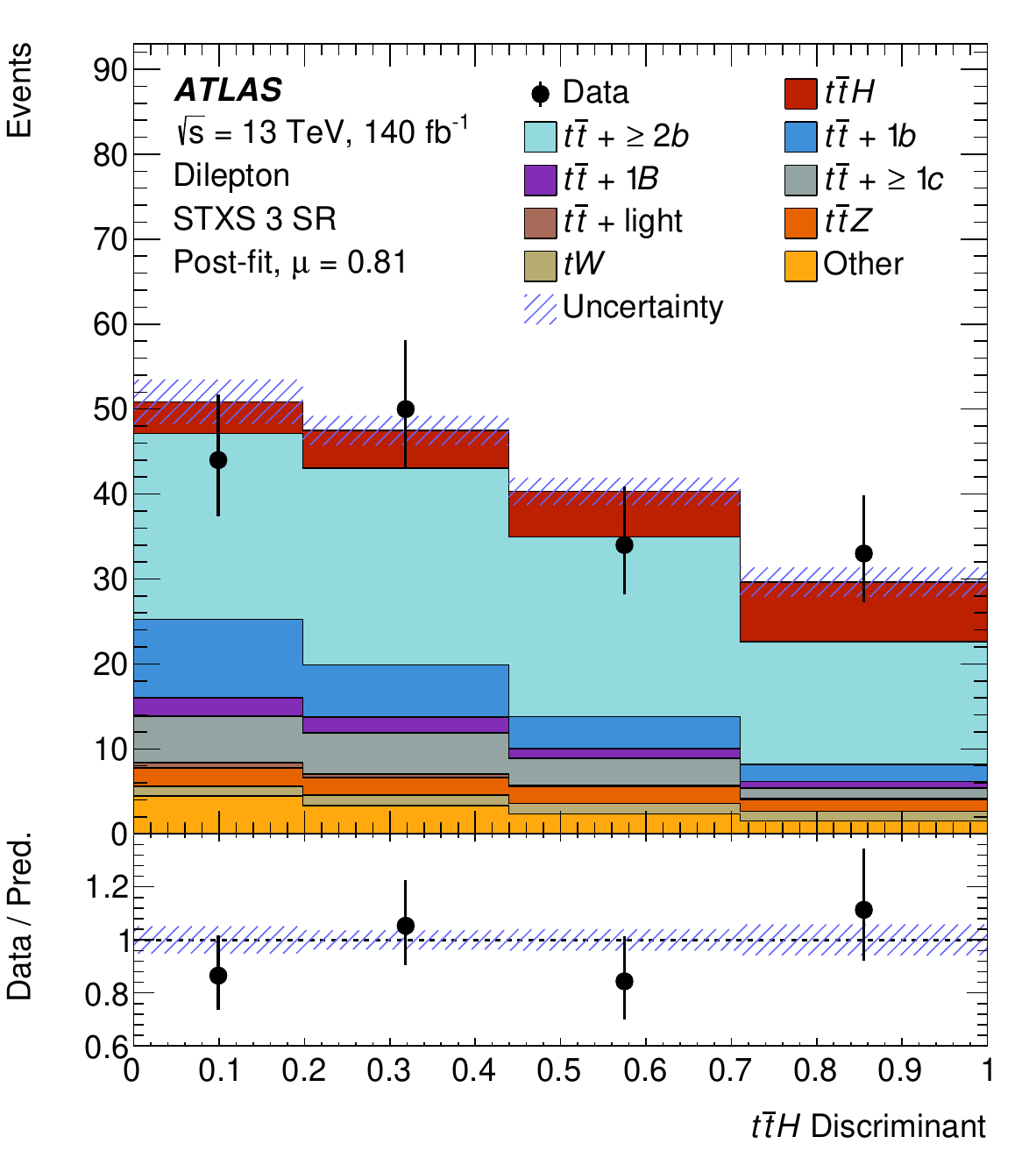}}\label{fig:dil_postfit_SRs_3}\\
\subfigure[]{\includegraphics[width=0.3\textwidth]{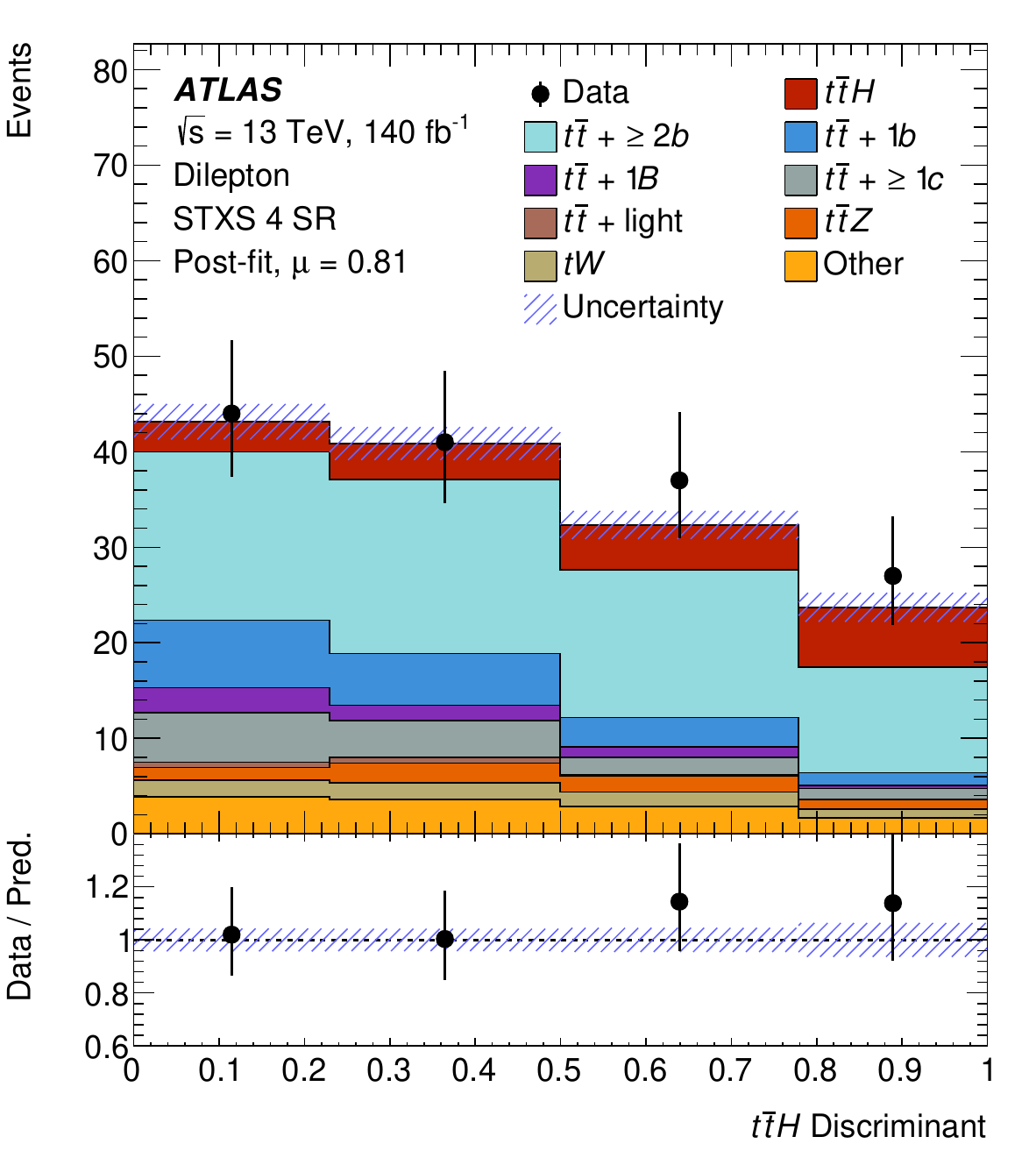}}\label{fig:dil_postfit_SRs_4}
\subfigure[]{\includegraphics[width=0.3\textwidth]{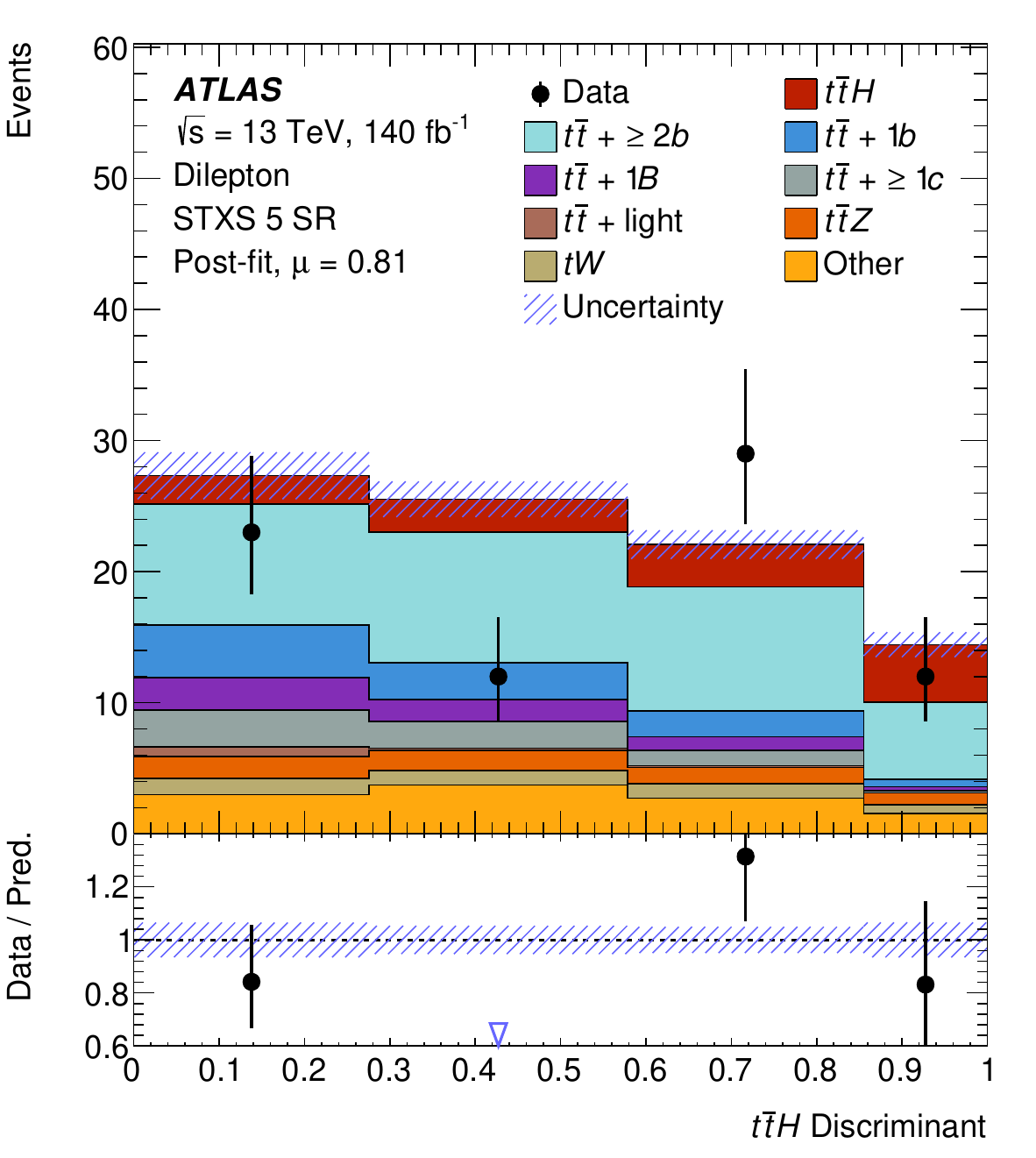}}\label{fig:dil_postfit_SRs_5}
\subfigure[]{\includegraphics[width=0.3\textwidth]{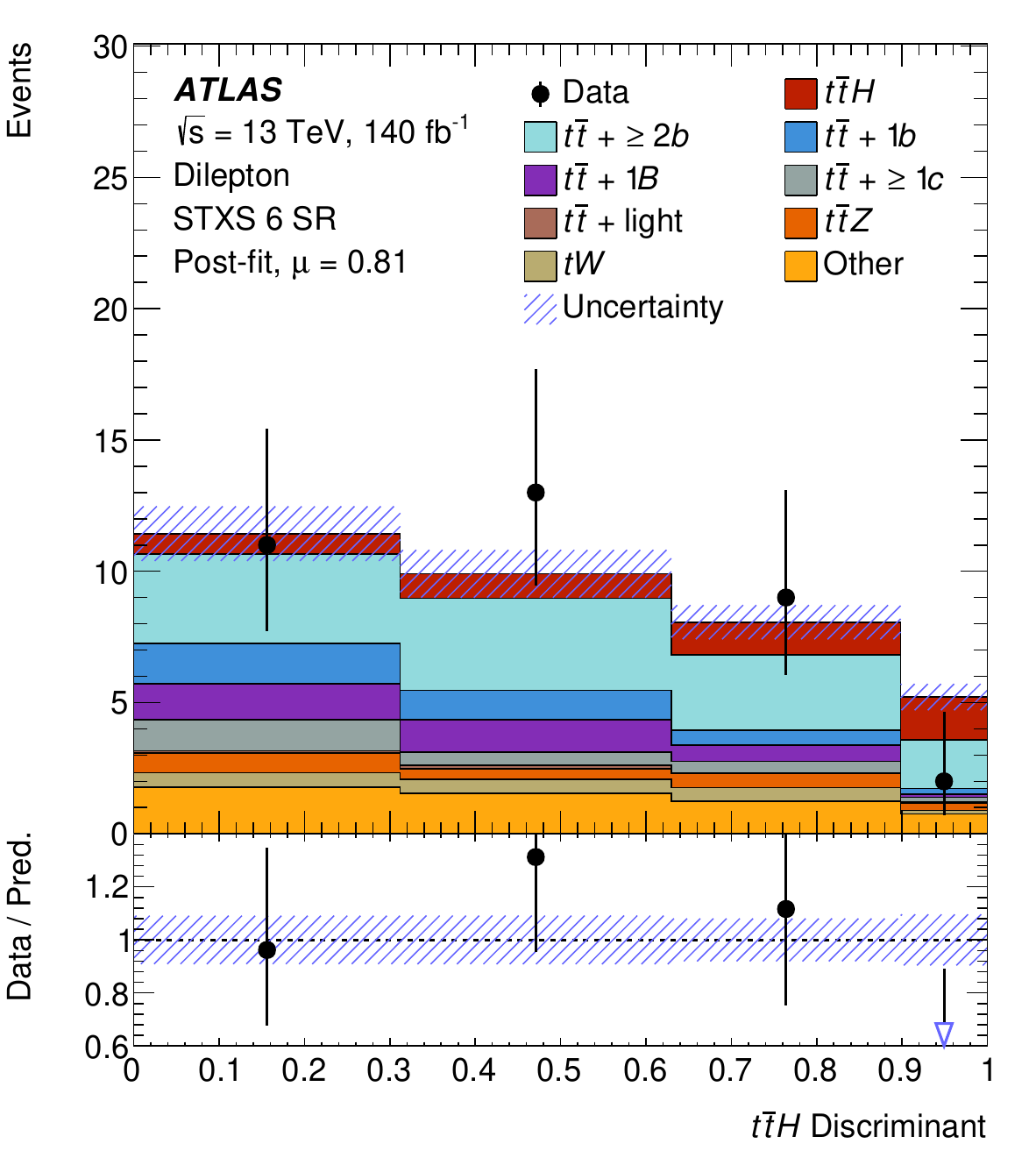}}\label{fig:dil_postfit_SRs_6}
\caption{
Post-fit data/prediction comparisons of the signal region distributions entering the fit in the dilepton channel.
The discriminants are rescaled to lie between zero and one using a logistic function.
The signal and background normalisation parameters and the nuisance parameters are set to their best-fit values. %
The uncertainty band includes all uncertainties and their correlations.
\label{fig:dil_postfit_SRs}
}
\end{figure}


\begin{figure}[t!]
\centering
\subfigure[]{\includegraphics[width=0.3\textwidth]{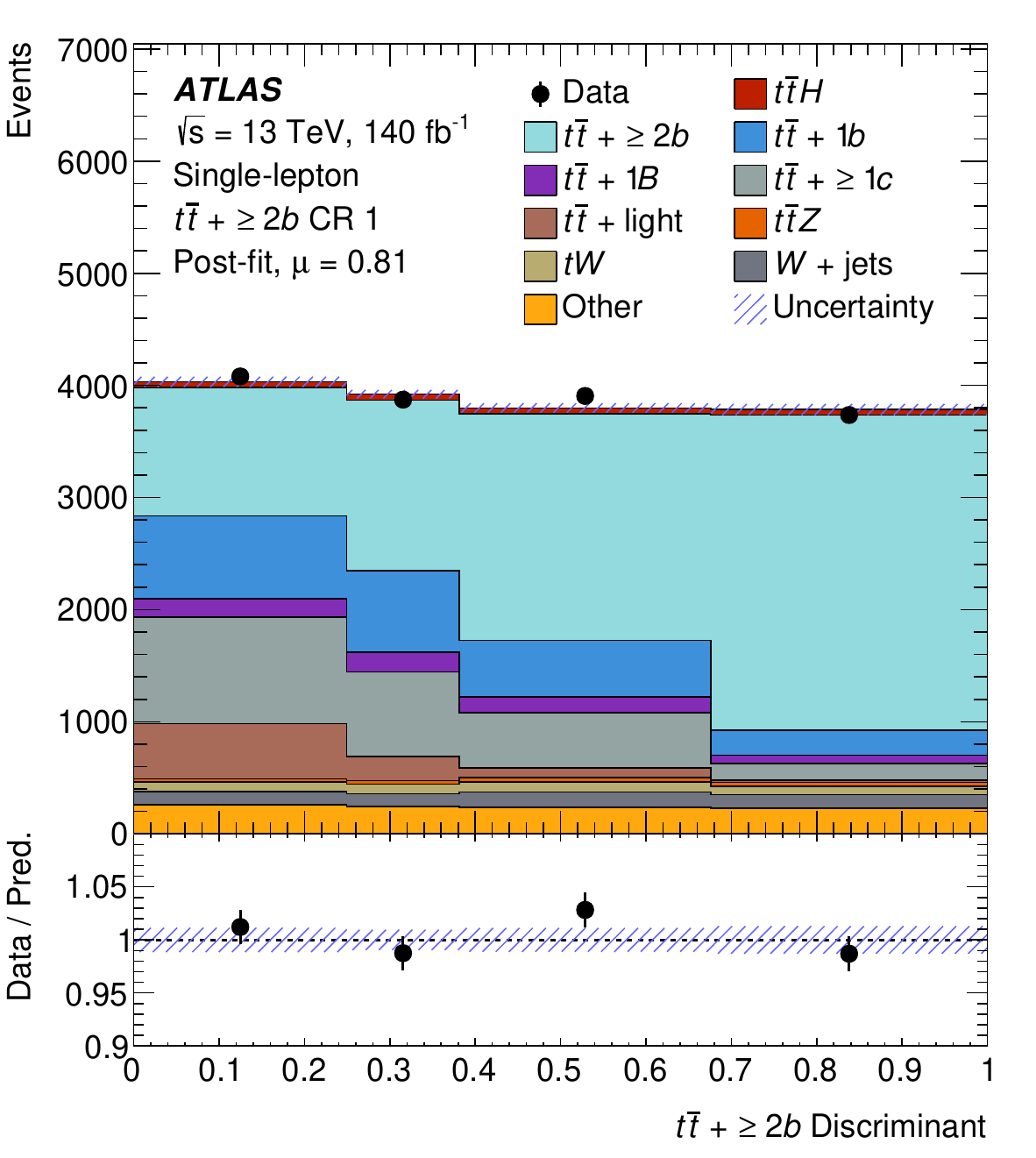}}\label{fig:lj_postfit_CRs_tt2b_1}
\subfigure[]{\includegraphics[width=0.3\textwidth]{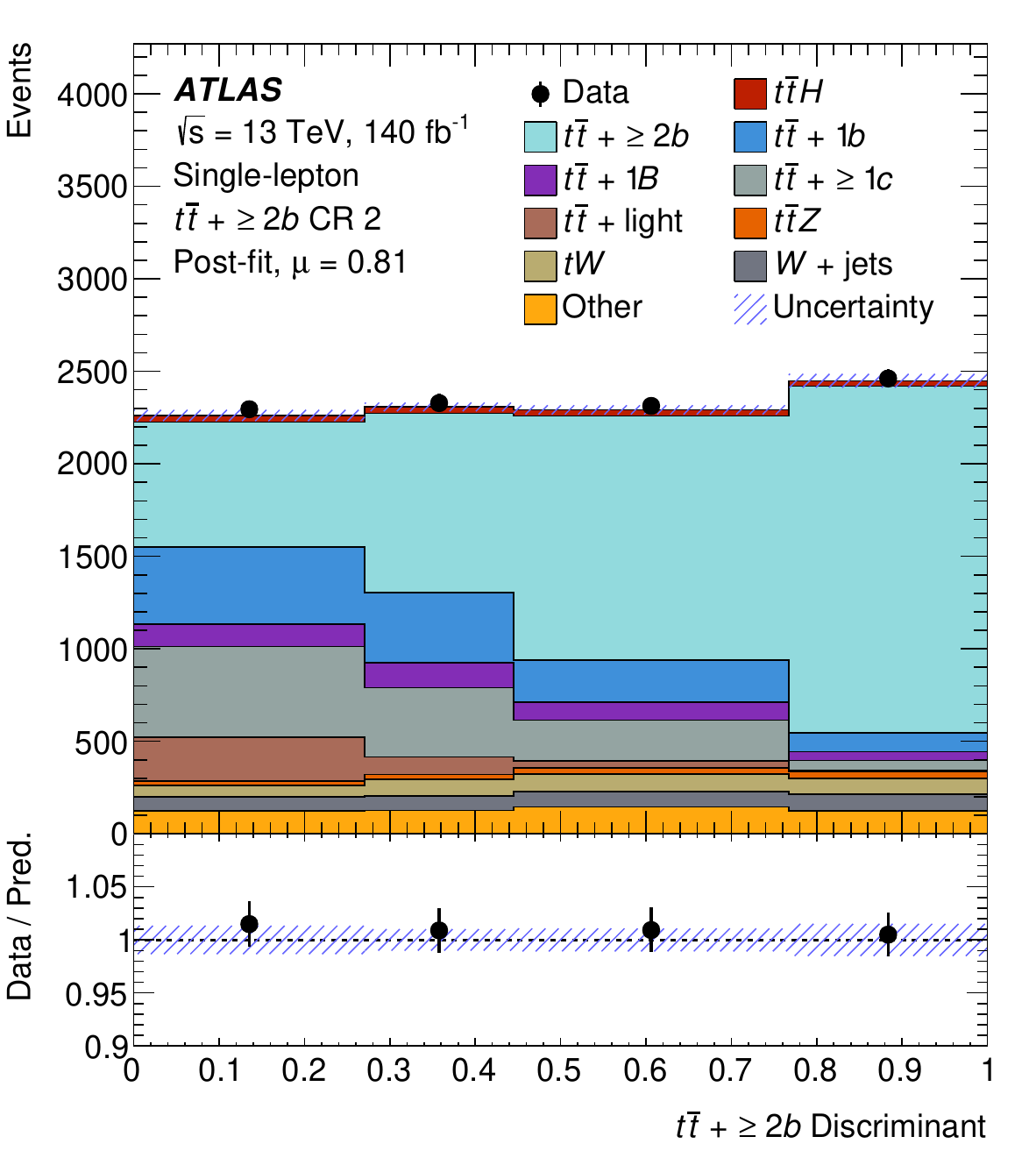}}\label{fig:lj_postfit_CRs_tt2b_2}
\subfigure[]{\includegraphics[width=0.3\textwidth]{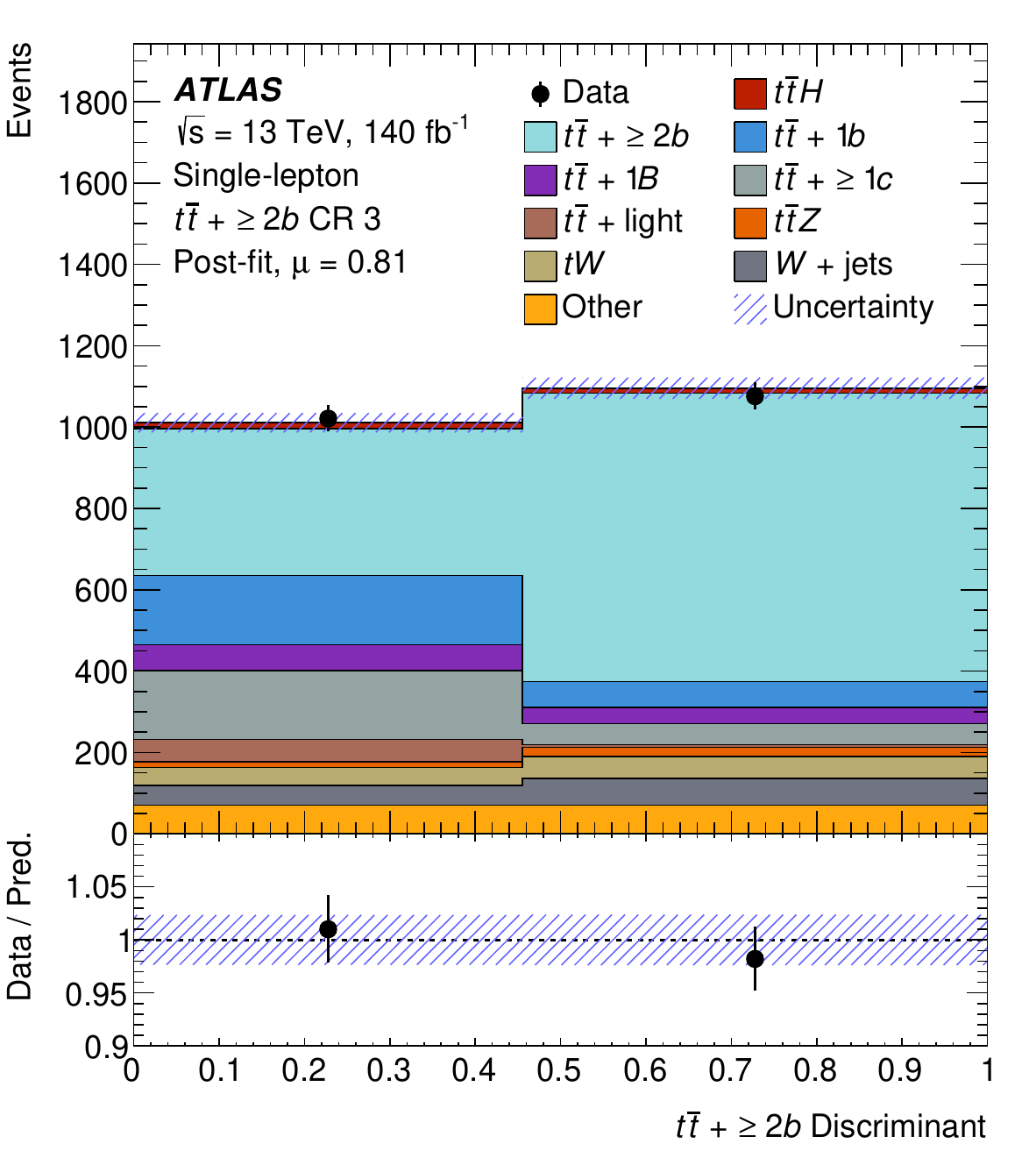}}\label{fig:lj_postfit_CRs_tt2b_3}
\subfigure[]{\includegraphics[width=0.3\textwidth]{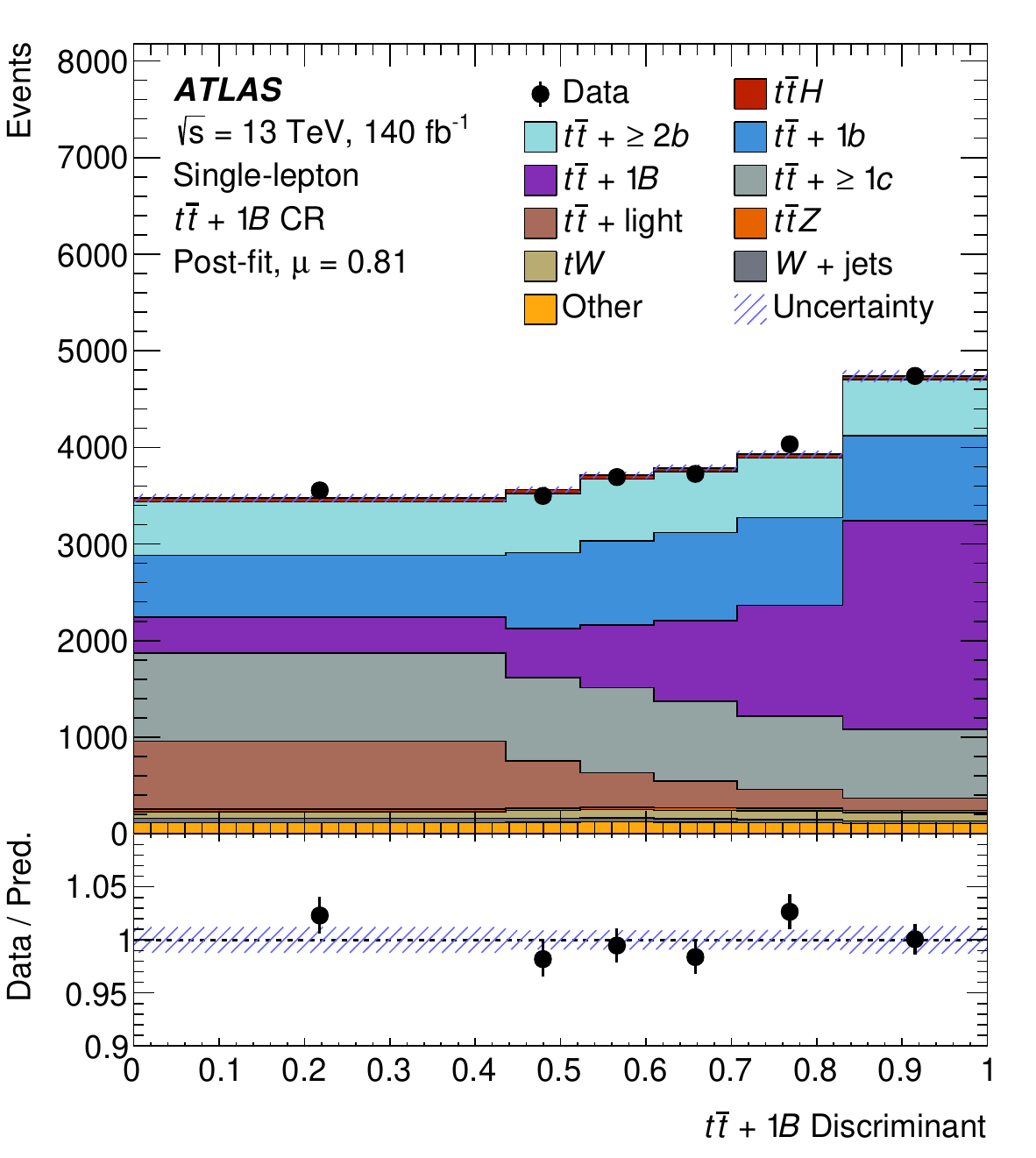}}\label{fig:lj_postfit_CRs_ttB}
\subfigure[]{\includegraphics[width=0.3\textwidth]{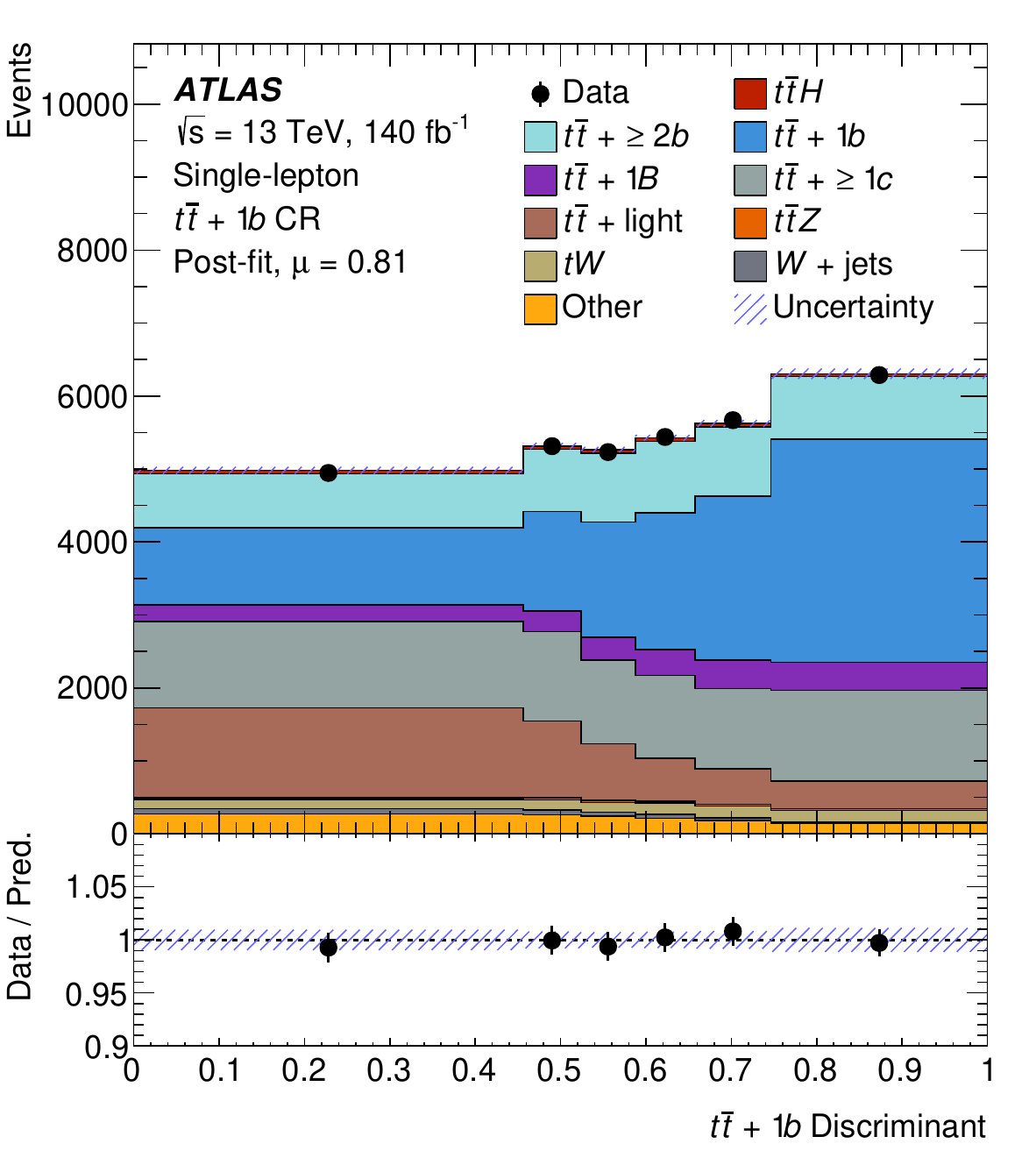}}\label{fig:lj_postfit_CRs_tt1b}\\
\subfigure[]{\includegraphics[width=0.3\textwidth]{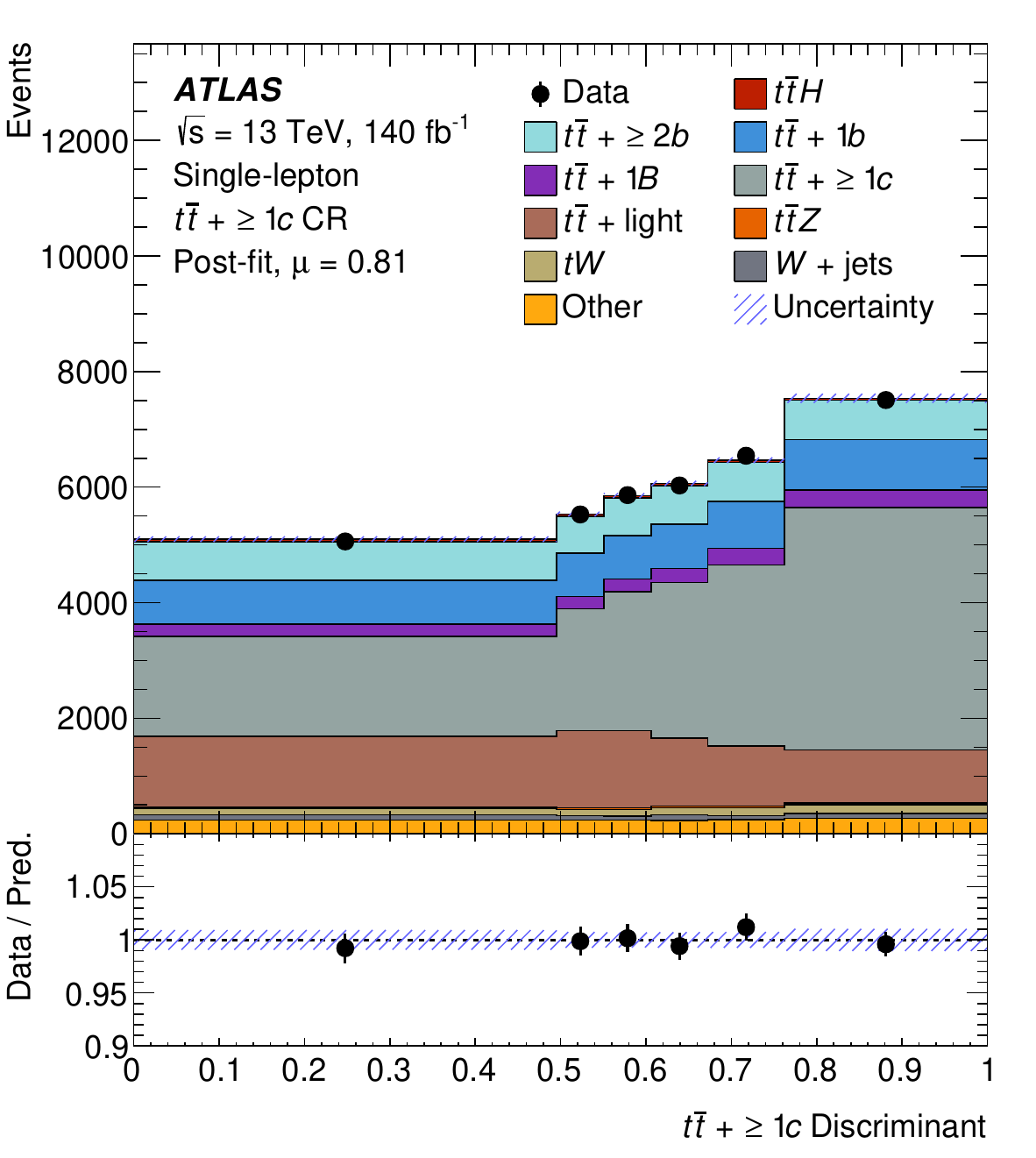}}\label{fig:lj_postfit_CRs_ttc}
\subfigure[]{\includegraphics[width=0.3\textwidth]{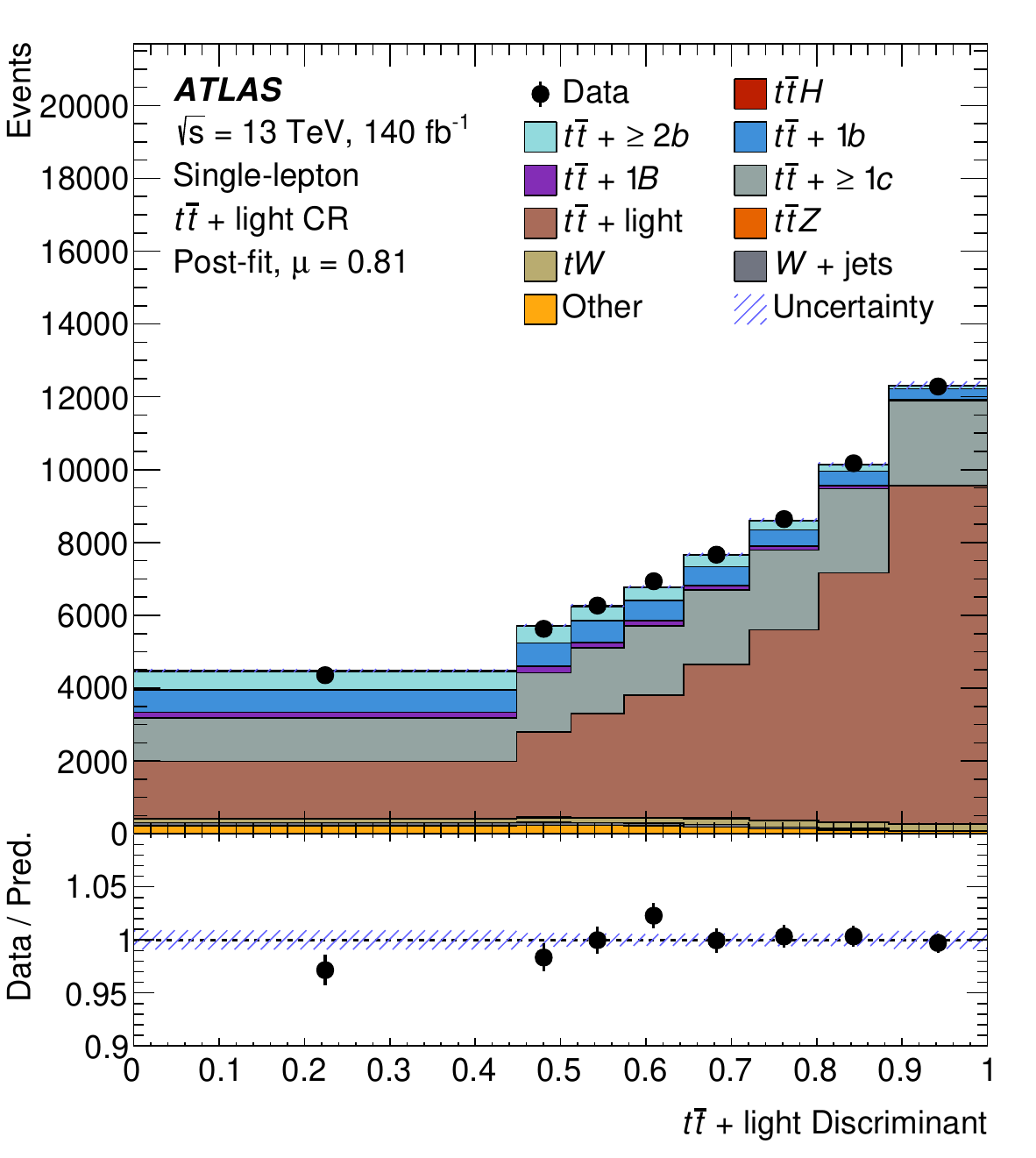}}\label{fig:lj_postfit_CRs_ttlight}
\subfigure[]{\includegraphics[width=0.3\textwidth]{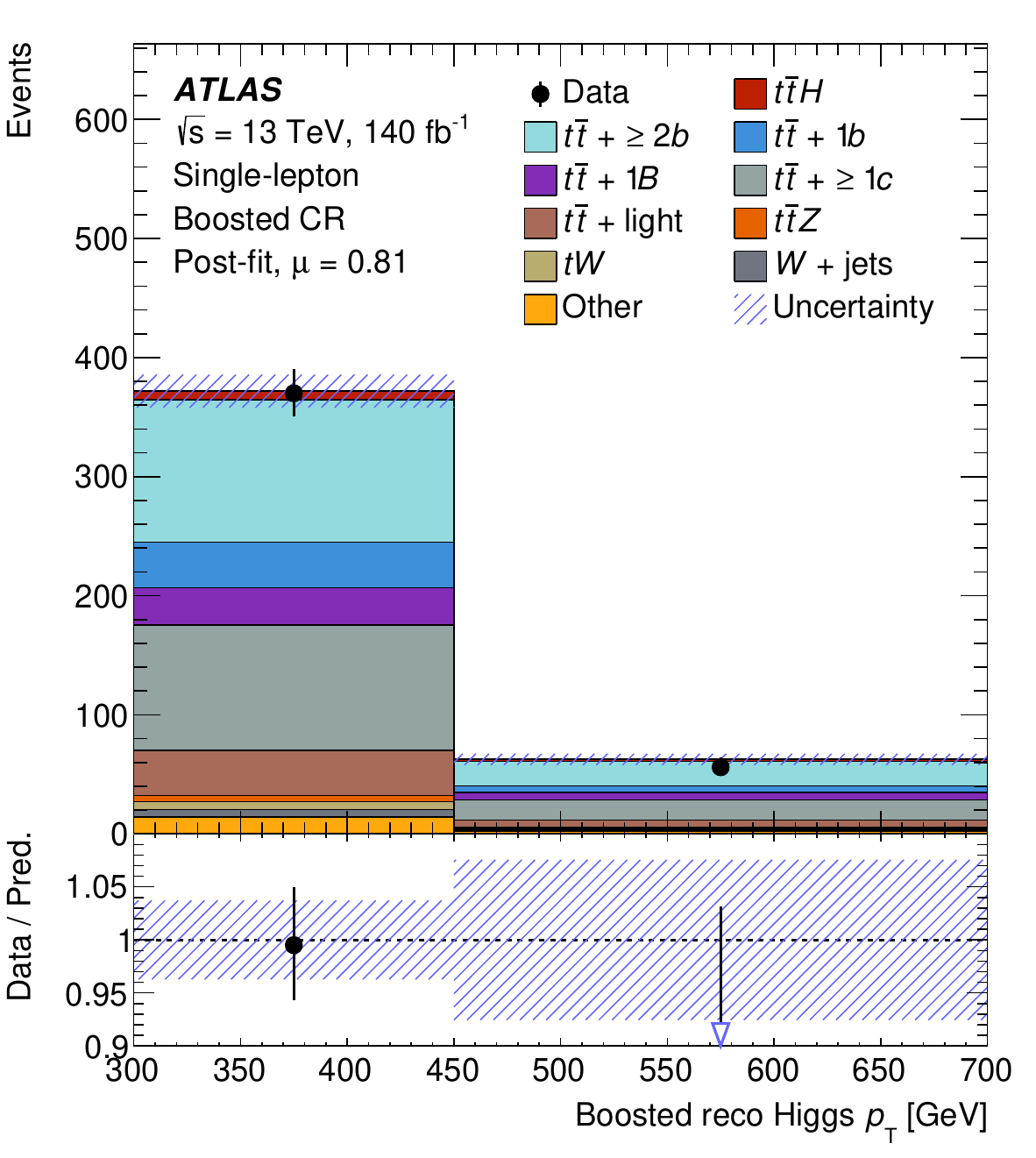}}\label{fig:lj_postfit_CRs_boost}
\caption{
Post-fit data/prediction comparisons of the control region distributions entering the fit in the single-lepton channel.
The discriminants are rescaled to lie between zero and one using a logistic function.
The signal and background normalisation parameters and the nuisance parameters are set to their best-fit values. %
The uncertainty band includes all uncertainties and their correlations.
For the reconstructed Higgs boson \pt in the boosted control region, the first (last) bin includes the underflow (overflow) contributions.
\label{fig:lj_postfit_CRs}
}
\end{figure}


\begin{figure}[t!]
\centering
\subfigure[]{\includegraphics[width=0.3\textwidth]{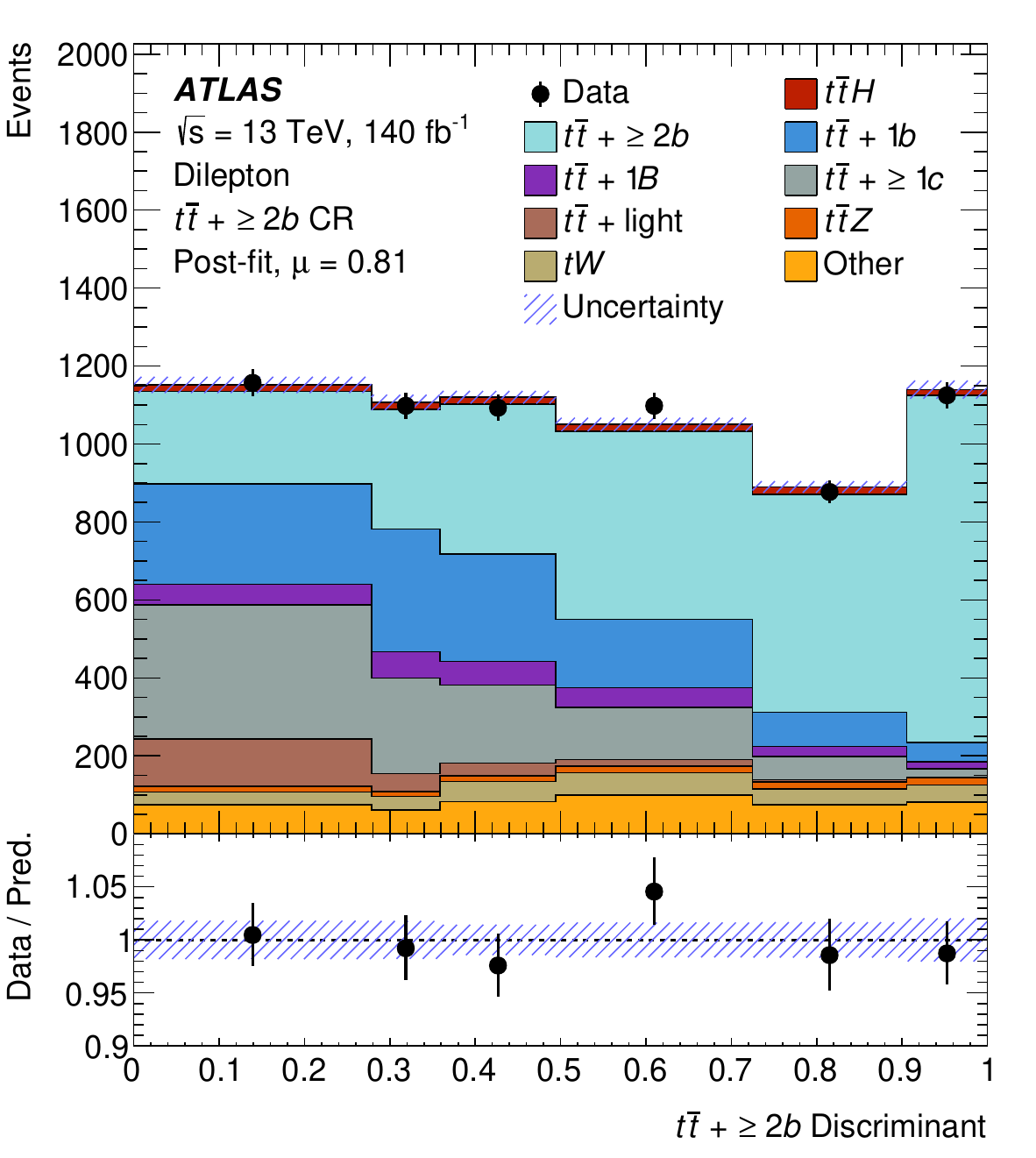}}\label{fig:dil_postfit_CRs_tt2b}
\subfigure[]{\includegraphics[width=0.3\textwidth]{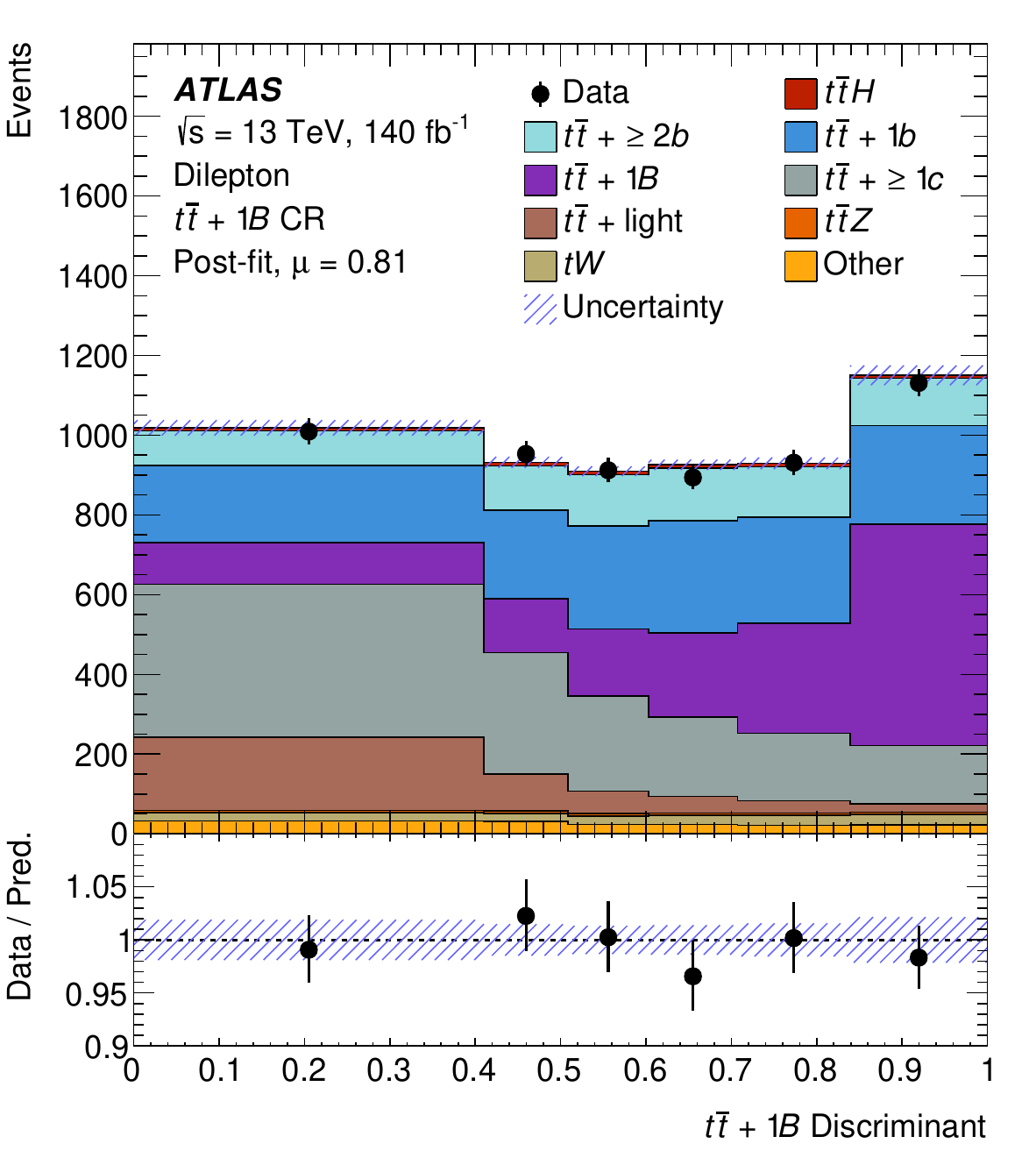}}\label{fig:dil_postfit_CRs_ttB}
\subfigure[]{\includegraphics[width=0.3\textwidth]{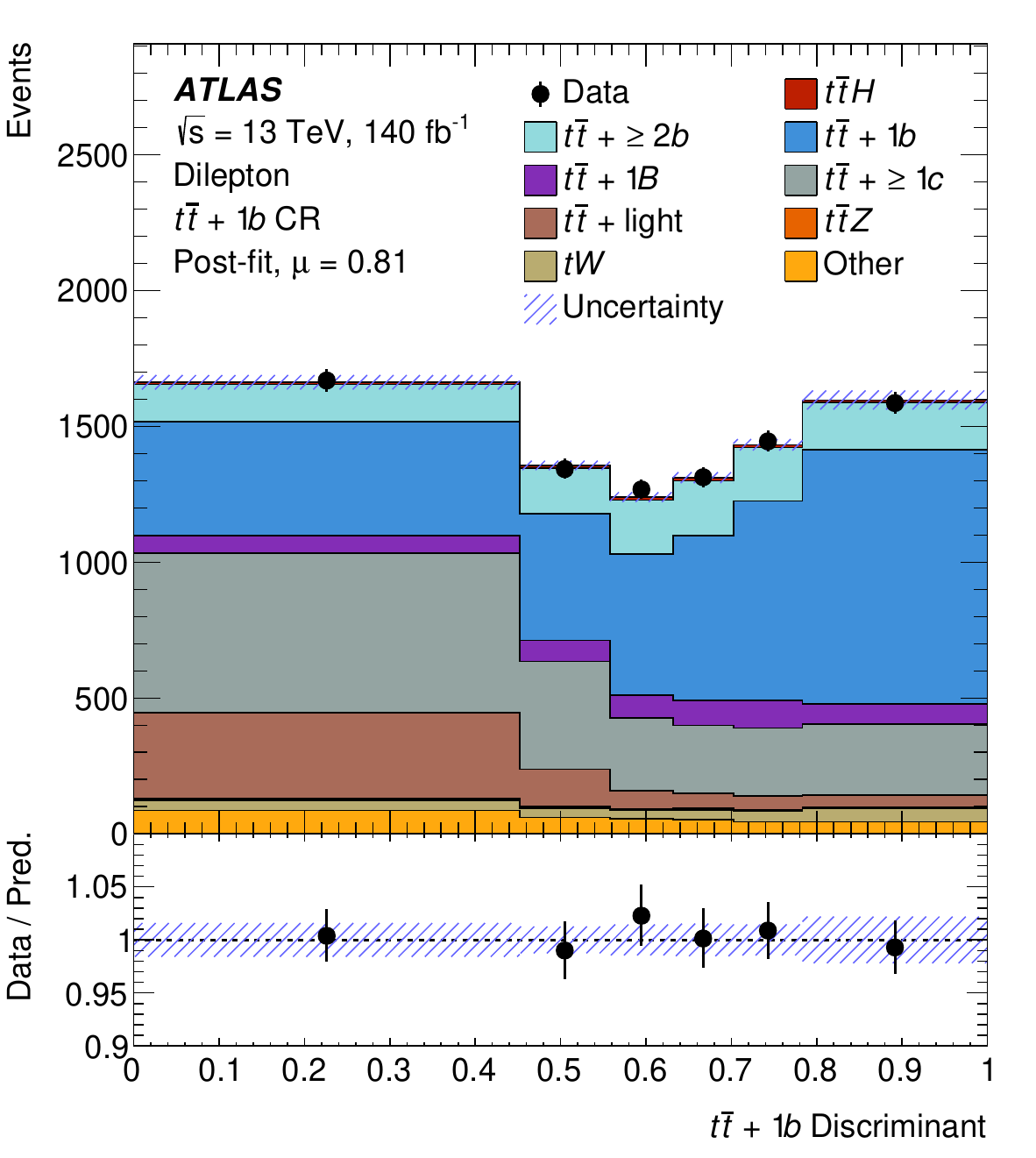}}\label{fig:dil_postfit_CRs_tt1b}\\
\subfigure[]{\includegraphics[width=0.3\textwidth]{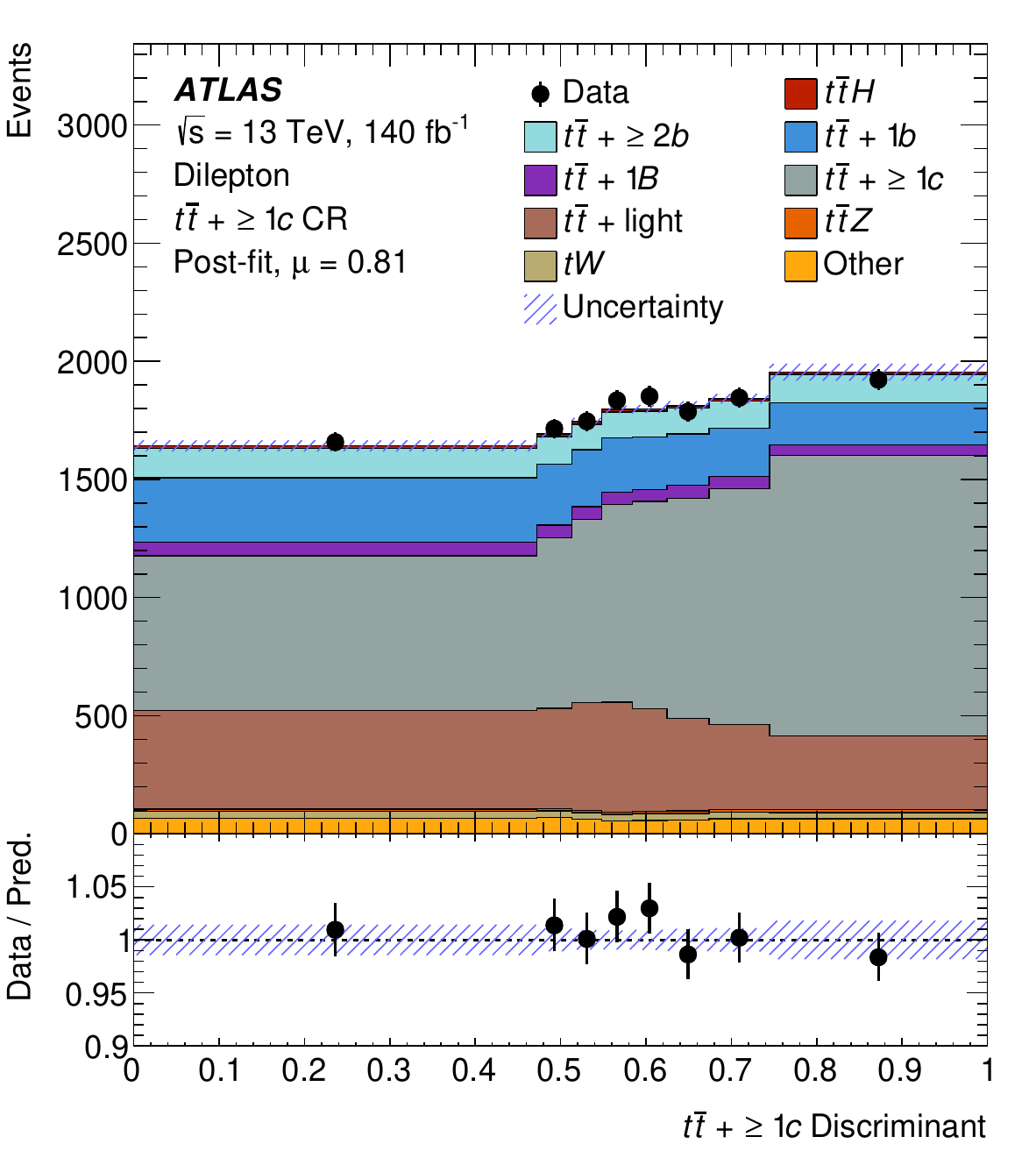}}\label{fig:dil_postfit_CRs_ttc}
\subfigure[]{\includegraphics[width=0.3\textwidth]{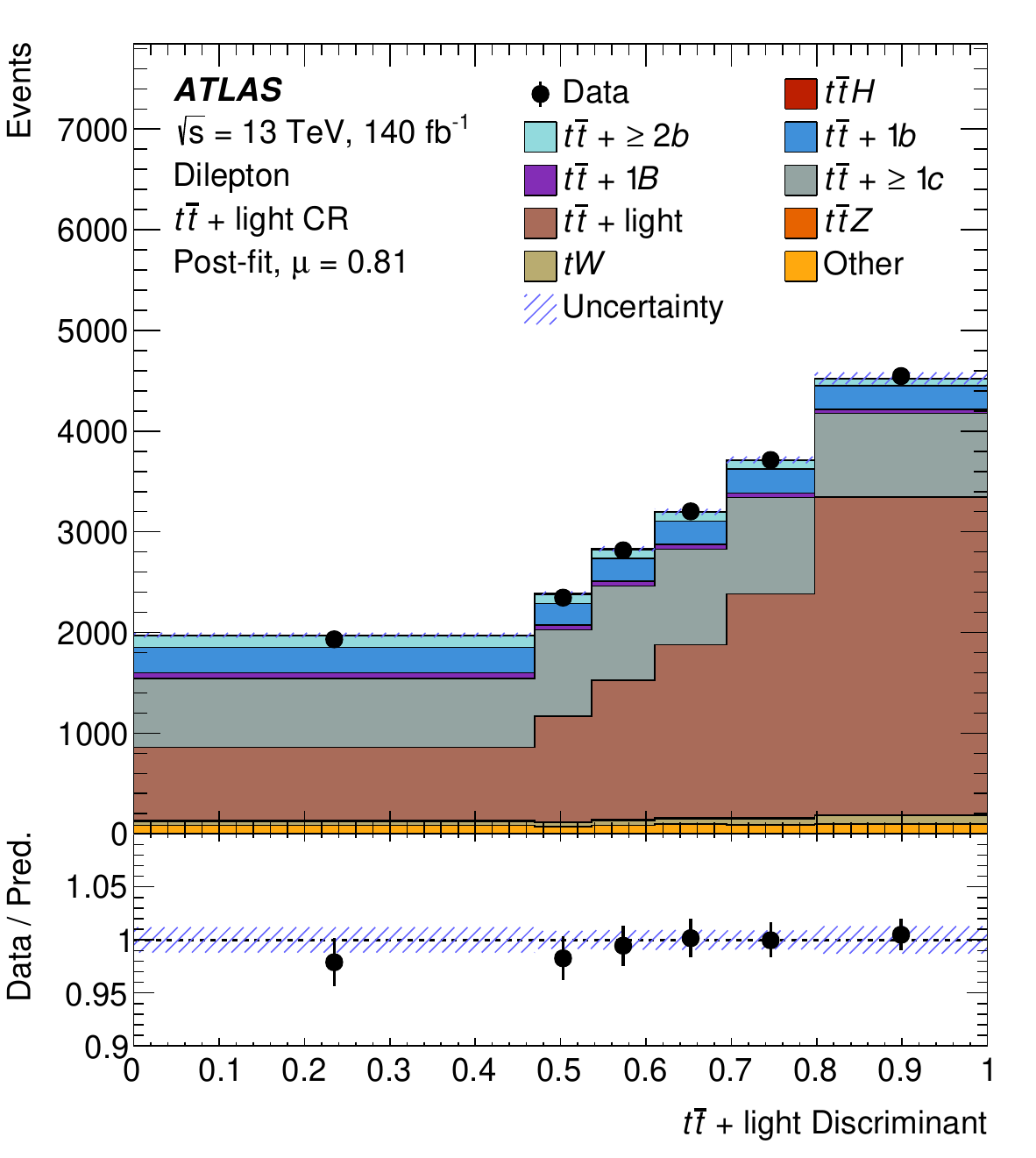}}\label{fig:dil_postfit_CRs_ttlight}
\caption{
Post-fit data/prediction comparisons of the control region distributions entering the fit in the dilepton channel.
The discriminants are rescaled to lie between zero and one using a logistic function.
The signal and background normalisation parameters and the nuisance parameters are set to their best-fit values. %
The uncertainty band includes all uncertainties and their correlations.
\label{fig:dil_postfit_CRs}
}
\end{figure}


\begin{figure}[t!]
\centering
\includegraphics[width=0.95\textwidth]{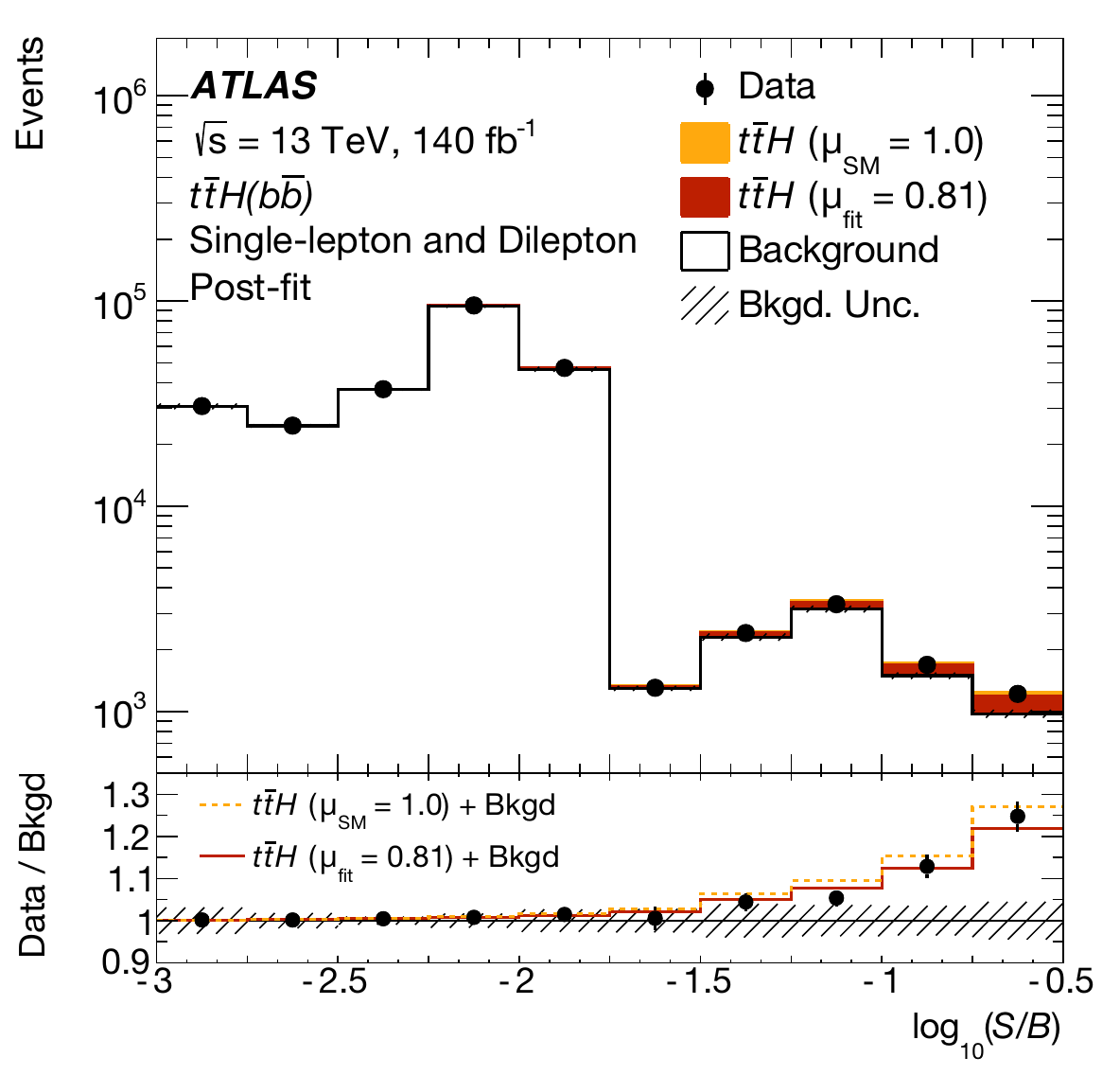}
\caption{Observed and expected event yields as a function of \mbox{$\log_{10}\left(S/B\right)$}, where $S$ and $B$ are the post-fit signal and total background yields, respectively. The bins in all fitted regions are ordered and grouped in bins of \mbox{$\log_{10}\left(S/B\right)$}. The signal is shown for the best-fit signal strength, $\mu=0.81$, or the SM prediction, $\mu=1.0$. The lower panel shows the ratio of the data to the post-fit background prediction, compared with the signal-plus-background prediction with the best-fit signal strength and the SM prediction. The shaded band represents the total uncertainty in the background prediction. The first (last) bin includes the underflow (overflow) contributions.}
\label{fig:SoverB}
\end{figure}

\FloatBarrier


%
\section{Conclusion}
\label{sec:conclusion}

The associated production of a Higgs boson with a pair of top quarks was measured in the $H\rightarrow b\Bar{b}$ decay channel, using the full Run 2 proton--proton collision dataset collected at
$\sqrt{s}=13$~\TeV by the ATLAS detector at the LHC, corresponding to 140 \ifb.
The measurement uses final states containing one or two leptons. The measured
inclusive cross-section assuming a Higgs boson mass of 125.09~\GeV is $\xsstatsyst$~fb with a relative uncertainty of 24\%, consistent with the SM prediction. The measurement is dominated by the systematic uncertainties arising from the \ttH signal modelling followed by the \ttjets background modelling.
The observed (expected) significance of the \ttH signal relative to the SM background-only hypothesis is 4.6 (5.4) standard deviations. The cross-section measurement is also performed in six bins of \ptH in the STXS framework with the highest bin probing Higgs bosons with \ptH above 450~\GeV. The uncertainty in the measurement for \ptH above 300~\GeV is dominated by the limited number of data events.

With respect to the previous iteration of the analysis, this measurement profits from looser selection requirements and improved \bjet identification that increase the \ttH signal acceptance. Control regions enriched in each of the \ttjets components are defined based on a more powerful multiclass neural network. Together with data-driven modelling corrections for the \ttcin and \ttlight components, and a dedicated MC simulation and systematic model for the \ttbin component, they provide improved signal sensitivity and better control over the background, such that  the modelling uncertainty in \ttbin is no longer the dominant contribution to the total systematic uncertainty.

This analysis is the most
precise \ttH cross-section measurement in a single decay channel, inclusively and in each \ptH bin.

\FloatBarrier


\FloatBarrier

\section*{Acknowledgements}
%

%
%

%
%

We thank CERN for the very successful operation of the LHC and its injectors, as well as the support staff at
CERN and at our institutions worldwide without whom ATLAS could not be operated efficiently.

The crucial computing support from all WLCG partners is acknowledged gratefully, in particular from CERN, the ATLAS Tier-1 facilities at TRIUMF/SFU (Canada), NDGF (Denmark, Norway, Sweden), CC-IN2P3 (France), KIT/GridKA (Germany), INFN-CNAF (Italy), NL-T1 (Netherlands), PIC (Spain), RAL (UK) and BNL (USA), the Tier-2 facilities worldwide and large non-WLCG resource providers. Major contributors of computing resources are listed in Ref.~\cite{ATL-SOFT-PUB-2023-001}.

We gratefully acknowledge the support of ANPCyT, Argentina; YerPhI, Armenia; ARC, Australia; BMWFW and FWF, Austria; ANAS, Azerbaijan; CNPq and FAPESP, Brazil; NSERC, NRC and CFI, Canada; CERN; ANID, Chile; CAS, MOST and NSFC, China; Minciencias, Colombia; MEYS CR, Czech Republic; DNRF and DNSRC, Denmark; IN2P3-CNRS and CEA-DRF/IRFU, France; SRNSFG, Georgia; BMBF, HGF and MPG, Germany; GSRI, Greece; RGC and Hong Kong SAR, China; ISF and Benoziyo Center, Israel; INFN, Italy; MEXT and JSPS, Japan; CNRST, Morocco; NWO, Netherlands; RCN, Norway; MNiSW, Poland; FCT, Portugal; MNE/IFA, Romania; MSTDI, Serbia; MSSR, Slovakia; ARIS and MVZI, Slovenia; DSI/NRF, South Africa; MICIU/AEI, Spain; SRC and Wallenberg Foundation, Sweden; SERI, SNSF and Cantons of Bern and Geneva, Switzerland; NSTC, Taipei; TENMAK, T\"urkiye; STFC/UKRI, United Kingdom; DOE and NSF, United States of America.

Individual groups and members have received support from BCKDF, CANARIE, CRC and DRAC, Canada; CERN-CZ, FORTE and PRIMUS, Czech Republic; COST, ERC, ERDF, Horizon 2020, ICSC-NextGenerationEU and Marie Sk{\l}odowska-Curie Actions, European Union; Investissements d'Avenir Labex, Investissements d'Avenir Idex and ANR, France; DFG and AvH Foundation, Germany; Herakleitos, Thales and Aristeia programmes co-financed by EU-ESF and the Greek NSRF, Greece; BSF-NSF and MINERVA, Israel; NCN and NAWA, Poland; La Caixa Banking Foundation, CERCA Programme Generalitat de Catalunya and PROMETEO and GenT Programmes Generalitat Valenciana, Spain; G\"{o}ran Gustafssons Stiftelse, Sweden; The Royal Society and Leverhulme Trust, United Kingdom.

In addition, individual members wish to acknowledge support from Armenia: Yerevan Physics Institute (FAPERJ); CERN: European Organization for Nuclear Research (CERN PJAS); Chile: Agencia Nacional de Investigaci\'on y Desarrollo (FONDECYT 1230812, FONDECYT 1230987, FONDECYT 1240864); China: Chinese Ministry of Science and Technology (MOST-2023YFA1605700), National Natural Science Foundation of China (NSFC - 12175119, NSFC 12275265, NSFC-12075060); Czech Republic: Czech Science Foundation (GACR - 24-11373S), Ministry of Education Youth and Sports (FORTE CZ.02.01.01/00/22\_008/0004632), PRIMUS Research Programme (PRIMUS/21/SCI/017); EU: H2020 European Research Council (ERC - 101002463); European Union: European Research Council (ERC - 948254, ERC 101089007), Horizon 2020 Framework Programme (MUCCA - CHIST-ERA-19-XAI-00), European Union, Future Artificial Intelligence Research (FAIR-NextGenerationEU PE00000013), Italian Center for High Performance Computing, Big Data and Quantum Computing (ICSC, NextGenerationEU); France: Agence Nationale de la Recherche (ANR-20-CE31-0013, ANR-21-CE31-0013, ANR-21-CE31-0022, ANR-22-EDIR-0002), Investissements d'Avenir Labex (ANR-11-LABX-0012); Germany: Baden-Württemberg Stiftung (BW Stiftung-Postdoc Eliteprogramme), Deutsche Forschungsgemeinschaft (DFG - 469666862, DFG - CR 312/5-2); Italy: Istituto Nazionale di Fisica Nucleare (ICSC, NextGenerationEU), Ministero dell'Università e della Ricerca (PRIN - 20223N7F8K - PNRR M4.C2.1.1); Japan: Japan Society for the Promotion of Science (JSPS KAKENHI JP22H01227, JSPS KAKENHI JP22H04944, JSPS KAKENHI JP22KK0227, JSPS KAKENHI JP23KK0245); Netherlands: Netherlands Organisation for Scientific Research (NWO Veni 2020 - VI.Veni.202.179); Norway: Research Council of Norway (RCN-314472); Poland: Ministry of Science and Higher Education (IDUB AGH, POB8, D4 no 9722), Polish National Agency for Academic Exchange (PPN/PPO/2020/1/00002/U/00001), Polish National Science Centre (NCN 2021/42/E/ST2/00350, NCN OPUS nr 2022/47/B/ST2/03059, NCN UMO-2019/34/E/ST2/00393, NCN \& H2020 MSCA 945339, UMO-2020/37/B/ST2/01043, UMO-2021/40/C/ST2/00187, UMO-2022/47/O/ST2/00148, UMO-2023/49/B/ST2/04085, UMO-2023/51/B/ST2/00920); Slovenia: Slovenian Research Agency (ARIS grant J1-3010); Spain: Generalitat Valenciana (Artemisa, FEDER, IDIFEDER/2018/048), Ministry of Science and Innovation (MCIN \& NextGenEU PCI2022-135018-2, MICIN \& FEDER PID2021-125273NB, RYC2019-028510-I, RYC2020-030254-I, RYC2021-031273-I, RYC2022-038164-I), PROMETEO and GenT Programmes Generalitat Valenciana (CIDEGENT/2019/027); Sweden: Carl Trygger Foundation (Carl Trygger Foundation CTS 22:2312), Swedish Research Council (Swedish Research Council 2023-04654, VR 2018-00482, VR 2022-03845, VR 2022-04683, VR 2023-03403, VR grant 2021-03651), Knut and Alice Wallenberg Foundation (KAW 2018.0157, KAW 2018.0458, KAW 2019.0447, KAW 2022.0358); Switzerland: Swiss National Science Foundation (SNSF - PCEFP2\_194658); United Kingdom: Leverhulme Trust (Leverhulme Trust RPG-2020-004), Royal Society (NIF-R1-231091); United States of America: U.S. Department of Energy (ECA DE-AC02-76SF00515), Neubauer Family Foundation.

%
%


%
%
%
%
%
%
%

%
\clearpage
\printbibliography
\clearpage
\input{atlas_authlist}
%

%
%

%

%
%
%
%

\end{document}

%% file: atlas_authlist.tex
 
\begin{flushleft}
\hypersetup{urlcolor=black}
{\Large The ATLAS Collaboration}

\bigskip

\AtlasOrcid[0000-0002-6665-4934]{G.~Aad}$^\textrm{\scriptsize 104}$,
\AtlasOrcid[0000-0001-7616-1554]{E.~Aakvaag}$^\textrm{\scriptsize 17}$,
\AtlasOrcid[0000-0002-5888-2734]{B.~Abbott}$^\textrm{\scriptsize 123}$,
\AtlasOrcid[0000-0002-0287-5869]{S.~Abdelhameed}$^\textrm{\scriptsize 119a}$,
\AtlasOrcid[0000-0002-1002-1652]{K.~Abeling}$^\textrm{\scriptsize 56}$,
\AtlasOrcid[0000-0001-5763-2760]{N.J.~Abicht}$^\textrm{\scriptsize 50}$,
\AtlasOrcid[0000-0002-8496-9294]{S.H.~Abidi}$^\textrm{\scriptsize 30}$,
\AtlasOrcid[0009-0003-6578-220X]{M.~Aboelela}$^\textrm{\scriptsize 45}$,
\AtlasOrcid[0000-0002-9987-2292]{A.~Aboulhorma}$^\textrm{\scriptsize 36e}$,
\AtlasOrcid[0000-0001-5329-6640]{H.~Abramowicz}$^\textrm{\scriptsize 155}$
% ...
\bigskip
\\

% ...
$^{17}$Department for Physics and Technology, University of Bergen, Bergen; Norway.\\
% ...
$^{30}$Physics Department, Brookhaven National Laboratory, Upton NY; United States of America.\\
% ...
$^{36}$$^{(a)}$Facult\'e des Sciences Ain Chock, Universit\'e Hassan II de Casablanca;$^{(b)}$Facult\'{e} des Sciences, Universit\'{e} Ibn-Tofail, K\'{e}nitra;$^{(c)}$Facult\'e des Sciences Semlalia, Universit\'e Cadi Ayyad, LPHEA-Marrakech;$^{(d)}$LPMR, Facult\'e des Sciences, Universit\'e Mohamed Premier, Oujda;$^{(e)}$Facult\'e des sciences, Universit\'e Mohammed V, Rabat;$^{(f)}$Institute of Applied Physics, Mohammed VI Polytechnic University, Ben Guerir; Morocco.\\
% ...
$^{45}$Physics Department, Southern Methodist University, Dallas TX; United States of America.\\
% ...
$^{50}$Fakult\"{a}t Physik , Technische Universit{\"a}t Dortmund, Dortmund; Germany.\\
% ...
$^{56}$II. Physikalisches Institut, Georg-August-Universit\"{a}t G\"ottingen, G\"ottingen; Germany.\\
% ...
$^{104}$CPPM, Aix-Marseille Universit\'e, CNRS/IN2P3, Marseille; France.\\
% ...
$^{119}$$^{(a)}$New York University Abu Dhabi, Abu Dhabi;$^{(b)}$United Arab Emirates University, Al Ain; United Arab Emirates.\\
% ...
$^{123}$Homer L. Dodge Department of Physics and Astronomy, University of Oklahoma, Norman OK; United States of America.\\
% ...
$^{155}$Raymond and Beverly Sackler School of Physics and Astronomy, Tel Aviv University, Tel Aviv; Israel.\\
% ...
$^{a}$ Also Affiliated with an institute covered by a cooperation agreement with CERN.\\
$^{b}$ Also at An-Najah National University, Nablus; Palestine.\\
$^{c}$ Also at Borough of Manhattan Community College, City University of New York, New York NY; United States of America.\\
$^{d}$ Also at Center for Interdisciplinary Research and Innovation (CIRI-AUTH), Thessaloniki; Greece.\\
$^{e}$ Also at CERN, Geneva; Switzerland.\\
$^{f}$ Also at CMD-AC UNEC Research Center, Azerbaijan State University of Economics (UNEC); Azerbaijan.\\
% ...

\end{flushleft}
